\documentclass[]{aastex}

\usepackage{emulateapj5}
\usepackage{onecolfloat}
\usepackage{graphicx} 
\usepackage{fancyheadings} 
\usepackage{ulem}
\usepackage{rotating}
\usepackage{lscape}

\newcommand{\tskip}{\tablevspace{1pt}}
\newcommand{\secdustdepl}{3}
\newcommand{\secdustobsc}{5}
\newcommand{\secZn}{2.1.6}

\newcommand{\ndla}{38~}

\newcommand{\delv}{\Delta v}

\newcommand{\lya}{Ly$\alpha$ }

\newcommand{\cm}[1]{\, {\rm cm^{#1}}}
\newcommand{\N}[1]{{N({\rm #1})}}

\newcommand{\rAA}{{\AA \enskip}}

\newcommand{\ltp}{\left ( \,}

\newcommand{\rtp}{\, \right  ) }

\newcommand{\smm}{\sum\limits}
\newcommand{\perd}{\;\;\; .}
\newcommand{\cmma}{\;\;\; ,}

\newcommand{\mkms}{{\rm \; km\;s^{-1}}}

\begin{document}

\twocolumn[%
\submitted{Accepted to the Astrophysical Journal Supplements: June 8, 2001}

\title{The UCSD HIRES/Keck~I Damped \lya Abundance Database: I. The Data}

\author{ Jason X. Prochaska\altaffilmark{1,2},
Arthur M. Wolfe\altaffilmark{1,3}, David Tytler\altaffilmark{1,3},
Scott Burles\altaffilmark{1,4}, Jeff Cooke\altaffilmark{1,3},
Eric Gawiser\altaffilmark{1,3}, David Kirkman\altaffilmark{1,3},
John M. O'Meara\altaffilmark{1,3}, \& Lisa Storrie-Lombardi\altaffilmark{1,5}}

\begin{abstract} 

We present new chemical abundance measurements of 16 damped \lya 
systems at $z>1.5$ and update our previous abundance analyses.  The 
entire database presented here was derived from HIRES observations on
the Keck~I telescope, reduced with the same software package, and
analysed with identical techniques.  Altogether, we present a large,
homogeneous database of chemical abundance measurements for protogalaxies
in the early universe, ideal for studying a number of important aspects
of galaxy formation.  In addition, we have established an online directory
for this database and will continuously update the results. 


\end{abstract}

]
\altaffiltext{1}{Visiting Astronomer, W.M. Keck Telescope.
The Keck Observatory is a joint facility of the University
of California and the California Institute of Technology.}
\altaffiltext{2}{The Observatories of the Carnegie Institute of Washington,
813 Santa Barbara St., Pasadena, CA 91101}
\altaffiltext{3}{Department of Physics, and Center for Astrophysics and 
Space Sciences, University of California, San Diego, C--0424, La Jolla, 
CA 92093-0424}
\altaffiltext{4}{Experimental Astrophysics Department, Fermi National
Accelerator Laboratory, MS 127, P.O. Box 500, 500 Wilson Rd., Batavia, IL
60637-1433}
\altaffiltext{5}{SIRTF Science Center, California Institute of 
Technology, MS 100-22, Pasadena, CA 91125}

\pagestyle{fancyplain}
\lhead[\fancyplain{}{\thepage}]{\fancyplain{}{PROCHASKA ET AL.}}
\rhead[\fancyplain{}{The UCSD HIRES/Keck~I Damped \lya Abundance Database I}]{\fancyplain{}{\thepage}}
\setlength{\headrulewidth=0pt}
\cfoot{}

\section{INTRODUCTION}
\label{sec-intro}

Since their discovery nearly twenty years ago \citep{wol86},
the damped \lya systems have provided an extraordinary means for
probing the properties of high redshift galaxies.  
For example, Wolfe and his collaborators have led a number of 
observing programs to trace the universal neutral baryonic content,
a study made possible by the fact that the damped \lya systems
are the dominant neutral gas reservoirs at every epoch 
\citep[e.g.][]{wol95,lzwt95,storr96,storr00}.  
Through accurate metallicity measurements, damped \lya
studies also provide an examination of the chemical enrichment history of
the universe (Pettini et al.\ 1994, 1997; Prochaska \& Wolfe 2000;
Prochaska, Gawiser, \& Wolfe 2001, hereafter PGW01).
In addition, the advent of high resolution echelle spectrographs on 10m class
telescopes have enabled researchers to trace the relative chemical
abundances of these systems, measurements which provide direct insight
into processes of nucleosynthesis and dust depletion in the early 
universe (Lu et al.\ 1996, hereafter L96; 
Prochaska \& Wolfe 1999, hereafter PW99; Pettini et al.\ 2000).

In this paper, we build on previous observations of the chemical abundances
of damped \lya systems.  In particular, we introduce new measurements
of over 15 damped \lya systems and revise the measurements of our previously
analysed systems (PW99).  The principle
goal of this paper is to provide the community with a uniform, homogeneous
database of $z>1.5$ damped \lya abundances.  To this end, we have
created a web site\footnote{http://kingpin.ucsd.edu/$\sim$hiresdla} 
where we will continuously update our observations and possibly include
measurements from throughout the community.  Our new observations
include spectra with wavelength coverage extending blueward of
\lya where elements like Ar, P, S, N, and O can be examined.  
In the second paper in this series
\citep[Paper II;][]{pro01a}, we address the implications
of the complete data set on chemical evolution,
dust depletion, dust obscuration, and nucleosynthesis.  In companion
papers, we address the N/O abundance of the damped \lya systems,
examine the kinematic characteristics of the full sample \citep{pro01b},
infer the star formation rate of the damped systems \citep{wp01},
and investigate the observational evidence for photoionization in these systems.

\section{OBSERVATIONS AND ANALYSIS}
\label{sec-obs}

All of the observations presented in this paper were acquired with the
High Resolution Echelle Spectrograph \citep[HIRES;][]{vogt94} on the
Keck~I 10m telescope.  The HIRES spectrograph
is mounted at the Nasmyth focus of Keck~I
and is equipped with an image rotator, two collomators
(red and blue sensitive), and a Tektronix 2048$\times$2048 CCD.
For each observation, we implemented either a $1.1''$ or $0.8''$ slit
which provides a resolution of FWHM~$\approx 8.4$ and 6.3 km/s 
respectively.  The HIRES
spectrograph affords $\approx 2500$\rAA of wavelength coverage 
per setting with continuous coverage below $\lambda \approx 5100$\AA.
Table~\ref{tab:jouobs} summarizes all of the new observations; refer to 
PW99 for previous observations.

\begin{table*}\footnotesize
\begin{center}
\caption{
{\sc JOURNAL OF NEW OBSERVATIONS \label{tab:jouobs}}}
\begin{tabular}{llcccccc}
\tableline
\tableline
QSO & Alt.\ name & $z_{em}$ & Wavelength & Date & Exposure & Resolution 
& SNR \\
& & &  (\AA) & & (s) & (km/s) & (pix$^{-1}$) \\
\tableline
J0255+00   & ...             & 4.02 & 5100--8160 & F99,F00 & 20200 & 6.3 & 15 \\
Q0336$-$01 & ...             & 3.22 & 3940--6390 & F99     & 5400  & 8.2 & 10 \\
Q0347$-$38 & ...             & 3.23 & 3600--5900 & F98     & 4500  & 8.2 &  6 \\
HS0741+4741&                 & 3.20 & 3600--5900 & F98,S00 & 10800 & 8.2 & 25 \\
           &                 &      & 5050--7470 & F00     & 5400  & 6.3 & 30 \\
Q0951$-$04 & BRI0951$-$0450  & 4.37 & 5720--8150 & F99     & 7200  & 8.2 & 10 \\
BRI 0952$-$0115 & ...        & 4.43 & 5700--8150 & S99     & 28800 & 8.2 & 15 \\
PSS0957+33 & ...             & 4.25 & 6440--8760 & F00     & 7200  & 6.3 & 15 \\
BRI 1108$-$0747 & ...        & 3.92 & 5950--8340 & F98,F99 & 12600 & 8.2 & 20 \\
Q1210+17   &                 & 2.54 & 3760--6170 & S00     & 7200  & 8.2 & 20 \\
Q1223+17   &                 & 2.92 & 4780--7160 & S98     & 19600 & 8.2 & 30 \\
           &                 &      & 3560--5900 & S98     & 5000  & 8.2 &  7 \\
BRI1346--0322&               & 3.99 & 4280--6600 & S00     & 7200  & 8.2 &  4 \\
PSS1443+27 &                 & 4.41 & 6070--8500 & S99     & 25200 & 8.2 & 20 \\
           &                 &      & 6790--9180 & S00     & 11000 & 8.2 & 15 \\
Q1759+75   & GB1759+7539     & 3.05 & 3500--5800 & S00     & 22200 & 8.2 & 30 \\
Q1946+7658 &                 & 2.99 & 3470--5055 & F98     & 47970 & 8.2 & 50 \\
Q2344+12   &                 & 4.30 & 3400--4985 & F98     & 4000  & 8.2 & 12 \\
Q2348--01  &                 & 3.01 & 5060--7480 & F99,F00 & 16200 & 8.2 & 15 \\
\tableline
\end{tabular}
\end{center}
\end{table*}

All of the data were reduced with the MAKEE package as tailored for HIRES
observations by 
T. Barlow\footnote{http://spider.ipac.caltech.edu/staff/tab/makee}.  
This package flat-fields
the exposures, optimally extracts the spectra from the 2-D images 
traced by a standard star or pinhole image, 
removes cosmic rays, and wavelength calibrates the spectra by cross-correlating
each object's Th-Ar calibrations with a large database of calibrated HIRES
data.  We then continuum fit each spectrum 
with an in-house routine similar to the IRAF task CONTINUUM.
For quasars with multiple exposures, the 
individual spectra were coadded (weighting by S/N and rejecting bad pixels) 
to produce a single 1-D spectrum with 2 km/s pixels.

\section{IONIC COLUMN DENSITIES}
\label{sec-ion}

In this section, we present ionic column density measurements for all of
the new systems as well as a number of transitions excluded in PW99.
All of the ionic column densities 
were derived with the apparent optical depth method 
\citep[AODM;][]{sav91}.  
This technique corrects for hidden saturation by 
comparing the apparent column density, $N_a$, for multiple transitions
from a single ion.  This technique
also gives an efficient means of calculating total column densities for each ion.
The analysis involves calculating $N_a(v)$
for each pixel from the optical depth equation
 
\begin{equation}
N_a(v) = {m_e c \over \pi e^2} {\tau_a(v) \over f \lambda} ,
\end{equation}
 
\noindent where $\tau_a(v) = \ln [I_i (v) / I_a (v)]$, $f$ is
the oscillator strength, $\lambda$ is the rest wavelength, and
$I_i$ and $I_a$ are the incident and measured intensity.  
By summing over the velocity profile of a given transition one calculates
the total column density,

\begin{equation} 
N_{T} = \smm N_a(v) \delv \cmma
\label{eq:AODMsig}
\end{equation}

\noindent and a $1\sigma$ error on
the column density through standard error propagation

\begin{equation}
\sigma^2(N_{T}) = \smm \ltp \frac{m_e c}{\pi e^2 f \lambda} \rtp^2
\frac{\sigma^2(I_a(v))}{I^2_a(v)} \delv^2 \perd
\end{equation}
In previous papers \citep{wol94,pro96,pro97a}, we
showed that the damped \lya profiles are not contaminated by hidden
saturation. Furthermore, we
demonstrated the total column densities derived with
the AODM agree very well with
line-profile fitting, which should give
a more accurate measure of the ionic column densities when
hidden saturation is negligible.
As the AODM is easier to apply to a large data set, we have chosen
to use this technique to measure the ionic column densities for the
damped \lya sample.  

\begin{table}[hb]\footnotesize
\begin{center}
\caption{ 
{\sc ATOMIC DATA \label{tab:fosc}}}
\begin{tabular}{lcccc}
\tableline
\tableline
Transition &$\lambda$ &$f$ & Ref \\
\tableline
HI-19 914  &  914.0390 & 0.00019700 &  1  \\
HI-18 914  &  914.2860 & 0.00023000 &  1  \\
HI-17 914  &  914.5760 & 0.00027000 &  1  \\
HI-16 914  &  914.9190 & 0.00032100 &  1  \\
HI-15 915  &  915.3290 & 0.00038600 &  1  \\
  NII 915  &  915.6120 & 0.14490000 &  1  \\
HI-14 915  &  915.8240 & 0.00046900 &  1  \\
HI-13 916  &  916.4290 & 0.00057700 &  1  \\
 PIII 917  &  917.1180 & 0.40490000 &  1  \\
HI-12 917  &  917.1806 & 0.00072260 &  1  \\
HI-11 918  &  918.1294 & 0.00092100 &  1  \\
\tableline
\end{tabular}
\end{center}
\tablerefs{Key to References -- 1:
\cite{morton91}; 2: \cite{howk00}; 3:
\cite{morton01}; 4: \cite{trp96}; 5:
\cite{fedchak99}; 6: \cite{verner94}; 7:
\cite{fedchak00}; 8: \cite{mullman98}; 9:
\cite{schect98}; 10: \cite{bergs96}; 11:
\cite{bergs93}; 12: \cite{wiese01}; 13:
\cite{bergs93b}; 14: \cite{bergs94}; 15:
\cite{verner96}; 16: \cite{raassen98}}
\tablecomments{The complete version of this table is in the electronic edition\\
of the Journal.  The printed edition contains only a sample}
\end{table}

Tables~\ref{tab:Q0000-2619_3.390}$-$\ref{tab:Q2359-02_2.154} 
present the results of the abundance measurements
including an estimate of the $1\sigma$ error.
For those transitions where the profile saturates, 
i.e.\ $I_a/I_i < 0.05$ in at least one pixel,
the column densities are listed as lower limits.
The values reported as upper limits are $3 \sigma$ limits except in the
cases where we set an upper limit due to significant line blending. 
We have ignored continuum
error in our analysis which may dominate the measurements
of very weak transitions especially those blueward of \lya emission. 
We estimate a systematic error of $\approx 10\%$ in most cases.
The 3$\sigma$ statistical limits are conservative, however,
and are likely to account for the continuum error in all but the noisiest
and/or crowded absorption regions.
In the following subsections,
we comment briefly on each damped \lya system,
plot the metal-line profiles, and tabulate column densities for 
each profile.  We adopt ionic column densities from these measurements
by calculating the weighted-mean. 
In the velocity plots, $v=0$ is chosen arbitrarily and
corresponds to the redshift listed in the figure caption. 
We indicate regions of blending, primarily
through blends with other metal-line systems
or the \lya forest, by plotting with dotted lines.
For those systems previously analysed in PW99, we report only upon the
changes made since publication.  For completeness, when we include the
measurement of a new transition (e.g.\ Ni~II 1317) we report other
measurements of the same ion and the new adopted ionic column density.

Throughout the paper, we adopt the wavelengths
and oscillator strengths presented in Table~\ref{tab:fosc}. 
When possible, we have adopted laboratory values for the oscillator
strengths. 
Since PW99 there have been several new measurements of $f$-values 
which impact the abundances of the damped \lya systems,  including
new Ni~II and Ti~II oscillator strengths which have significantly
revised the abundances of these elements.
Most importantly, however, is the adoption of new oscillator
strengths for the majority of Fe~II transitions and in particular
the Fe~II $\lambda 1608,1611$ transitions.
In this paper, we adopt $f(1608) = 0.0580$ from the laboratory measurement
of \cite{bergs96} and $f(1611) = 0.00136$ from \cite{raassen98}.
In general, these values revise the abundance of Fe$^+$ downward by
$\approx 0.1$~dex. 
Finally, we adopt solar meteoritic abundances from \cite{grvss96}.

\begin{figure}[ht]
\begin{center}
\includegraphics[height=3.8in, width=2.8in,angle=-90]{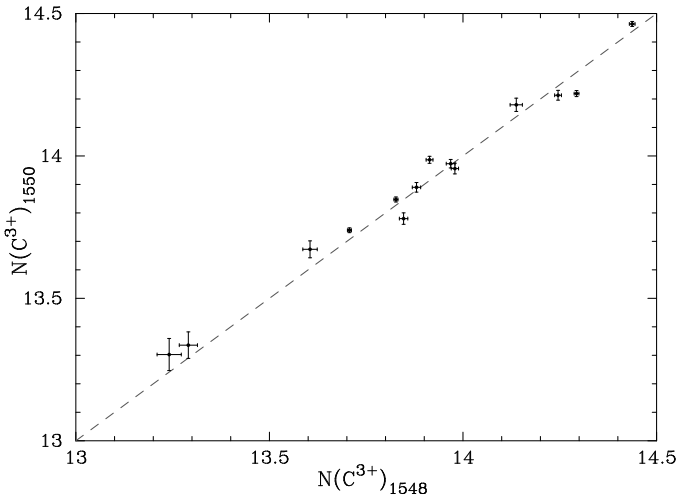}
\caption{A comparison of C$^{3+}$ column densities measured independently
from C~IV 1548 and 1550 in the same system using the apparent 
optical depth method (AODM).  The dashed line traces the line of equality.  
The good agreement between the two values over a large range in
$\N{C^{3+}}$ indicates the AODM is an accurate method for measuring
column densities in unblended transitions.}
\label{fig:civtst}
\end{center}
\end{figure}

To assess the accuracy of our measurements and the reliability of the
error analysis, one can compare the column density measurements from
two transitions for the same ion with very accurately known relative
oscillator strengths.  The majority of our damped \lya systems exhibit
absorption from the high-ion C$^{+3}$ \citep{wol00a} which exhibits
a pair of resonance absorption lines at $\lambda \approx 1550$~\rAA due
to the spin-orbit coupling of the 2p electronic level.  Because the 
physics of spin-orbit coupling is well understood, we have high confidence
that the relative oscillator strengths of the $\lambda\lambda 1548,1550$ 
transitions is 2:1.  Figure~\ref{fig:civtst} plots the $\N{1550}$ value
vs.\ $\N{1548}$ for all of the damped \lya systems where the transitions
are unsaturated and unblended.  One notes that the
agreement in the values is excellent even out to $\N{C^{+3}} \approx 14.5$
where saturation could affect the stronger C~IV 1548 transition.  In short,
Figure~\ref{fig:civtst} demonstrates that the
AODM provides a reasonably accurate measure of the column densities and
$1\sigma$ errors for our analysis. 

We now comment on the individual systems noting revisions from 
previous works where applicable.  In the figures, we have dotted out 
identified line blends.

\begin{figure}[ht]
\begin{center}
\includegraphics[height=2.3in, width=2.5in]{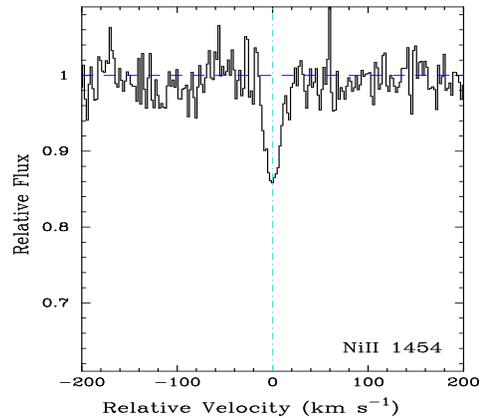}
\caption{Velocity plot of the new metal-line transition for the 
damped \lya system at $z = 3.390$ toward Q0000$-$26.
The vertical line at $v=0$ corresponds to $z = 3.3901$.}
\label{fig:q0000}
\end{center}
\end{figure}

\begin{table}[hb]\footnotesize
\begin{center}
\caption{ {\sc
IONIC COLUMN DENSITIES: Q0000-2619, $z = 3.390$ \label{tab:Q0000-2619_3.390}}}
\begin{tabular}{lcccc}
\tableline
\tableline
Ion & $\lambda$ & AODM & $N_{\rm adopt}$ & [X/H] \\
\tableline
HI &1215 & $21.410  \pm 0.080  $ \\
Ni II &1454&$13.365 \pm  0.045$&$13.365 \pm  0.045$&$-2.295 \pm  0.092$\\  
Ni II &1751&$<13.689$\\  
\tableline
\end{tabular}
\end{center}
\end{table}

\subsection{Q0000$-$26, $z$ = 3.390}
\label{subsec-0000}

We presented metal abundances for this system in PW99, but missed the
Ni~II 1454 profile (Figure~\ref{fig:q0000}).  Furthermore, we have identified
a telluric blend with the Ni~II 1751 profile which lead to an overestimate of
Ni for this system.  The Ni~II 1454 column density is in excellent 
agreement with the Ni abundance derived by \cite{molaro00} and further
clouds the nucleosynthetic interpretation of this system \citep{molaro01}.
Owing to the higher signal-to-noise of the UVES spectrum, we now adopt
the Fe$^+$ column density from \cite{molaro00} but revise it downward to
$\N{Fe^+} = 14.75 \pm 0.03$ due to the new Fe~II 1611 oscillator 
strength.  This places the Fe$^+$ column density in reasonably good 
agreement with the $\N{Ni^+}$ measurement and 
further supports the notion that the $\alpha$-elements (O, Si) are 
enhanced in this system.  We note in passing that the $\N{Fe^+}$ value 
derived from our HIRES spectrum of Fe~II 1611 still significantly
exceeds the UVES measurement for reasons we do not fully appreciate.

\begin{figure}[ht]
\begin{center}
\includegraphics[height=3.7in, width=3.0in]{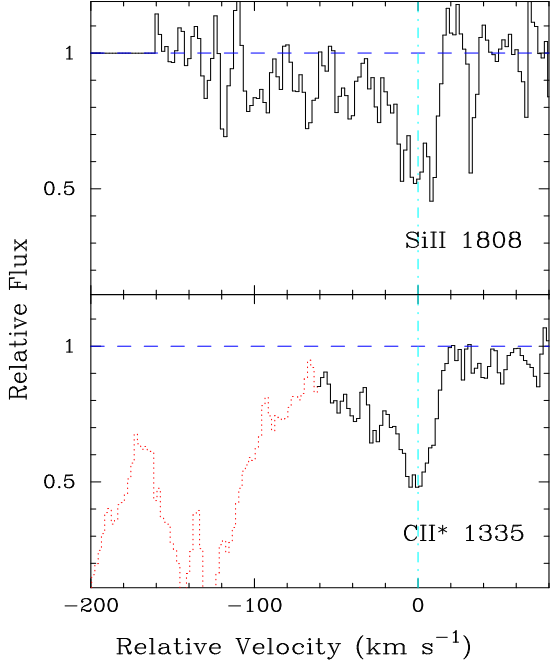}
\caption{Velocity plot of the C~II$^*$ 1335 profile for the
damped \lya system at $z = 3.439$ toward BR0019$-$15.  The
Si~II 1808 profile is shown for comparison.
The vertical line at $v=0$ corresponds to $z = 3.4388$.}
\label{fig:0019}
\end{center}
\end{figure}

\subsection{BR0019$-$15, $z$ = 3.439}
\label{subsec-0019}

This system was analysed in PW99.  Since publication, 
we have identified the C~II$^*$ 1335 profile (Figure~\ref{fig:0019}) 
and revised 
the Fe~II 1608 column density to a lower limit because the
profile is mildly saturated.
This revision accounts for the large Ni/Fe ratio reported in PW99.

\begin{table}[ht]\footnotesize
\begin{center}
\caption{ {\sc
IONIC COLUMN DENSITIES: BR0019-15, $z = 3.439$ \label{tab:BR0019-15_3.439}}}
\begin{tabular}{lcccc}
\tableline
\tableline
Ion & $\lambda$ & AODM & $N_{\rm adopt}$ & [X/H] \\
\tableline
HI &1215 & $20.920  \pm 0.100  $ \\
C  II &1335&$13.838 \pm  0.018$\\  
Fe II &1608&$>14.789$&$>14.789$&$>-1.631$\\  
Ni II &1709&$13.607 \pm  0.105$&$13.683 \pm  0.040$&$-1.487 \pm  0.108$\\  
Ni II &1741&$13.701 \pm  0.043$\\  
\tableline
\end{tabular}
\end{center}
\end{table}

\begin{figure}[ht]
\begin{center}
\includegraphics[height=4.5in, width=3.5in]{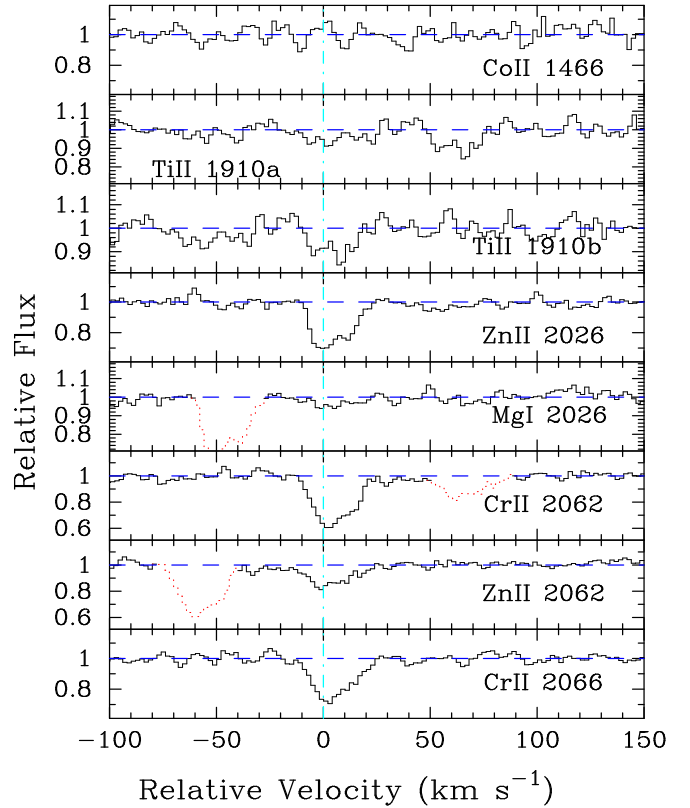}
\caption{Velocity plot of several new metal-line transitions for the 
damped \lya system at $z = 2.309$ toward PH~957.
The vertical line at $v=0$ corresponds to $z = 2.309$.}
\label{fig:ph957}
\end{center}
\end{figure}

\subsection{PH957, $z$ = 2.309}
\label{subsec-ph957}

This system was carefully studied in \cite{wol94} and
subsequently in PW99.  We present new limits on $\N{Co^+}$ and
$\N{Ti^+}$, the latter which places a tight constraint on the
Ti/Fe ratio ([Ti/Fe]~$<-2.04$~dex).
We also present a measurement of $\N{Mg^0}$ from the
Mg~I 2026 transition.   
Figure~\ref{fig:ph957} presents the new
profiles as well as several Zn~II and Cr~II transitions which provide
clarification with respect to the identification of Mg~I 2026. 
Table~\ref{tab:PH957_2.309} lists the new values.

\begin{table}[ht]\footnotesize
\begin{center}
\caption{ {\sc
IONIC COLUMN DENSITIES: PH957, $z = 2.309$ \label{tab:PH957_2.309}}}
\begin{tabular}{lcccc}
\tableline
\tableline
Ion & $\lambda$ & AODM & $N_{\rm adopt}$ & [X/H] \\
\tableline
HI &1215 & $21.400  \pm 0.050  $ \\
Mg I  &2026&$12.344 \pm  0.126$\\  
Ti II &1910&$<12.207$&$<12.207$&$<-2.133$\\  
Co II &1466&$<13.164$&$<13.164$&$<-1.146$\\  
\tableline
\end{tabular}
\end{center}
\end{table}

\begin{figure}[ht]
\begin{center}
\includegraphics[height=4.5in, width=3.5in]{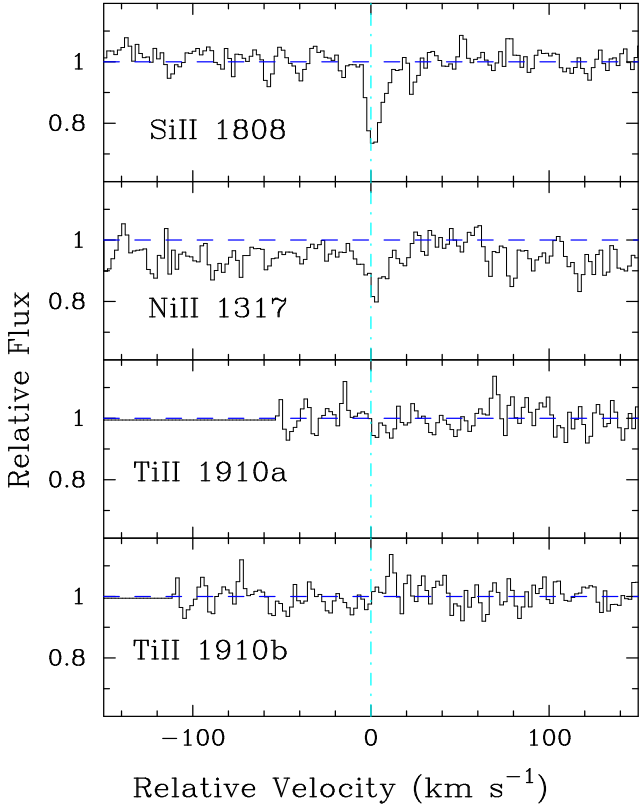}
\caption{Velocity plot of several new metal-line transitions for the 
damped \lya system at $z = 2.141$ toward Q0149$+$33.  For
comparison, we also plot the Si~II 1808 profile.
The vertical line at $v=0$ corresponds to $z = 2.140755$.}
\label{fig:q0149}
\end{center}
\end{figure}

\subsection{Q0149$+$33, $z$ = 2.140}
\label{subsec-0149}

We have several changes to report on this system since PW99.
Figure~\ref{fig:q0149} presents the Ni~II 1317 and Ti~II 1910 profiles 
which were overlooked in PW99.  
We have also revised the Si~II 1304 and Si~II 1526
column densities to lower limits and base the Si abundance solely on
the unsaturated Si~II 1808 profile.  For Cr, we now include the Cr~II 2062
profile in our analysis.
Finally, we warn that the Fe~II 1608
column density might be considered a lower limit for $\N{Fe^+}$ which would
explain its underabundance relative to Cr and Ni.
As noted in PW99, this system exhibits a super-solar Cr/Zn ratio 
([Cr/Zn]~$+0.22 \pm 0.1$). 
Because of its low [Zn/H] and $\N{HI}$ values, this system has 
special significance in terms of dust depletion (Paper~II, $\S$~\secdustdepl).

\begin{table}[ht]\footnotesize
\begin{center}
\caption{ {\sc
IONIC COLUMN DENSITIES: Q0149+33, $z = 2.141$ \label{tab:Q0149+33_2.141}}}
\begin{tabular}{lcccc}
\tableline
\tableline
Ion & $\lambda$ & AODM & $N_{\rm adopt}$ & [X/H] \\
\tableline
HI &1215 & $20.500  \pm 0.100  $ \\
Si II &1304&$>14.432$&$14.572 \pm  0.047$&$-1.488 \pm  0.110$\\  
Si II &1526&$>14.339$\\  
Si II &1808&$14.572 \pm  0.047$\\  
Ti II &1910&$<12.169$&$<12.169$&$<-1.271$\\  
Cr II &2056&$12.793 \pm  0.044$&$12.720 \pm  0.035$&$-1.450 \pm  0.106$\\  
Cr II &2062&$12.520 \pm  0.090$\\  
Cr II &2066&$12.841 \pm  0.062$\\  
Ni II &1317&$13.092 \pm  0.071$&$13.169 \pm  0.036$&$-1.581 \pm  0.106$\\  
Ni II &1370&$13.192 \pm  0.103$\\  
Ni II &1703&$<13.885$\\  
Ni II &1709&$13.184 \pm  0.088$\\  
Ni II &1741&$13.250 \pm  0.064$\\  
Ni II &1751&$13.179 \pm  0.090$\\  
\tableline
\end{tabular}
\end{center}
\end{table}

\begin{figure}[ht]
\begin{center}
\includegraphics[height=4.5in, width=3.5in]{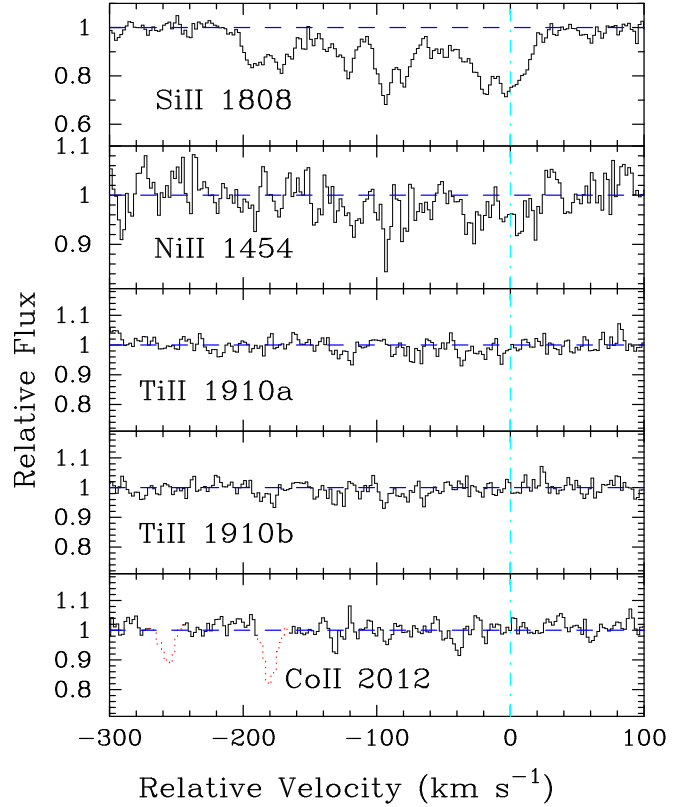}
\caption{Velocity plot of several new metal-line transitions for the 
damped \lya system at $z = 2.463$ toward Q0201$+$36.  
For comparison, we also plot the Si~II 1808 profile.
The vertical line at $v=0$ corresponds to $z = 2.4628$.}
\label{fig:0201}
\end{center}
\end{figure}

\subsection{Q0201$+$36, $z$ = 2.463}
\label{subsec-0201p36}

This system was studied at length in \cite{pro96} and we now include
an upper limit on the Ti~II 1910 transition and a measurement for
the Ni~II 1454 profile (Figure~\ref{fig:0201}).  
To provide the best comparison with other
objects in the complete sample, we adopt AODM column densities 
for all of the transitions
(Table~\ref{tab:Q0201+36_2.463}).

\begin{table}[ht]\footnotesize
\begin{center}
\caption{ {\sc
IONIC COLUMN DENSITIES: Q0201+36, $z = 2.463$ \label{tab:Q0201+36_2.463}}}
\begin{tabular}{lcccc}
\tableline
\tableline
Ion & $\lambda$ & AODM & $N_{\rm adopt}$ & [X/H] \\
\tableline
HI &1215 & $20.380  \pm 0.045  $ \\
C  IV &1550&$14.612 \pm  0.005$\\  
Al II &1670&$>14.133$&$>14.133$&$>-0.737$\\  
Al III&1862&$13.601 \pm  0.007$\\  
Si II &1808&$15.534 \pm  0.010$&$15.534 \pm  0.010$&$-0.406 \pm  0.046$\\  
Si IV &1393&$>14.071$\\  
Ti II &1910&$<12.196$&$<12.196$&$<-1.124$\\  
Cr II &2056&$13.266 \pm  0.030$&$13.248 \pm  0.029$&$-0.802 \pm  0.054$\\  
Cr II &2066&$13.132 \pm  0.094$\\  
Fe II &1608&$15.010 \pm  0.004$&$15.010 \pm  0.004$&$-0.870 \pm  0.045$\\  
Co II &2012&$<12.957$&$<12.957$&$<-0.333$\\  
Ni II &1454&$13.669 \pm  0.073$&$14.022 \pm  0.010$&$-0.608 \pm  0.046$\\  
Ni II &1709&$14.080 \pm  0.021$\\  
Ni II &1741&$14.021 \pm  0.013$\\  
Ni II &1751&$13.984 \pm  0.021$\\  
\tableline
\end{tabular}
\end{center}
\end{table}

\break

\begin{figure*}
\begin{center}
\includegraphics[height=8.5in, width=6.0in]{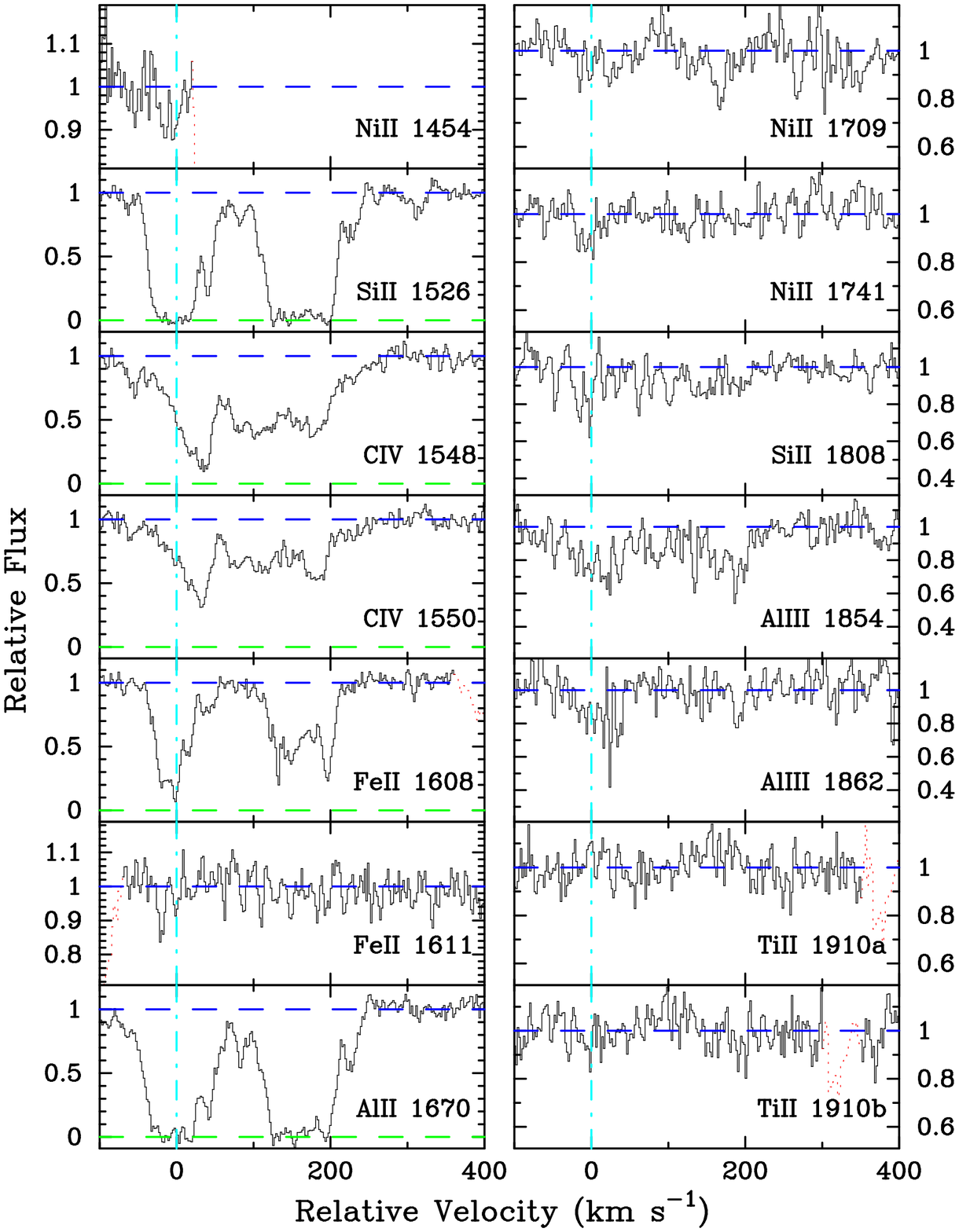}
\caption{Velocity plot of the metal-line transitions for the 
damped \lya system at $z = 3.253$ toward J0255+00.  
The vertical line at $v=0$ corresponds to $z = 3.252931$.}
\label{fig:0255A}
\end{center}
\end{figure*}

\begin{figure*}[ht]
\begin{center}
\includegraphics[height=8.5in, width=6.0in]{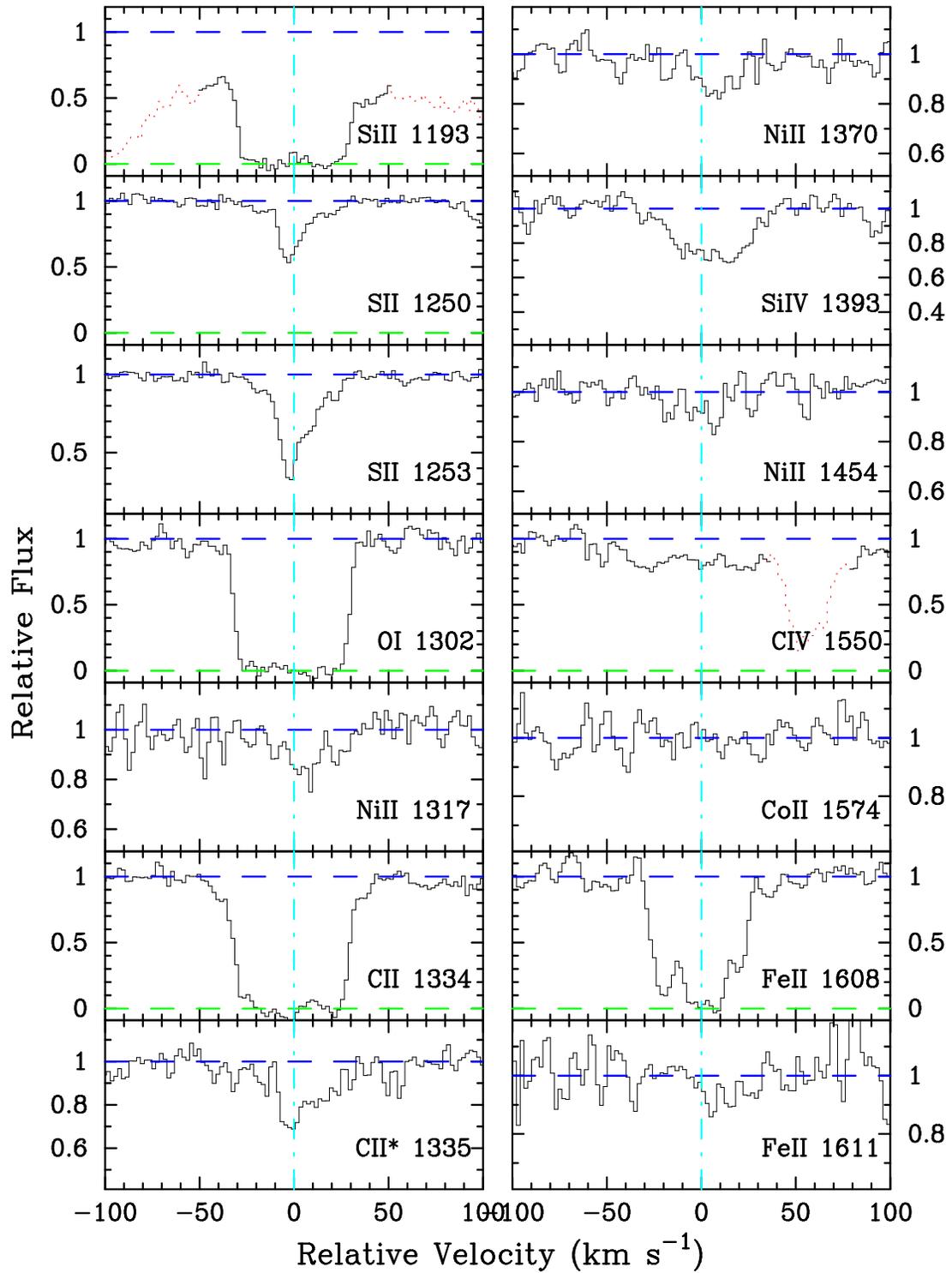}
\caption{Velocity plot of the metal-line transitions for the 
damped \lya system at $z = 3.915$ toward J0255+00.  
The vertical line at $v=0$ corresponds to $z = 3.914617$.}
\label{fig:0255B}
\end{center}
\end{figure*}

\clearpage

\subsection{J0255$+$00, $z$=3.253 and $z$=3.915}
\label{subsec-0255}

The two damped \lya systems observed toward this
faint SDSS quasar \citep[$r=19.1$;][]{fan99} 
were identified as part of a program
designed to survey $z>3$ damped \lya systems \citep{wolfe01}.  
We measured the $\N{HI}$ column densities of the two systems
with an LRIS spectrum and then acquired HIRES observations
at wavelengths redward of the \lya forest.   Figures~\ref{fig:0255A} and
\ref{fig:0255B} present the metal-line profiles for the two systems and
Tables~\ref{tab:J0255+00_3.253} and 
\ref{tab:J0255+00_3.915} list the ionic column densities.
For the system at $z=3.915$, $\N{Fe^+}$ is well constrained by the lower
and upper limits from the Fe~II 1608 and Fe~II 1611 transitions respectively
and we have adopted a column density by averaging the two limits:
$\log \N{Fe^+} = 14.75 \pm 0.08$~dex.

\begin{table}[ht]\footnotesize
\begin{center}
\caption{ {\sc
IONIC COLUMN DENSITIES: J0255+00, $z = 3.253$ \label{tab:J0255+00_3.253}}}
\begin{tabular}{lcccc}
\tableline
\tableline
Ion & $\lambda$ & AODM & $N_{\rm adopt}$ & [X/H] \\
\tableline
HI &1215 & $20.700  \pm 0.100  $ \\
C  IV &1548&$14.437 \pm  0.007$\\  
C  IV &1550&$14.463 \pm  0.009$\\  
Al II &1670&$>13.879$&$>13.879$&$>-1.311$\\  
Al III&1854&$13.277 \pm  0.025$\\  
Al III&1862&$12.977 \pm  0.102$\\  
Si II &1526&$>15.119$&$15.323 \pm  0.038$&$-0.937 \pm  0.107$\\  
Si II &1808&$15.323 \pm  0.038$\\  
Ti II &1910&$<12.805$&$<12.805$&$<-0.835$\\  
Fe II &1608&$14.764 \pm  0.010$&$14.764 \pm  0.010$&$-1.436 \pm  0.100$\\  
Fe II &1611&$<14.832$\\  
Ni II &1709&$13.771 \pm  0.075$&$13.608 \pm  0.066$&$-1.342 \pm  0.120$\\  
Ni II &1741&$13.473 \pm  0.114$\\  
\tableline
\end{tabular}
\end{center}
\end{table}

\begin{table}[ht]\footnotesize
\begin{center}
\caption{ {\sc
IONIC COLUMN DENSITIES: J0255+00, $z = 3.915$ \label{tab:J0255+00_3.915}}}
\begin{tabular}{lcccc}
\tableline
\tableline
Ion & $\lambda$ & AODM & $N_{\rm adopt}$ & [X/H] \\
\tableline
HI &1215 & $21.300  \pm 0.050  $ \\
C  II &1334&$>14.732$&$>14.732$&$>-3.118$\\  
C  II &1335&$13.442 \pm  0.040$\\  
O  I  &1302&$>15.167$&$>15.167$&$>-3.003$\\  
Si II &1193&$>14.193$&$>14.193$&$>-2.667$\\  
Si IV &1393&$12.856 \pm  0.028$\\  
S  II &1250&$14.763 \pm  0.021$&$14.721 \pm  0.011$&$-1.779 \pm  0.051$\\  
S  II &1253&$14.707 \pm  0.013$\\  
Fe II &1608&$>14.707$&$14.750 \pm  0.088$&$-2.050 \pm  0.101$\\  
Fe II &1611&$<14.809$\\  
Co II &1574&$<13.212$&$<13.212$&$<-0.998$\\  
Ni II &1317&$13.315 \pm  0.052$&$13.271 \pm  0.037$&$-2.279 \pm  0.062$\\  
Ni II &1370&$13.213 \pm  0.058$\\  
Ni II &1454&$13.387 \pm  0.104$\\  
\tableline
\end{tabular}
\end{center}
\end{table}

\subsection{Q0336$-$01, $z$=3.062}

This LBQS quasar is one of the few damped \lya systems at $z>3$
with $\N{HI} > 10^{21} \cm{-2}$.  Although most of the transitions that we 
examine lie within the \lya forest,  we have carefully 
avoided lines which are clearly blended with forest clouds.  Unfortunately,
we only place a lower limit on $\N{Si^+}$ although we do report 
a reasonably secure value for $\N{S^+}$.  The system also provides an
accurate measurement of Ar~I and a reasonable estimate of P~II.
Finally, we note a very large $\N{O^0}$ lower limit derived from the
saturated O~I 988 transition.  This limit may be influenced by blending
in the \lya forest, but we have no reason to believe this is the case
at present.

\begin{table}[ht]\footnotesize
\begin{center}
\caption{ {\sc
IONIC COLUMN DENSITIES: Q0336-01, $z = 3.062$ \label{tab:Q0336-01_3.062}}}
\begin{tabular}{lcccc}
\tableline
\tableline
Ion & $\lambda$ & AODM & $N_{\rm adopt}$ & [X/H] \\
\tableline
HI &1215 & $21.200  \pm 0.100  $ \\
C  II &1334&$>14.958$&$>14.958$&$>-2.792$\\  
C  II &1335&$14.041 \pm  0.025$\\  
C  IV &1548&$14.138 \pm  0.016$\\  
C  IV &1550&$14.180 \pm  0.023$\\  
O  I  & 988&$>16.940$&$>16.940$&$>-1.130$\\  
O  I  &1302&$>15.389$\\  
Si II &1020&$>15.141$&$>15.141$&$>-1.619$\\  
Si II &1193&$>14.322$\\  
Si II &1304&$>14.926$\\  
Si IV &1393&$13.767 \pm  0.015$\\  
Si IV &1402&$13.656 \pm  0.032$\\  
P  II &1152&$13.133 \pm  0.075$&$13.133 \pm  0.075$&$-1.597 \pm  0.125$\\  
S  II &1250&$15.118 \pm  0.022$&$14.994 \pm  0.011$&$-1.406 \pm  0.101$\\  
S  II &1259&$14.970 \pm  0.012$\\  
Ar I  &1048&$>13.860$&$14.346 \pm  0.065$&$-1.374 \pm  0.119$\\  
Ar I  &1066&$14.346 \pm  0.065$\\  
Fe II &1081&$14.879 \pm  0.055$&$14.905 \pm  0.033$&$-1.795 \pm  0.105$\\  
Fe II &1125&$14.920 \pm  0.046$\\  
Fe II &1142&$14.936 \pm  0.099$\\  
Fe II &1144&$>14.719$\\  
Ni II &1317&$<13.389$&$13.469 \pm  0.057$&$-1.981 \pm  0.115$\\  
Ni II &1370&$13.424 \pm  0.065$\\  
Ni II &1454&$13.880 \pm  0.091$\\  
\tableline
\end{tabular}
\end{center}
\end{table}

\begin{figure*}
\begin{center}
\includegraphics[height=8.5in, width=6.0in]{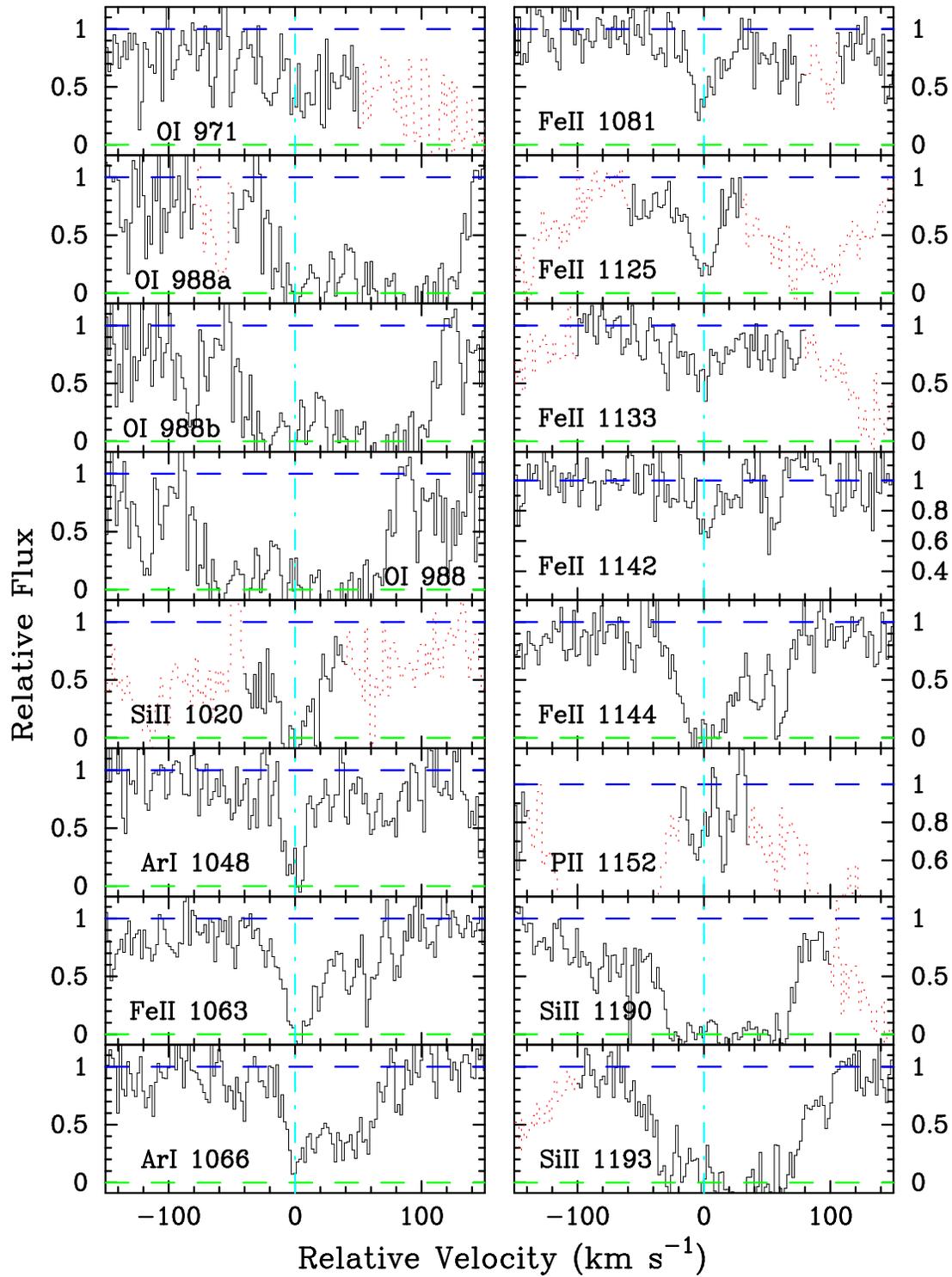}
\caption{Velocity plot of the metal-line transitions for the 
damped \lya system at $z = 3.062$ toward Q0336--01. 
The vertical line at $v=0$ corresponds to $z = 3.062078$.}
\label{fig:0336}
\end{center}
\end{figure*}

\begin{figure*}
\begin{center}
\includegraphics[height=8.5in, width=6.0in]{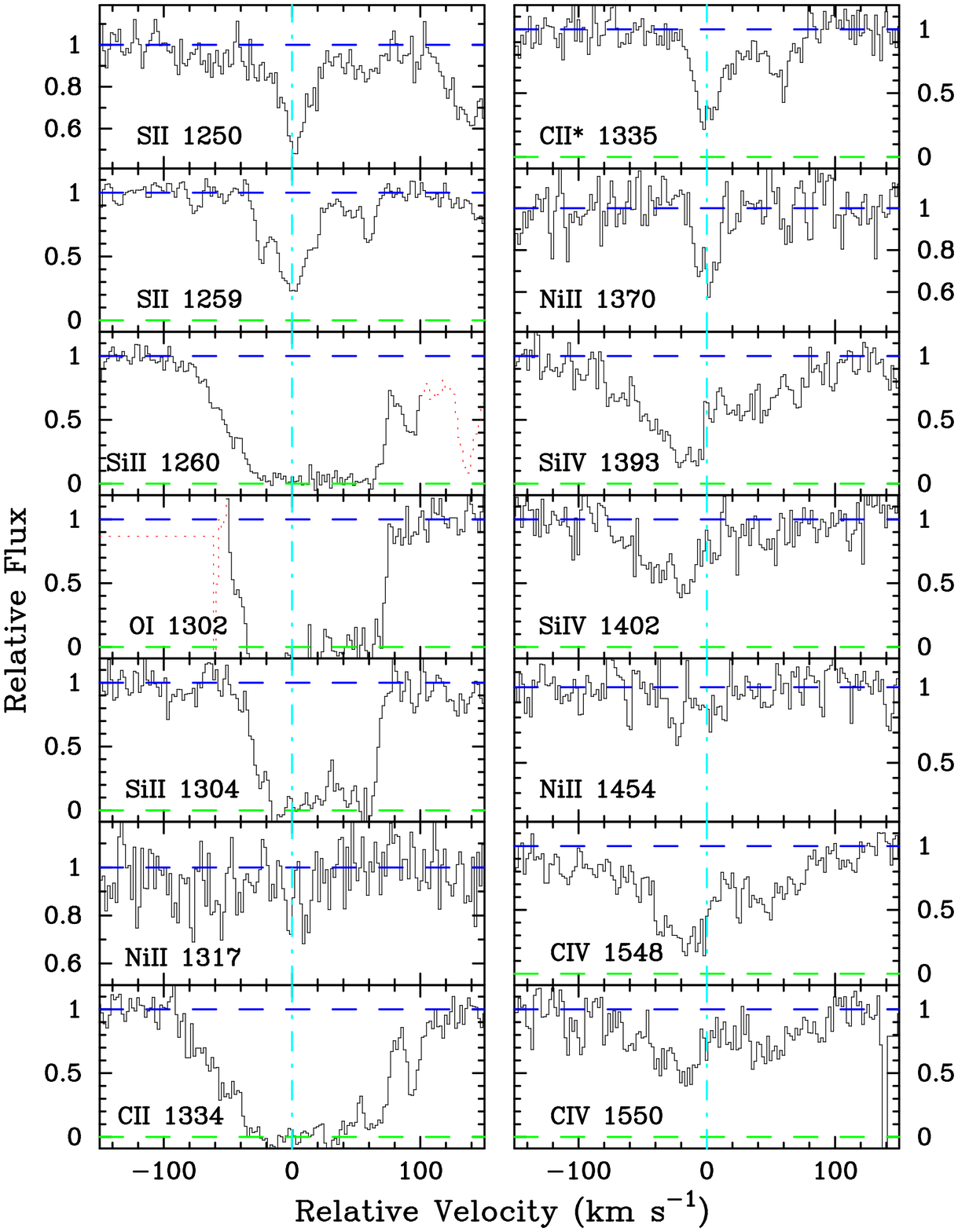}
\end{center}
\end{figure*}

\begin{figure*}
\begin{center}
\includegraphics[height=8.5in, width=6.0in]{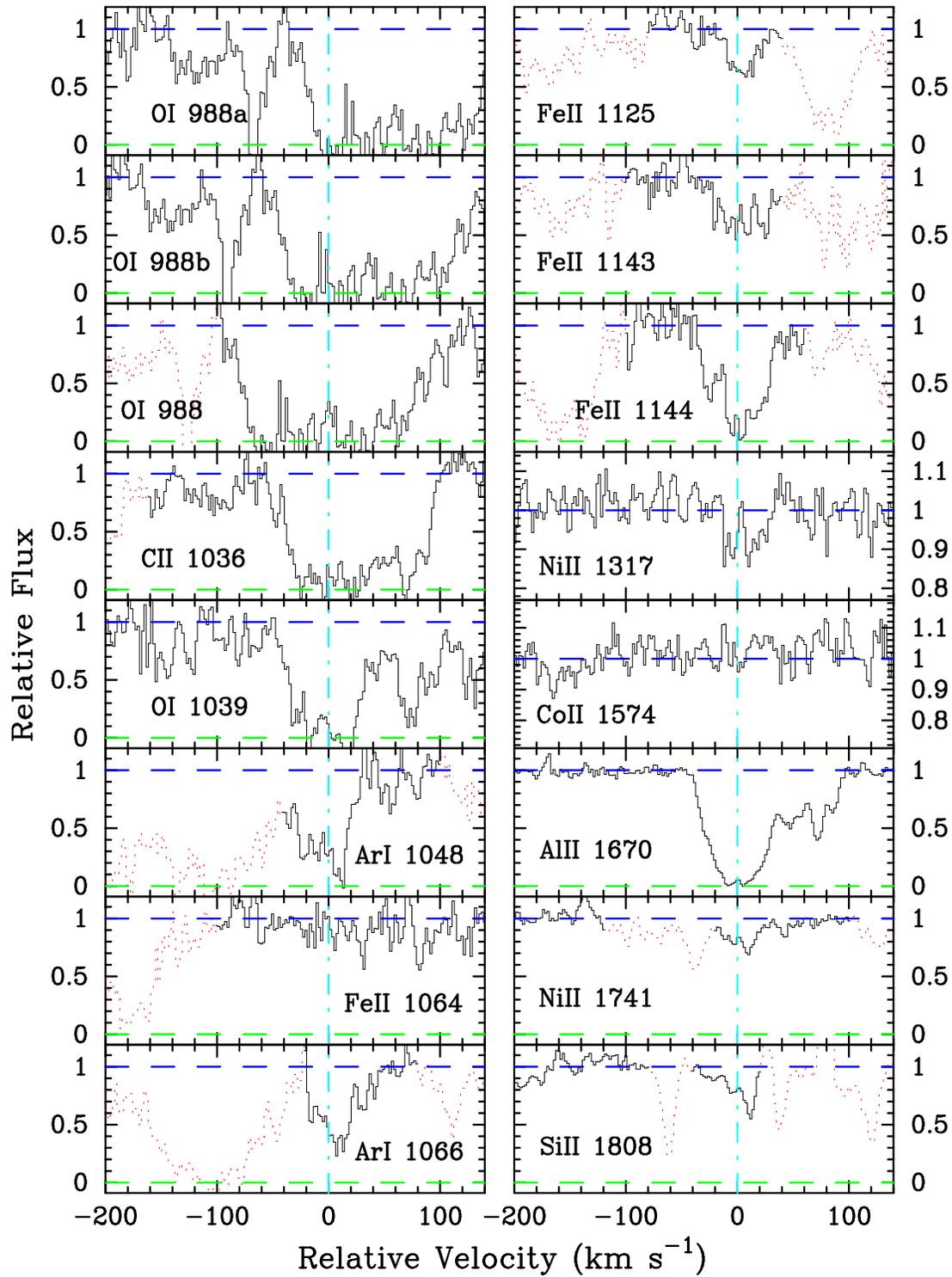}
\caption{Velocity plot of several new metal-line transitions for the 
damped \lya system at $z = 3.025$ toward Q0347--38.  
The vertical line at $v=0$ corresponds to $z = 3.0247$.}
\label{fig:0347}
\end{center}
\end{figure*}

\clearpage

\subsection{Q0347$-$38, $z$ = 3.025}
\label{subsec-0347}
Although this system was analysed in PW99, we report a number of 
measurements and limits based on the original data and 
new observations taken with a second, blue setup. 
Table~\ref{tab:Q0347-38_3.025} presents the ionic column densities and 
Figure~\ref{fig:0347} plots the new profiles.
We also reclassify the sulfur abundance as 
an upper limit because this transition is blended with a
\lya forest cloud.

\begin{table}[hb]\footnotesize
\begin{center}
\caption{ {\sc
IONIC COLUMN DENSITIES: Q0347-38, $z = 3.025$ \label{tab:Q0347-38_3.025}}}
\begin{tabular}{lcccc}
\tableline
\tableline
Ion & $\lambda$ & AODM & $N_{\rm adopt}$ & [X/H] \\
\tableline
HI &1215 & $20.800  \pm 0.100  $ \\
C  II &1036&$>14.930$&$>15.065$&$>-2.285$\\  
C  II &1334&$>15.066$\\  
O  I  &1039&$>15.953$&$>15.953$&$>-1.717$\\  
O  I  &1302&$>15.449$\\  
Al II &1670&$>13.408$&$>13.408$&$>-1.882$\\  
Si II &1260&$>14.329$&$15.017 \pm  0.026$&$-1.343 \pm  0.103$\\  
Si II &1304&$>14.889$\\  
Si II &1808&$15.017 \pm  0.026$\\  
S  II &1259&$<14.760$&$<14.760$&$<-1.240$\\  
Ar I  &1048&$>14.063$&$14.282 \pm  0.035$&$-1.038 \pm  0.106$\\  
Ar I  &1066&$14.282 \pm  0.035$\\  
Fe II &1063&$14.763 \pm  0.143$&$14.503 \pm  0.007$&$-1.797 \pm  0.100$\\  
Fe II &1125&$14.453 \pm  0.056$\\  
Fe II &1143&$14.712 \pm  0.040$\\  
Fe II &1608&$14.501 \pm  0.007$\\  
Fe II &1611&$<14.447$\\  
Co II &1574&$<13.196$&$<13.196$&$<-0.514$\\  
Ni II &1317&$13.034 \pm  0.093$&$13.383 \pm  0.031$&$-1.667 \pm  0.105$\\  
Ni II &1370&$13.115 \pm  0.099$\\  
Ni II &1741&$14.056 \pm  0.019$\\  
\tableline
\end{tabular}
\end{center}
\end{table}

\begin{table}[hb]\footnotesize
\begin{center}
\caption{ {\sc
IONIC COLUMN DENSITIES: Q0458-02, $z = 2.040$ \label{tab:Q0458-02_2.040}}}
\begin{tabular}{lcccc}
\tableline
\tableline
Ion & $\lambda$ & AODM & $N_{\rm adopt}$ & [X/H] \\
\tableline
HI &1215 & $21.650  \pm 0.090  $ \\
C  I  &1656&$<12.453$\\  
Mg I  &2026&$13.117 \pm  0.031$\\  
Ti II &1910&$<12.495$&$<12.495$&$<-2.095$\\  
Co II &2012&$<13.092$&$<13.093$&$<-1.467$\\  
Ni II &1317&$14.257 \pm  0.024$&$14.181 \pm  0.018$&$-1.719 \pm  0.092$\\  
Ni II &1370&$14.315 \pm  0.071$\\  
Ni II &1454&$14.187 \pm  0.114$\\  
Ni II &1703&$13.987 \pm  0.090$\\  
Ni II &1709&$14.158 \pm  0.034$\\  
Ni II &1741&$14.195 \pm  0.032$\\  
Ni II &1751&$14.170 \pm  0.033$\\  
Zn II &2026&$13.134 \pm  0.020$&$13.134 \pm  0.020$&$-1.186 \pm  0.092$\\  
Zn II &2062&$>13.031$\\  
\tableline
\end{tabular}
\end{center}
\end{table}

\begin{figure}[ht]
\begin{center}
\includegraphics[height=4.5in, width=3.5in]{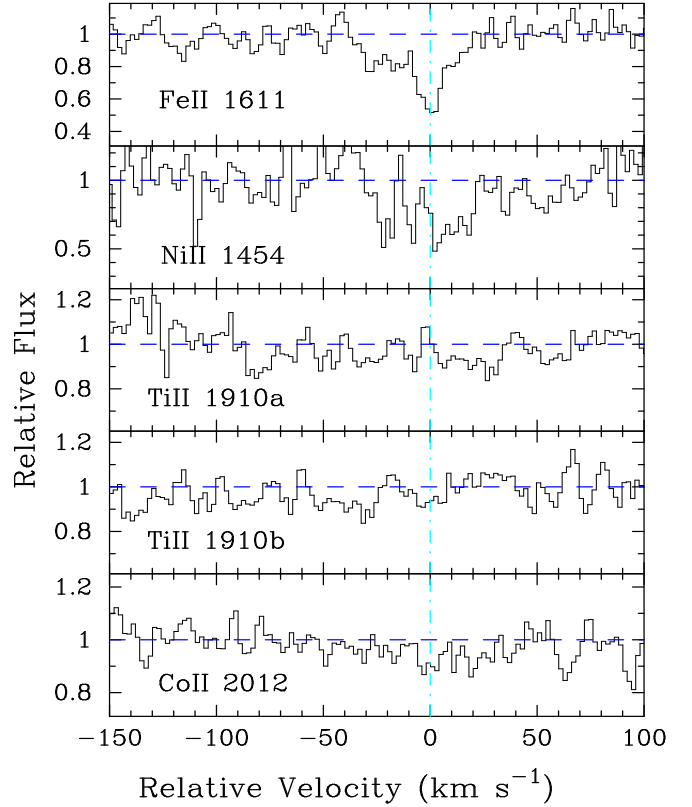}
\caption{Velocity plot of several new metal-line transitions for the 
damped \lya system at $z = 2.040$ toward Q0458--02.  
The vertical line at $v=0$ corresponds to $z = 2.03955$.}
\label{fig:0458}
\end{center}
\end{figure}

\subsection{Q0458$-$02, $z$ = 2.040}
\label{subsec-0458}

In reviewing this system, we identified a number of transitions
missed in PW99.  
Figure~\ref{fig:0458} presents the 
metal-line profiles and Table~\ref{tab:Q0458-02_2.040} 
shows the column densities.
Most of these only yield upper limits
on ionic column densities, but given the very large HI column density
of this system these limits are valuable.  For example, the limit on
Ti~II 1910 implies [Ti/Fe]~$< -0.45$~dex which is highly suggestive of 
dust depletion.  As discussed in Paper~II, $\S$~\secZn, we have
revised the measurement of $\N{Zn^+}$ downward by 0.02~dex due to
mild blending between the Zn~II 2026 and Mg~I 2026 profiles.

\subsection{HS0741$+$47, $z$ = 3.017}

\begin{figure*}
\begin{center}
\includegraphics[height=8.5in, width=6.0in]{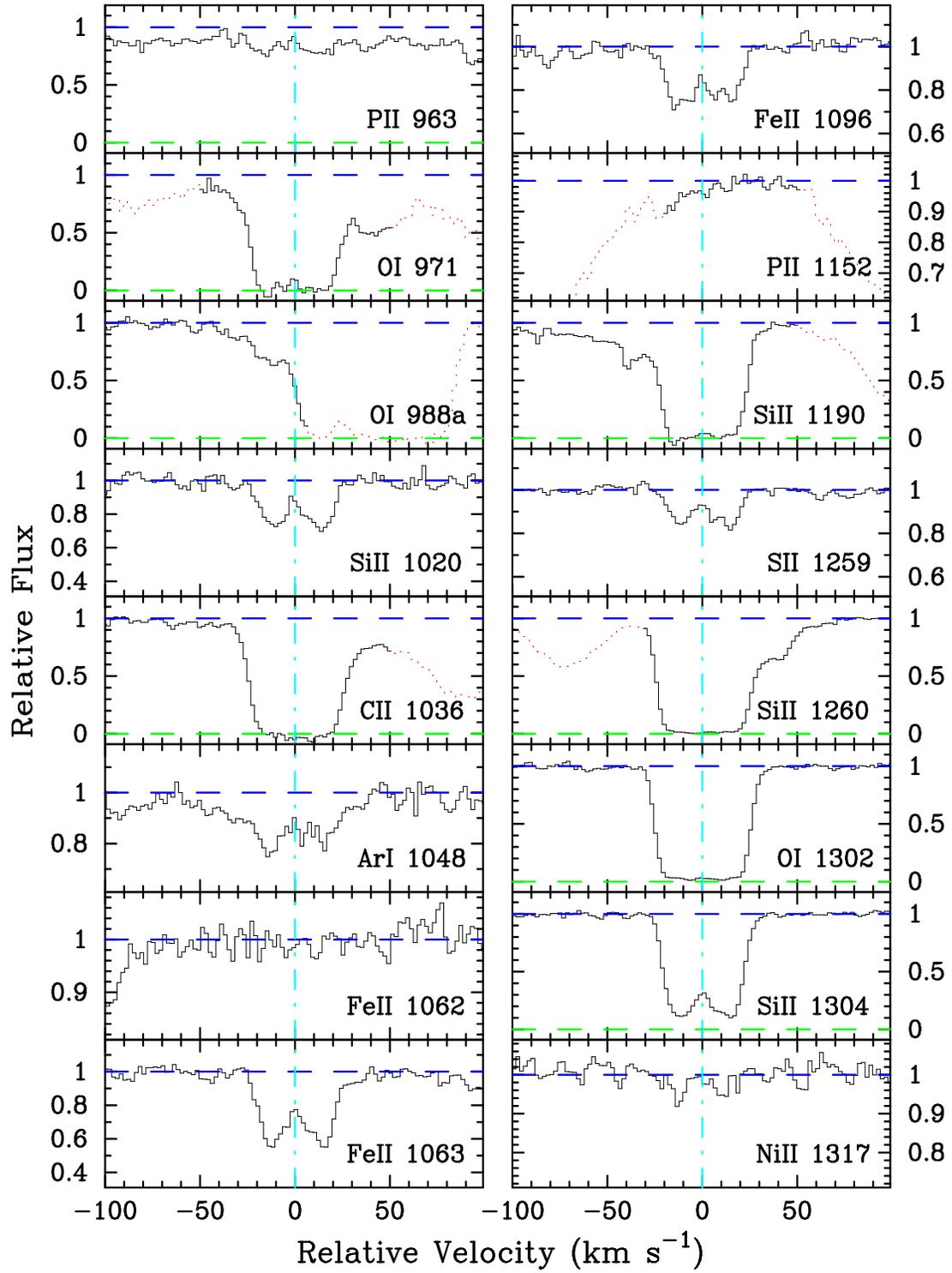}
\caption{Velocity plot of the metal-line transitions for the 
damped \lya system at $z = 3.017$ toward HS0741+4741.  
The vertical line at $v=0$ corresponds to $z = 3.017399$.}
\label{fig:0741}
\end{center}
\end{figure*}

The very bright QSO HS0741+47 is taken from the Hamburg ESO
QSO survey \citep{hagen99}.  
Our wavelength coverage of the damped
\lya system spans from 3600$-$7500\rAA and we have identified 
over 20 metal-line profiles.  
Figure~\ref{fig:0741} plots the velocity profiles and 
Table~\ref{tab:HS0741+4741_3.017} presents all of the measurements.  
Unfortunately,
our observations do not include the Zn~II and Cr~II transitions, 
although the $\N{Si^+}$ value 
indicates it would require very high S/N to obtain a measurement.

\clearpage

\begin{figure*}
\begin{center}
\includegraphics[height=8.5in, width=6.0in]{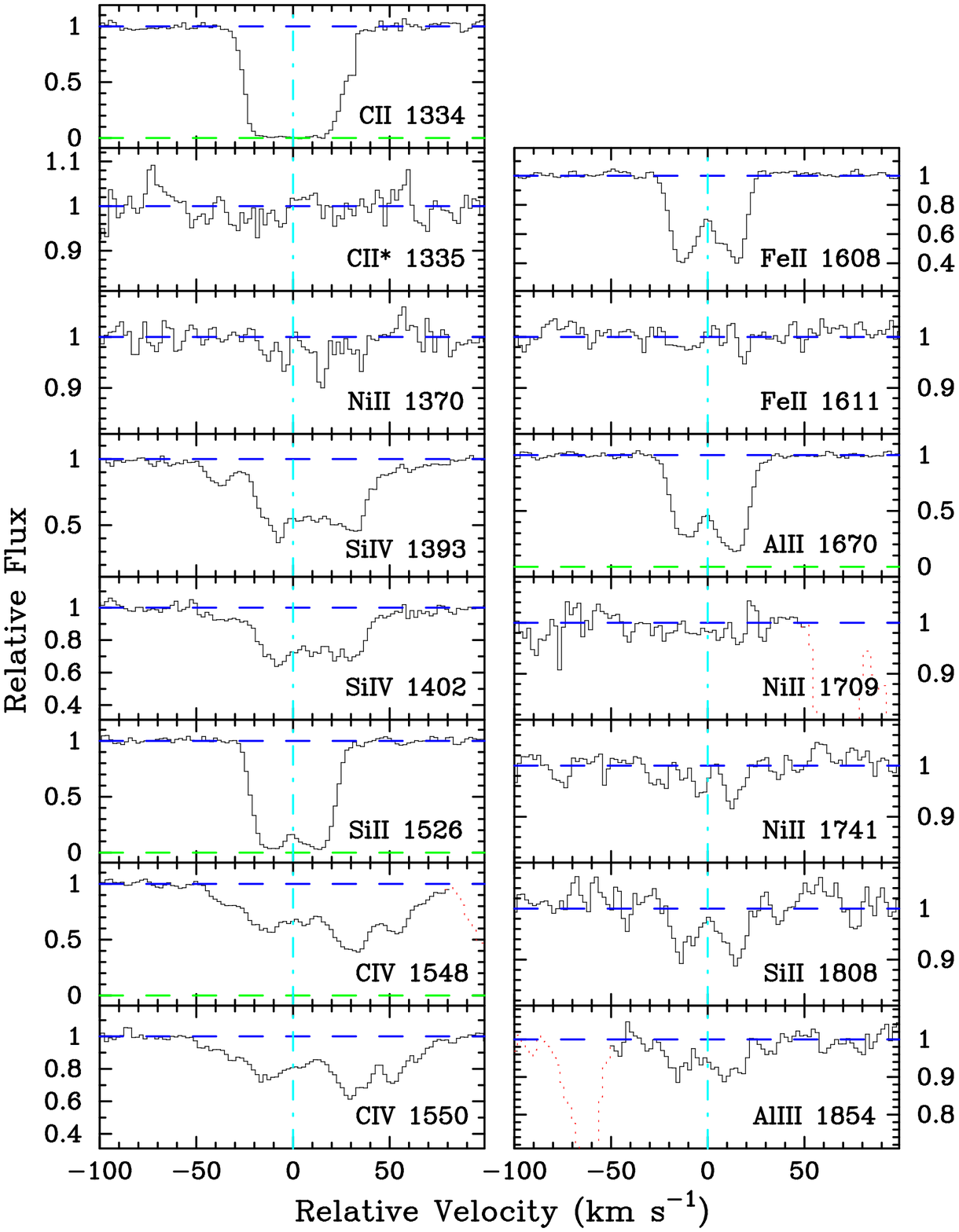}
\end{center}
\end{figure*}

\begin{figure*}
\begin{center}
\includegraphics[height=8.5in, width=6.0in]{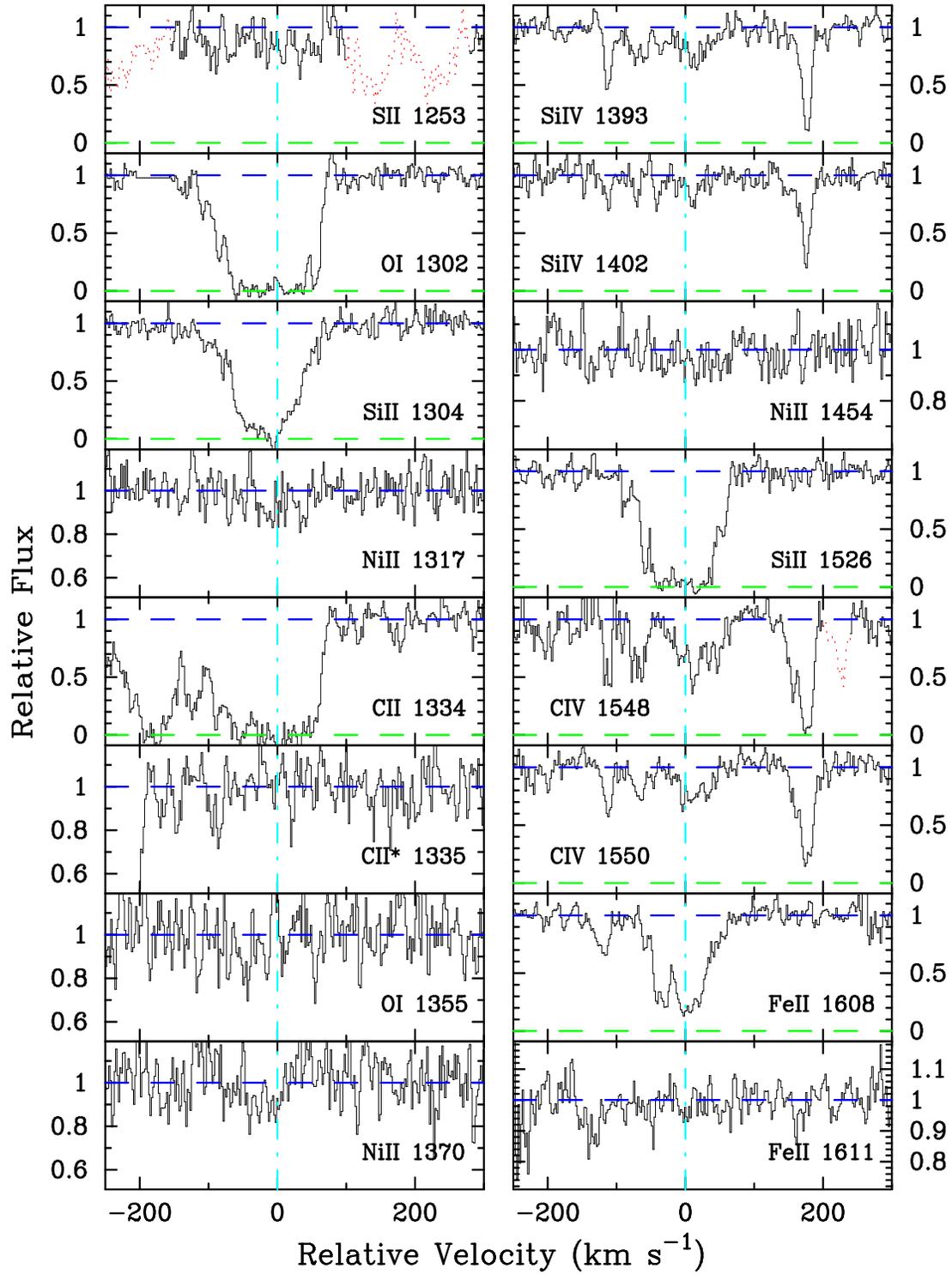}
\caption{Velocity plot of the metal-line transitions for the 
damped \lya system at $z = 2.465$ toward Q0836+11.  
The vertical line at $v=0$ corresponds to $z = 2.46527$.}
\label{fig:0836}
\end{center}
\end{figure*}

\begin{figure*}
\begin{center}
\includegraphics[height=8.5in, width=6.0in]{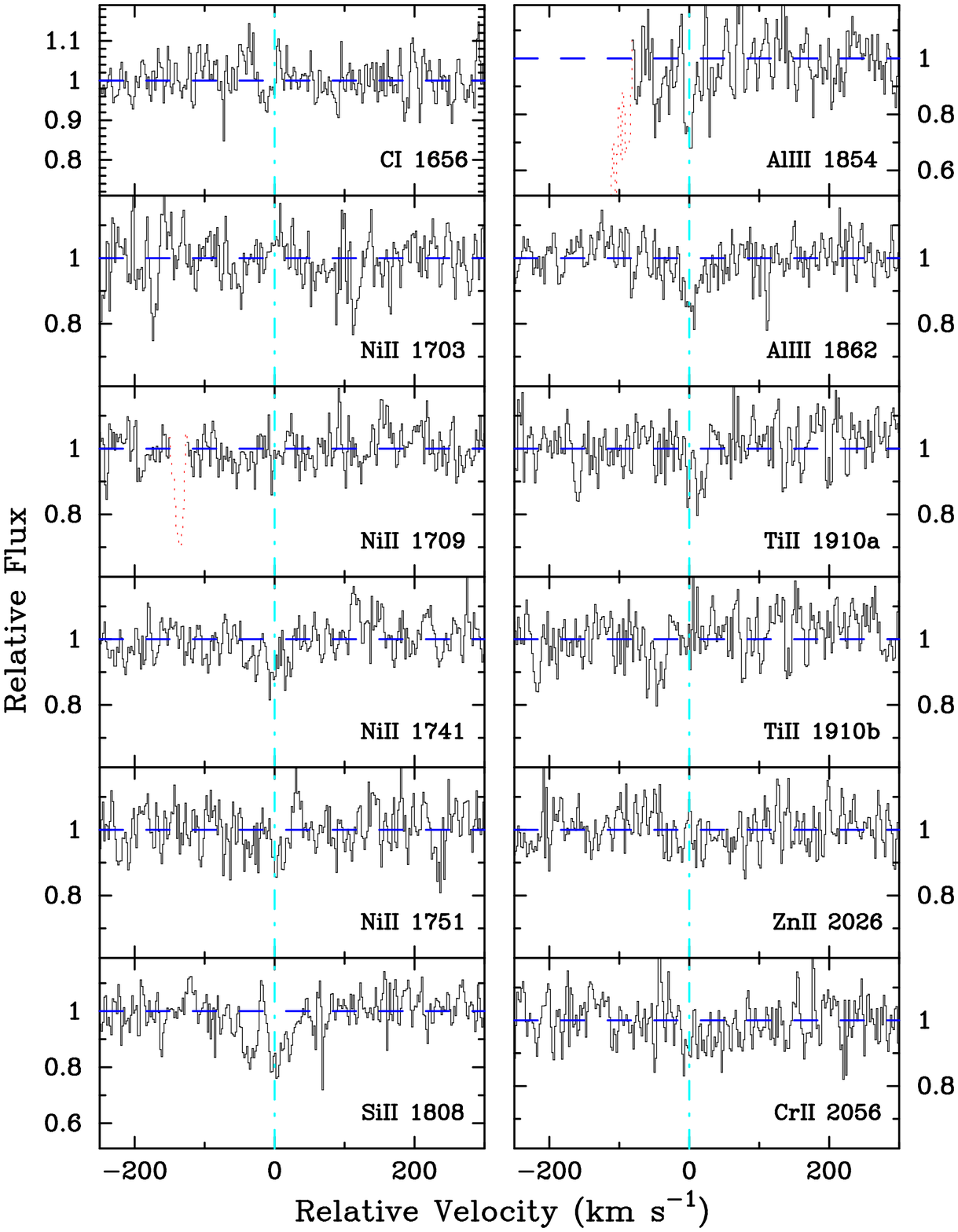}
\end{center}
\end{figure*}

\clearpage

\begin{table}[ht]\footnotesize
\begin{center}
\caption{ {\sc
IONIC COLUMN DENSITIES: HS0741+4741, $z = 3.017$ \label{tab:HS0741+4741_3.017}}}
\begin{tabular}{lcccc}
\tableline
\tableline
Ion & $\lambda$ & AODM & $N_{\rm adopt}$ & [X/H] \\
\tableline
HI &1215 & $20.480  \pm 0.100  $ \\
C  II &1036&$>14.864$&$>14.864$&$>-2.166$\\  
C  II &1334&$>14.737$\\  
C  II &1335&$<12.554$\\  
C  IV &1548&$13.827 \pm  0.005$\\  
C  IV &1550&$13.847 \pm  0.009$\\  
O  I  & 971&$>15.711$&$>15.711$&$>-1.639$\\  
O  I  &1302&$>15.229$\\  
Al II &1670&$12.823 \pm  0.005$&$12.824 \pm  0.005$&$-2.146 \pm  0.100$\\  
Al III&1854&$12.161 \pm  0.043$\\  
Si II &1020&$14.162 \pm  0.019$&$14.354 \pm  0.003$&$-1.686 \pm  0.100$\\  
Si II &1260&$>13.937$\\  
Si II &1304&$14.368 \pm  0.003$\\  
Si II &1526&$>14.469$\\  
Si II &1808&$14.395 \pm  0.051$\\  
Si IV &1393&$13.382 \pm  0.006$\\  
Si IV &1402&$13.376 \pm  0.011$\\  
P  II &1152&$<12.080$&$<12.080$&$<-1.930$\\  
S  II &1259&$14.000 \pm  0.016$&$14.000 \pm  0.016$&$-1.680 \pm  0.101$\\  
Ar I  &1048&$13.166 \pm  0.020$&$13.166 \pm  0.020$&$-1.834 \pm  0.102$\\  
Fe II &1063&$14.075 \pm  0.010$&$14.052 \pm  0.005$&$-1.928 \pm  0.100$\\  
Fe II &1096&$14.068 \pm  0.015$\\  
Fe II &1608&$14.041 \pm  0.006$\\  
Fe II &1611&$<14.082$\\  
Co II &1574&$<12.958$&$<12.958$&$<-0.432$\\  
Ni II &1317&$12.658 \pm  0.110$&$12.758 \pm  0.049$&$-1.972 \pm  0.111$\\  
Ni II &1370&$12.736 \pm  0.081$\\  
Ni II &1454&$<12.833$\\  
Ni II &1709&$12.745 \pm  0.144$\\  
Ni II &1741&$12.943 \pm  0.078$\\  
\tableline
\end{tabular}
\end{center}
\end{table}

\begin{table}[ht]\footnotesize
\begin{center}
\caption{ {\sc
IONIC COLUMN DENSITIES: Q0836+11, $z = 2.465$ \label{tab:Q0836+11_2.465}}}
\begin{tabular}{lcccc}
\tableline
\tableline
Ion & $\lambda$ & AODM & $N_{\rm adopt}$ & [X/H] \\
\tableline
HI &1215 & $20.580  \pm 0.100  $ \\
C  II &1334&$>15.026$&$>15.026$&$>-2.104$\\  
C  II &1335&$<13.121$\\  
C  IV &1548&$>14.218$\\  
O  I  &1302&$>15.485$&$>15.485$&$>-1.965$\\  
O  I  &1355&$<18.186$\\  
Al III&1854&$12.459 \pm  0.093$\\  
Al III&1862&$12.634 \pm  0.076$\\  
Si II &1304&$>14.918$&$14.987 \pm  0.045$&$-1.153 \pm  0.110$\\  
Si II &1526&$>14.850$\\  
Si II &1808&$14.987 \pm  0.045$\\  
Si IV &1393&$13.557 \pm  0.024$\\  
Si IV &1402&$13.592 \pm  0.043$\\  
S  II &1253&$<14.660$&$<14.660$&$<-1.120$\\  
Ti II &1910&$<12.537$&$<12.538$&$<-0.982$\\  
Cr II &2056&$<12.898$&$<12.898$&$<-1.352$\\  
Cr II &2066&$<13.230$\\  
Fe II &1608&$14.677 \pm  0.011$&$14.677 \pm  0.011$&$-1.403 \pm  0.101$\\  
Fe II &1611&$<14.784$\\  
Co II &1466&$<13.383$&$<13.383$&$<-0.107$\\  
Co II &1574&$<13.442$\\  
Ni II &1317&$13.364 \pm  0.105$&$13.388 \pm  0.065$&$-1.442 \pm  0.119$\\  
Ni II &1370&$<13.339$\\  
Ni II &1454&$<13.365$\\  
Ni II &1703&$<14.058$\\  
Ni II &1709&$<13.275$\\  
Ni II &1741&$13.406 \pm  0.083$\\  
Ni II &1751&$<13.393$\\  
Zn II &2026&$<12.119$&$<12.119$&$<-1.131$\\  
\tableline
\end{tabular}
\end{center}
\end{table}

\subsection{Q0836$+$11, $z$ = 2.465}

This damped \lya system is drawn from the LBQS survey \citep{wol95}
and we adopt the $\N{HI}$ value obtained from their analysis.  Our observations
cover a large number of transitions, many of which only provide upper
limits owing to the relatively poor S/N (Figure~\ref{fig:0836},
Table~\ref{tab:Q0836+11_2.465}).  In passing, we note 
that some of the Ni~II upper limits are in
contradiction with our adopted $\N{Ni^+}$ value.  This might reflect
an error in the relative $f$-values but more likely reflects the 
large error in the adopted value.

\begin{table}[ht]\footnotesize
\begin{center}
\caption{ {\sc
IONIC COLUMN DENSITIES: Q0841+12, $z = 2.375$ \label{tab:Q0841+12_2.375}}}
\begin{tabular}{lcccc}
\tableline
\tableline
Ion & $\lambda$ & AODM & $N_{\rm adopt}$ & [X/H] \\
\tableline
HI &1215 & $20.950  \pm 0.087  $ \\
Co II &1466&$<13.047$&$<12.990$&$<-0.870$\\  
Co II &1941&$<12.990$\\  
Ni II &1454&$13.415 \pm  0.075$&$13.523 \pm  0.030$&$-1.677 \pm  0.092$\\  
Ni II &1741&$13.560 \pm  0.040$\\  
Ni II &1751&$13.547 \pm  0.055$\\  
\tableline
\end{tabular}
\end{center}
\end{table}

\begin{figure}[ht]
\begin{center}
\includegraphics[height=4.0in, width=3.0in]{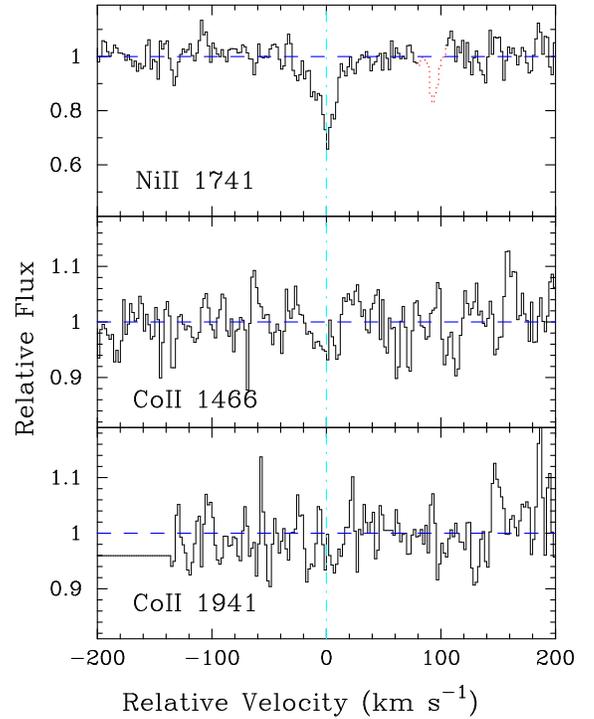}
\caption{Velocity plot of two Co~II profiles for the 
damped \lya system at $z = 2.375$ toward Q0841+12.  
For comparison, we also plot the Ni~II 1741 profile.
The vertical line at $v=0$ corresponds to $z = 2.374518$.}
\label{fig:0841A}
\end{center}
\end{figure}

\begin{figure}[ht]
\begin{center}
\includegraphics[height=4.3in, width=3.3in]{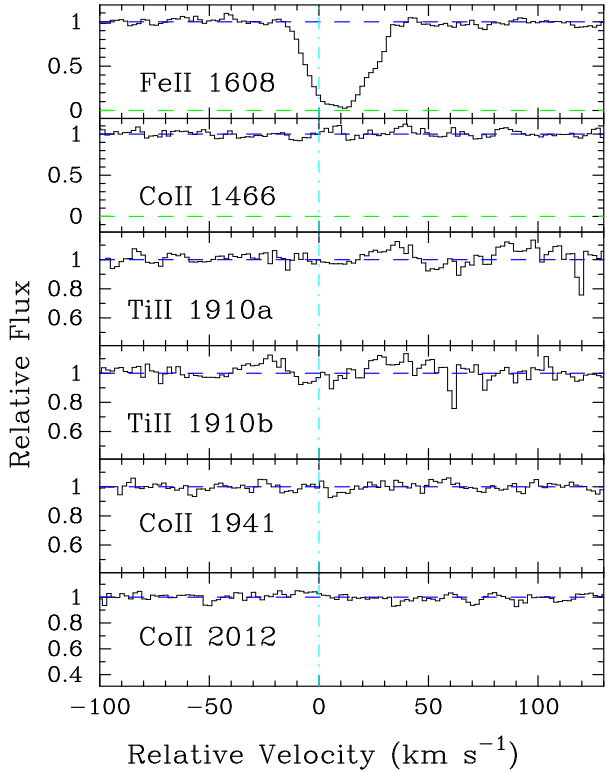}
\caption{Velocity plot of the new metal-line transitions for the 
damped \lya system at $z = 2.476$ toward Q0841+12.  
For comparison, we also plot the Fe~II 1608 profile.
The vertical line at $v=0$ corresponds to $z = 2.476219$.}
\label{fig:0841B}
\end{center}
\end{figure}

\subsection{Q0841+12, $z$ = 2.375 and $z$ = 2.476}

We augment the measurements presented in PW99 with a few additional
Ni~II transitions and several Co~II and Ti~II upper limits.  These
transitions are plotted in Figures~\ref{fig:0841A} and \ref{fig:0841B}
and tabulated in Tables~\ref{tab:Q0841+12_2.375} and \ref{tab:Q0841+12_2.476}.
As discussed in PW99, we based the Fe abundance for the $z=2.476$ system
on the saturated Fe~II 1608 profile.  We now choose to report the Fe~II 1608
column density as a lower limit on $\N{Fe^+}$ and adopt an Fe 
abundance based on averaging the lower and upper limits from 
Fe~II 1608 and Fe~II 1611 respectively: $\log \N{Fe^+} = 14.53 \pm 0.05$.

\begin{table}[ht]\footnotesize
\begin{center}
\caption{ {\sc
IONIC COLUMN DENSITIES: Q0841+12, $z = 2.476$ \label{tab:Q0841+12_2.476}}}
\begin{tabular}{lcccc}
\tableline
\tableline
Ion & $\lambda$ & AODM & $N_{\rm adopt}$ & [X/H] \\
\tableline
HI &1215 & $20.780  \pm 0.097  $ \\
Ti II &1910&$<12.158$&$<12.158$&$<-1.562$\\  
Fe II &1608&$>14.517$\\  
Fe II &1611&$<14.543$\\  
Co II &1466&$<13.206$&$<12.726$&$<-0.964$\\  
Co II &1941&$<12.783$\\  
Co II &2012&$<12.726$\\  
Ni II &1709&$13.415 \pm  0.060$&$13.355 \pm  0.040$&$-1.675 \pm  0.105$\\  
Ni II &1741&$13.321 \pm  0.054$\\  
Ni II &1751&$<13.348$\\  
\tableline
\end{tabular}
\end{center}
\end{table}

\break

\begin{figure}[ht]
\begin{center}
\includegraphics[height=3.5in, width=2.7in]{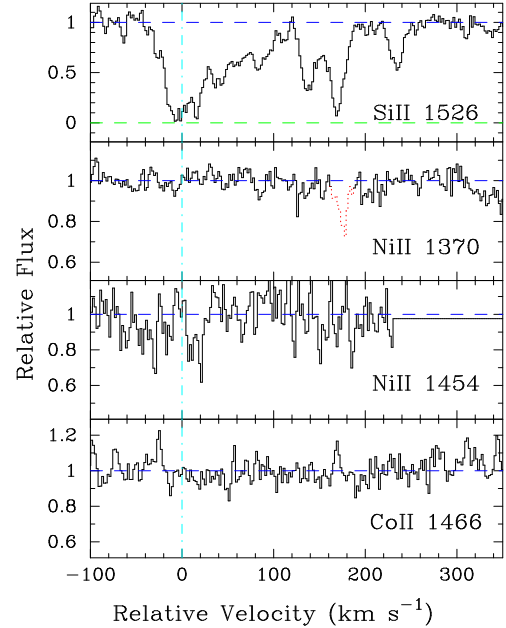}
\caption{Velocity plot of the new metal-line transitions for the 
damped \lya system at $z = 3.857$ toward BRI0951--04. 
For comparison, we also plot the Si~II 1526 profile.
The vertical line at $v=0$ corresponds to $z = 3.856689$.}
\label{fig:0951A}
\end{center}
\end{figure}

\begin{figure}[ht]
\begin{center}
\includegraphics[height=4.3in, width=3.3in]{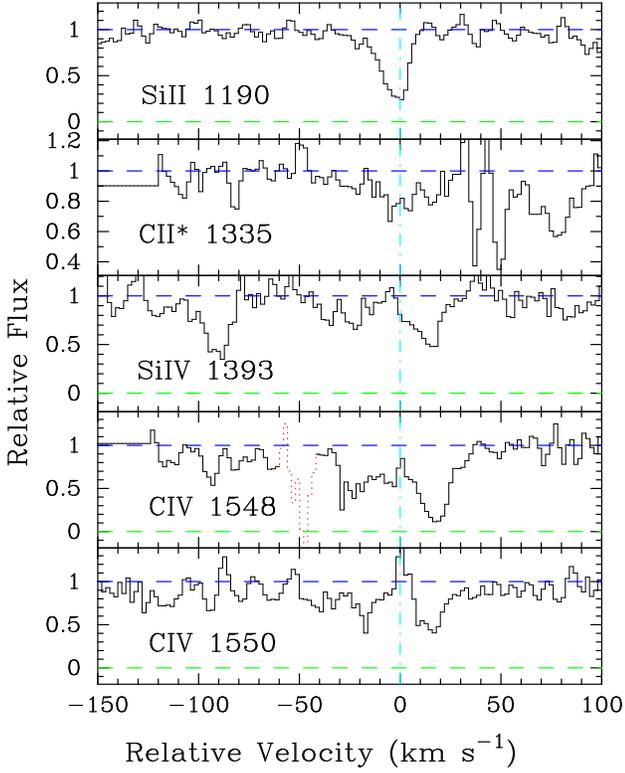}
\caption{Velocity plot of the new metal-line transitions for the 
damped \lya system at $z = 4.203$ toward BRI0951--04. 
The vertical line at $v=0$ corresponds to $z = 4.202896$.}
\label{fig:0951B}
\end{center}
\end{figure}

\subsection{BRI0951$-$04, $z$ = 3.857 and $z$ = 4.203}

Our combined spectrum now includes a second setup with significant 
coverage blueward of the \lya peak.  Unfortunately, the new data adds
only a few unblended transitions to the analysis 
(Figures~\ref{fig:0951A},\ref{fig:0951B} and 
Tables~\ref{tab:BRI0951-04_3.857},\ref{tab:BRI0951-04_4.203}).
With respect to PW99, we now suspect that the feature at $+177$~km/s
in the $z=3.857$ Ni~II 1370 profile is unrelated to that transition and obtain
an upper limit on $\N{Ni^+}$.  This value is in much better agreement with
the Fe and Al abundances.
In terms of the system at $z=4.203$, we still have no reliable estimate
of the Fe-peak abundance.  The combination of poor S/N and low HI
column density have resulted in the non-detection of Fe~II 1608 and
our observations did not cover C~II 1334 or Al~II 1670.
Finally, we revise the oxygen abundance to account for the
saturated OI~1302 profile.

\begin{table}[ht]\footnotesize
\begin{center}
\caption{ {\sc
IONIC COLUMN DENSITIES: BRI0951-04, $z = 3.857$ \label{tab:BRI0951-04_3.857}}}
\begin{tabular}{lcccc}
\tableline
\tableline
Ion & $\lambda$ & AODM & $N_{\rm adopt}$ & [X/H] \\
\tableline
HI &1215 & $20.600  \pm 0.100  $ \\
Co II &1466&$<13.597$&$<13.597$&$< 0.087$\\  
Ni II &1370&$<12.977$&$<12.977$&$<-1.873$\\  
Ni II &1454&$<13.927$\\  
\tableline
\end{tabular}
\end{center}
\end{table}

\begin{table}[ht]\footnotesize
\begin{center}
\caption{ {\sc
IONIC COLUMN DENSITIES: BRI0951-04, $z = 4.203$ \label{tab:BRI0951-04_4.203}}}
\begin{tabular}{lcccc}
\tableline
\tableline
Ion & $\lambda$ & AODM & $N_{\rm adopt}$ & [X/H] \\
\tableline
HI &1215 & $20.400  \pm 0.100  $ \\
C  IV &1548&$$\\  
C  IV &1550&$13.945 \pm  0.052$\\  
O  I  &1302&$>14.596$&$>14.596$&$>-2.674$\\  
Si II &1190&$13.417 \pm  0.042$&$13.342 \pm  0.030$&$-2.618 \pm  0.104$\\  
Si II &1526&$13.302 \pm  0.056$\\  
Si IV &1393&$12.918 \pm  0.063$\\  
\tableline
\end{tabular}
\end{center}
\end{table}

\subsection{BRI0952$-$01, $z$ = 4.024}

\begin{figure*}
\begin{center}
\includegraphics[height=8.5in, width=6.0in]{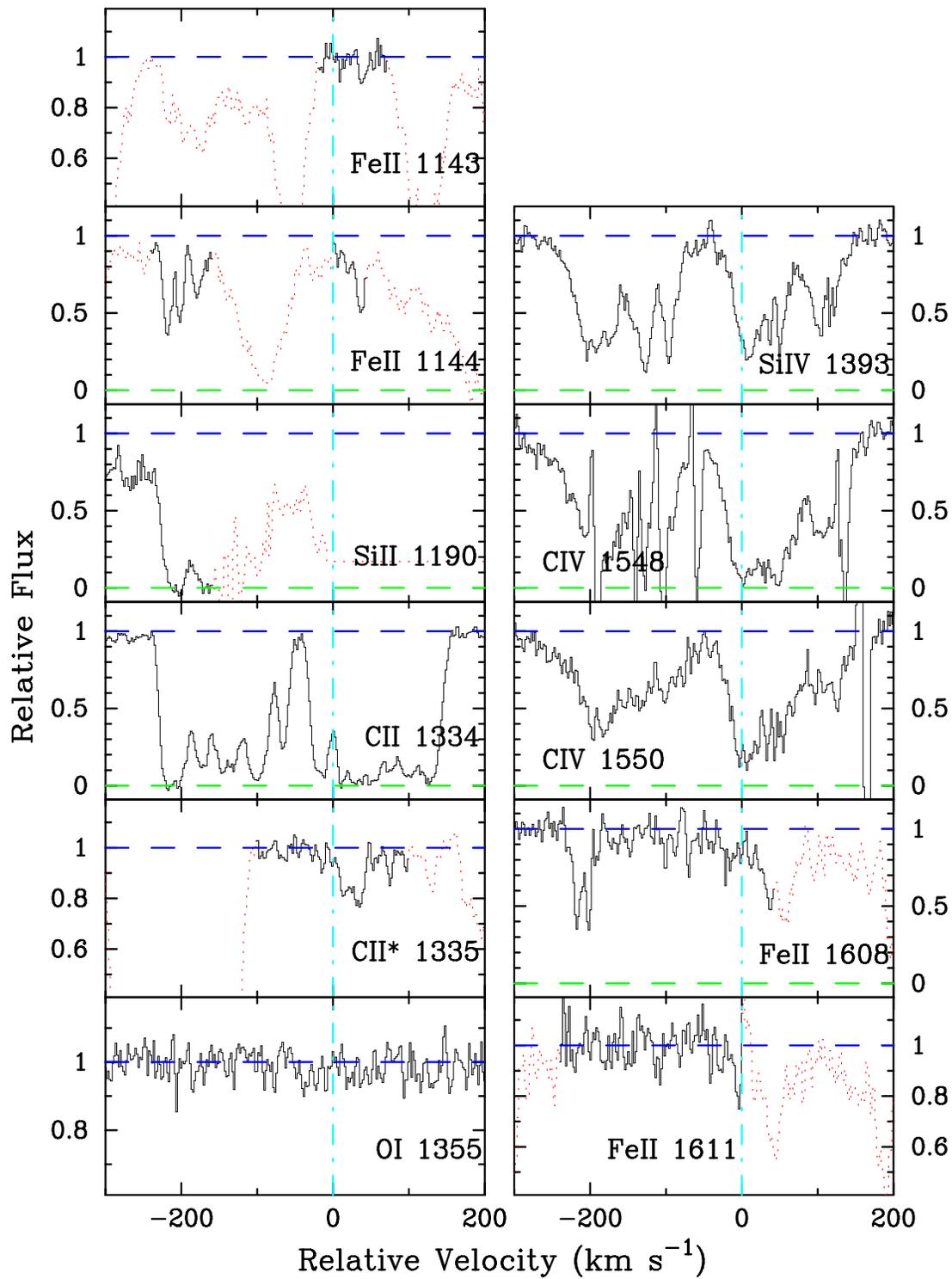}
\caption{Velocity plot of the metal-line transitions for the 
damped \lya system at $z = 4.024$ toward BRI0952--01.
The vertical line at $v=0$ corresponds to $z = 4.024433.$}
\label{fig:0952}
\end{center}
\end{figure*}

This damped \lya system was identified by \cite{storr96} and confirmed
with a follow-up LRIS spectrum by \cite{storr00}.  We adopt the 
$\N{HI}$ value from the latter analysis.  Figure~\ref{fig:0952} presents
the velocity profiles covered by our single setup. 
Unfortunately, a
mis-estimate of the absorption redshift coupled with several line
blends have limited our ionic column density measurements of this system
(Table~\ref{tab:BRI0952-01_4.024}).  
As reported in \cite{pro00}, we estimate the Fe$^+$ column
density by combining the unblended features observed for the Fe~II 1144
and 1608 profiles.  With the updated oscillator strengths, we find
$\N{Fe^+} = 14.187 \pm 0.07$.  

\begin{table}[ht]\footnotesize
\begin{center}
\caption{ {\sc
IONIC COLUMN DENSITIES: BRI0952-01, $z = 4.024$ \label{tab:BRI0952-01_4.024}}}
\begin{tabular}{lcccc}
\tableline
\tableline
Ion & $\lambda$ & AODM & $N_{\rm adopt}$ & [X/H] \\
\tableline
HI &1215 & $20.550  \pm 0.100  $ \\
C  II &1334&$>15.312$&$>15.312$&$>-1.788$\\  
C  II &1335&$13.549 \pm  0.024$\\  
C  IV &1550&$14.796 \pm  0.009$\\  
Si IV &1393&$14.134 \pm  0.006$\\  
Fe II &1144&$>13.864$&$14.187 \pm  0.076$&$-1.863 \pm  0.126$\\  
Fe II &1608&$>13.746$&\\
Co II &1574&$<13.750$&$<13.750$&$< 0.290$\\  
Ni II &1454&$<13.439$&$<13.439$&$<-1.361$\\  
\tableline
\end{tabular}
\end{center}
\end{table}

\subsection{PSS0957$+$33, $z$=3.279 and $z$ = 4.178}

\begin{figure*}
\begin{center}
\includegraphics[height=8.5in, width=6.0in]{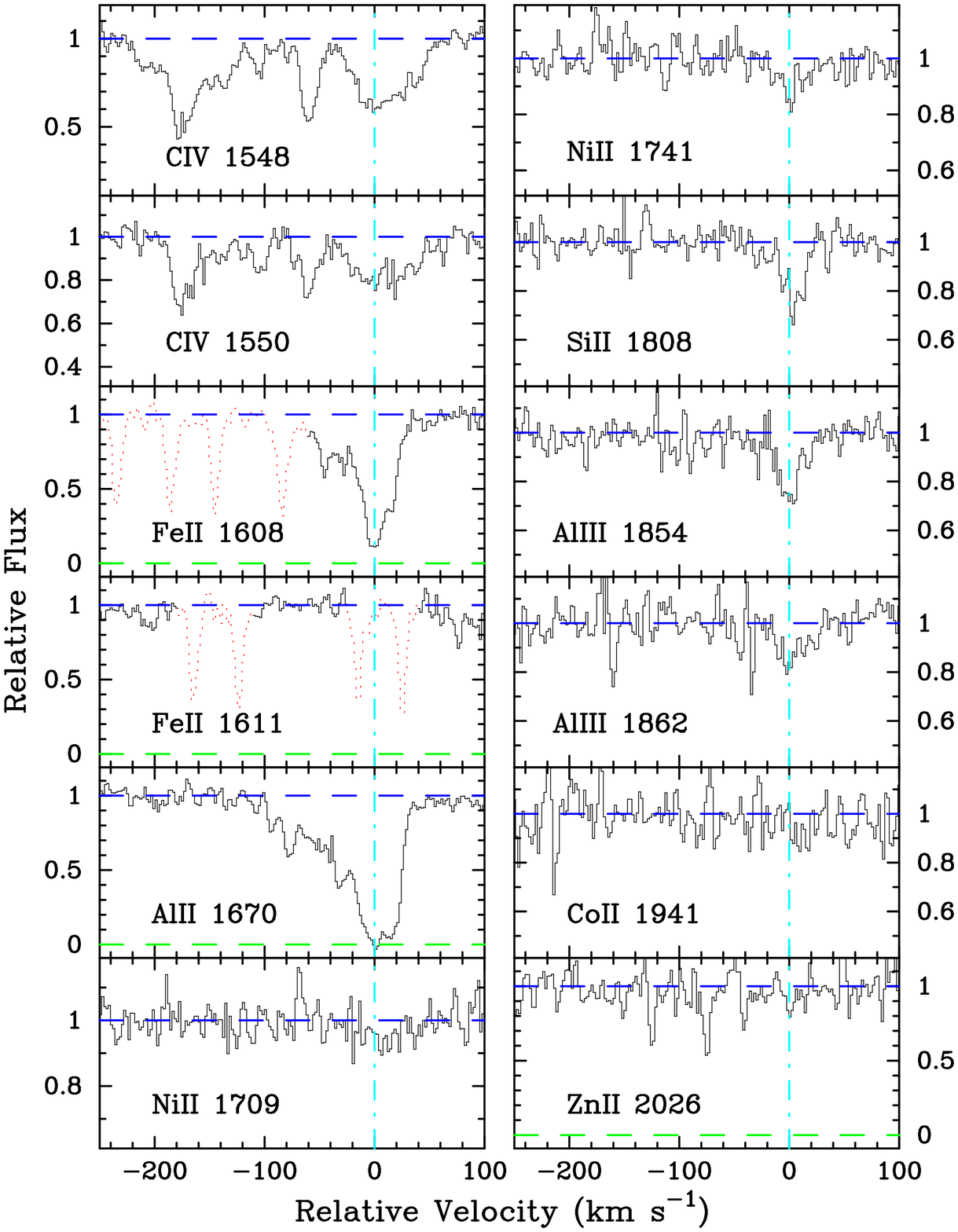}
\caption{Velocity plot of the metal-line transitions for the 
damped \lya system at $z = 3.279$ toward PSS0957+33.
The vertical line at $v=0$ corresponds to $z = 3.279576.$}
\label{fig:0957A}
\end{center}
\end{figure*}

\begin{figure*}
\begin{center}
\includegraphics[height=8.5in, width=6.0in]{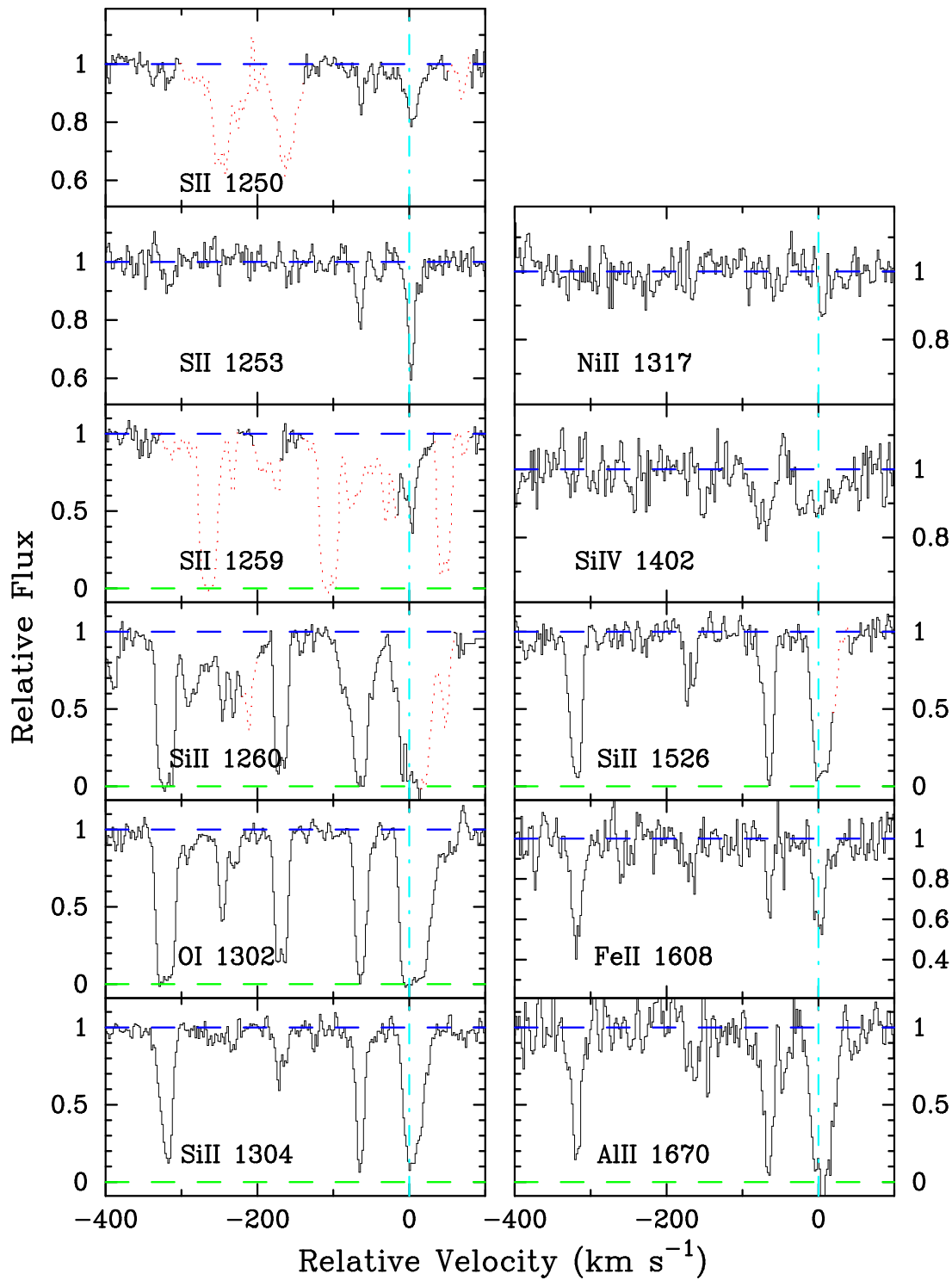}
\caption{Velocity plot of the metal-line transitions for the 
damped \lya system at $z = 4.178$ toward PSS0957+33.
The vertical line at $v=0$ corresponds to $z = 4.179825.$}
\label{fig:0957B}
\end{center}
\end{figure*}

The two damped \lya systems toward this PSS quasar \citep{djg98} 
were discovered during the first night of our ongoing ESI project 
designed to discover and measure the metallicity of $z>3$ damped systems
(PGW01).
Given the apparent brightness of this quasar we chose
to obtain a HIRES spectrum.  Figures~\ref{fig:0957A} and \ref{fig:0957B}
present the metal-line transition identified in our HIRES spectrum
and Tables~\ref{tab:PSS0957+33_3.279} and \ref{tab:PSS0957+33_4.178} 
give the ionic column densities.  

\begin{table}[ht]\footnotesize
\begin{center}
\caption{ {\sc
IONIC COLUMN DENSITIES: PSS0957+33, $z = 3.279$ \label{tab:PSS0957+33_3.279}}}
\begin{tabular}{lcccc}
\tableline
\tableline
Ion & $\lambda$ & AODM & $N_{\rm adopt}$ & [X/H] \\
\tableline
HI &1215 & $20.320  \pm 0.080  $ \\
C  IV &1548&$13.979 \pm  0.009$\\  
C  IV &1550&$13.956 \pm  0.019$\\  
Al II &1670&$>13.322$&$>13.322$&$>-1.488$\\  
Al III&1854&$12.514 \pm  0.048$\\  
Al III&1862&$12.578 \pm  0.087$\\  
Si II &1808&$14.880 \pm  0.053$&$14.880 \pm  0.053$&$-1.000 \pm  0.096$\\  
Fe II &1608&$14.367 \pm  0.016$&$14.367 \pm  0.016$&$-1.453 \pm  0.082$\\  
Co II &1941&$<13.285$&$<13.285$&$< 0.055$\\  
Ni II &1709&$13.283 \pm  0.122$&$13.318 \pm  0.070$&$-1.252 \pm  0.106$\\  
Ni II &1741&$13.339 \pm  0.085$\\  
Zn II &2026&$<12.127$&$<12.127$&$<-0.863$\\  
\tableline
\end{tabular}
\end{center}
\end{table}

\begin{table}[ht]\footnotesize
\begin{center}
\caption{ {\sc
IONIC COLUMN DENSITIES: PSS0957+33, $z = 4.178$ \label{tab:PSS0957+33_4.178}}}
\begin{tabular}{lcccc}
\tableline
\tableline
Ion & $\lambda$ & AODM & $N_{\rm adopt}$ & [X/H] \\
\tableline
HI &1215 & $20.500  \pm 0.100  $ \\
O  I  &1302&$>15.344$&$>15.344$&$>-2.026$\\  
Al II &1670&$>13.256$&$>13.256$&$>-1.734$\\  
Si II &1304&$14.556 \pm  0.012$&$14.556 \pm  0.012$&$-1.504 \pm  0.101$\\  
Si II &1526&$>14.488$\\  
Si IV &1402&$13.155 \pm  0.049$\\  
S  II &1250&$>14.647$&$14.392 \pm  0.060$&$-1.308 \pm  0.117$\\  
S  II &1253&$14.392 \pm  0.060$\\  
Fe II &1608&$14.129 \pm  0.045$&$14.129 \pm  0.045$&$-1.871 \pm  0.110$\\  
Co II &1466&$<13.648$&$<13.648$&$< 0.238$\\  
Co II &1574&$<13.907$\\  
Ni II &1317&$<12.910$&$<12.910$&$<-1.840$\\  
\tableline
\end{tabular}
\end{center}
\end{table}

To test the metallicities obtained with the ESI spectrum, we can
compare the HIRES values with the
Fe$^+$ column densities adopted in PGW01 after correcting for the
new oscillator strengths.  For the system at
$z=3.279$, the Fe~II 1608 column densities are in excellent agreement
but because this transition is blended with telluric absorption it might
only provide an upper limit on $\N{Fe^+}$.  In PGW01, we derived the
Fe$^+$ column density from Fe~II 2344 which is redward of our HIRES coverage.
We now suspect this value was an underestimate of $\N{Fe^+}$ as it implies
[Ni/Fe]~$> +0.3$~dex.  For now, we adopt an Fe$^+$ column density based
on the Fe~II 1608 profile.

The system at $z=4.178$ presents a more worrisome picture regarding
column densities derived from ESI echellette observations.  Comparing
the column densities for Fe~II 1608, we find that we underestimated 
$\N{Fe^+}$ by $\approx 0.3$~dex with the ESI spectrum.  It is somewhat
puzzling given that the profile is not very strong (unlike the Si~II profiles
for example) and the quality of the ESI data is high.
The strongest features are relatively narrow so the difference is probably
an effect of the lower resolution.  
Another puzzling aspect of the $z=4.178$ systems is the S$^+$
column density derived from S~II 1250 and 1253.  Although the S~II 1250
profile is partially blended with an unrelated C~IV system 
($z_{abs} = 3.181$), we are confident that the absorption at
$-100 \mkms <v<20 \mkms$ is free of contamination from the C~IV system.
In this case, $\N{S~II~1250}$ provides a lower limit on 
$\log \N{S^+} > 14.65$.  This lower limit, however, is well in excess 
of the $\N{S^+}$ value derived from the unblended S~II 1253 transition.
The component at $v=0$~km/s in the S~II 1250 profile is wider than its
counterpart in the S~II 1253 profile, but there is no identifiable blend.
Perhaps the difference suggests an extreme case of hidden saturation
\citep{sav91}.  If it is line saturation, this helps explain why the
ESI data significantly underestimates the column densities in this case
and it also raises the possibility that the abundances derived from the
HIRES observations are underestimates.  
This would be particularly surprising given that the spectra has a 
resolution of $R=47000$.  Unfortunately, our HIRES spectrum
did not cover any other pair of unsaturated transitions from the same ion
to further test this issue.
Furthermore, the difference even exists in a comparison 
of S~II 1250 and 1253 with respect to the feature at $v \approx -75$~km/s.
Perhaps this is all the result of errors in the S~II oscillator 
strengths, but it would have likely been identified by researchers
who study the ISM \citep[e.g.][]{howk99}.

\subsection{BRI1108$-$07, $z = 3.608$}

\begin{figure*}
\begin{center}
\includegraphics[height=8.5in, width=6.0in]{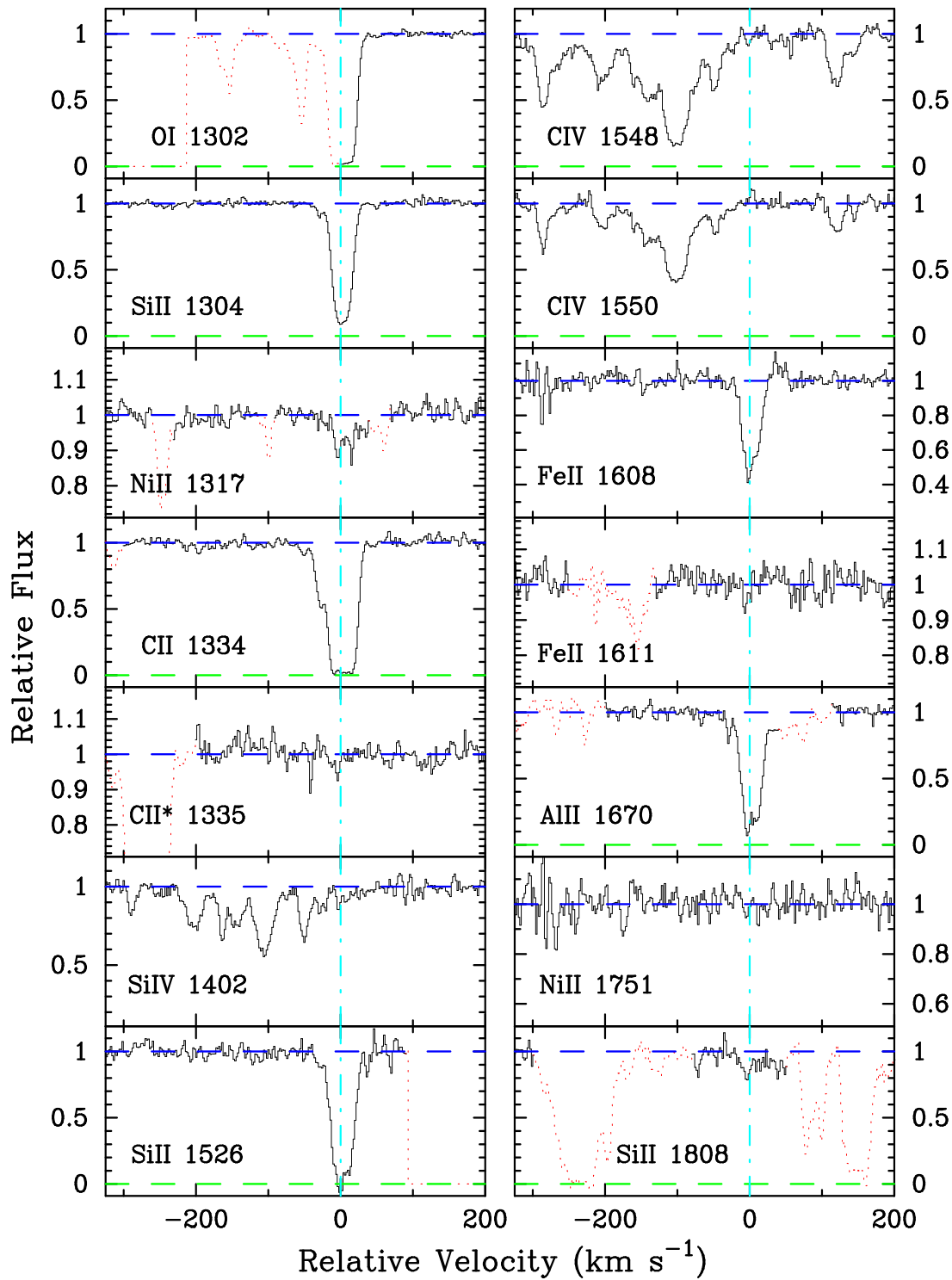}
\caption{Velocity plot of the metal-line transitions for the 
damped \lya system at $z = 3.608$ toward BRI1108--07.
The vertical line at $v=0$ corresponds to $z = 3.607619$.}
\label{fig:1108}
\end{center}
\end{figure*}

This damped \lya system was discovered and confirmed by \cite{storr96}. 
The quasar is relatively bright and we obtained
a reasonably high S/N HIRES spectrum.  
Figure~\ref{fig:1108} presents the velocity profiles and 
Table~\ref{tab:BRI1108-07_3.608} lists the ionic column densities.
The Fe$^+$ and Si$^+$ column densities are well measured and indicate
a very large Si/Fe ratio, perhaps indicative of substantial Type~II SN
enrichment.

\begin{table}[ht]\footnotesize
\begin{center}
\caption{ {\sc
IONIC COLUMN DENSITIES: BRI1108-07, $z = 3.608$ \label{tab:BRI1108-07_3.608}}}
\begin{tabular}{lcccc}
\tableline
\tableline
Ion & $\lambda$ & AODM & $N_{\rm adopt}$ & [X/H] \\
\tableline
HI &1215 & $20.500  \pm 0.100  $ \\
C  II &1334&$>14.675$&$>14.675$&$>-2.375$\\  
C  II &1335&$<12.341$\\  
C  IV &1548&$14.293 \pm  0.006$\\  
C  IV &1550&$14.219 \pm  0.011$\\  
O  I  &1302&$>14.873$&$>14.873$&$>-2.497$\\  
Al II &1670&$12.822 \pm  0.015$&$12.822 \pm  0.015$&$-2.168 \pm  0.101$\\  
Si II &1304&$14.262 \pm  0.004$&$14.262 \pm  0.004$&$-1.798 \pm  0.100$\\  
Si II &1526&$>14.260$\\  
Si II &1808&$<14.665$\\  
Si IV &1402&$13.685 \pm  0.010$\\  
Fe II &1608&$13.884 \pm  0.014$&$13.884 \pm  0.014$&$-2.116 \pm  0.101$\\  
Fe II &1611&$<14.269$\\  
Ni II &1317&$<13.136$&$<13.136$&$<-1.614$\\  
Ni II &1751&$<13.180$\\  
\tableline
\end{tabular}
\end{center}
\end{table}

\subsection{Q1210+17, $z = 1.892$}

\begin{figure*}
\begin{center}
\includegraphics[height=8.5in, width=6.0in]{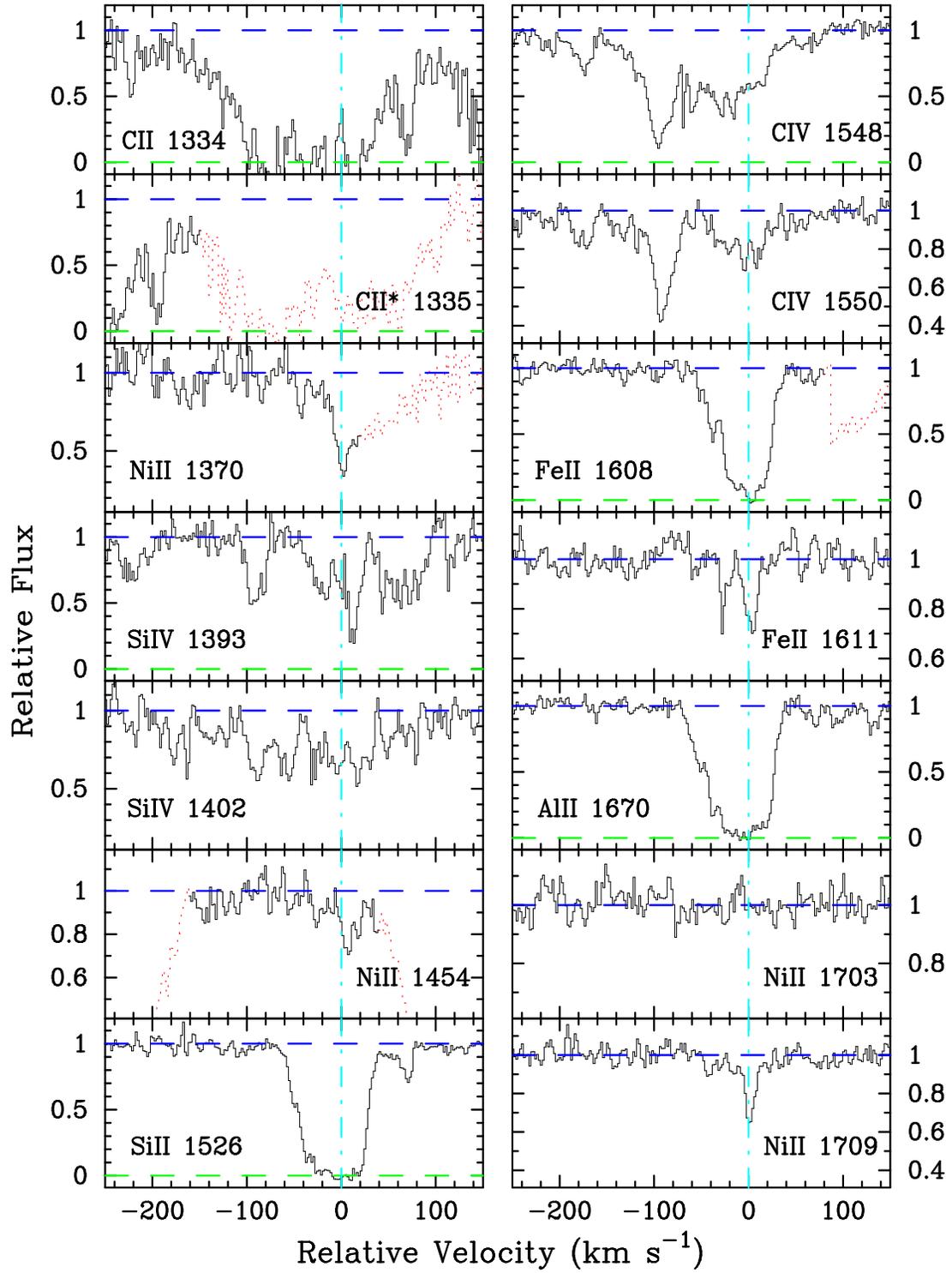}
\caption{Velocity plot of the metal-line transitions for the 
damped \lya system at $z = 1.892$ toward Q1210+17.
The vertical line at $v=0$ corresponds to $z = 1.891755$.}
\label{fig:1210}
\end{center}
\end{figure*}

This system is a member of the LBQS sample and we have adopted the 
$\N{HI}$ value from their analysis.  We plot all of the transitions
covered by our observations in Figure~\ref{fig:1210} and list the column
densities in Table~\ref{tab:Q1210+17_1.892}.  
This damped system exhibits a relatively
low Zn/Fe ratio which suggests it is largely free of dust depletion.
In passing, we note a remarkable similarity of the 
relative abundances of Si, Ni, Cr, Fe, and Zn with the same pattern observed
by \cite{molaro00} for the damped system toward Q0000$-$26, albeit at
a much higher metallicity.

\begin{table}[ht]\footnotesize
\begin{center}
\caption{ {\sc
IONIC COLUMN DENSITIES: Q1210+17, $z = 1.892$ \label{tab:Q1210+17_1.892}}}
\begin{tabular}{lcccc}
\tableline
\tableline
Ion & $\lambda$ & AODM & $N_{\rm adopt}$ & [X/H] \\
\tableline
HI &1215 & $20.600  \pm 0.100  $ \\
C  IV &1548&$<14.230$\\  
C  IV &1550&$14.053 \pm  0.017$\\  
Al II &1670&$>13.440$&$>13.440$&$>-1.650$\\  
Al III&1854&$12.999 \pm  0.017$\\  
Al III&1862&$13.005 \pm  0.021$\\  
Si II &1526&$>14.780$&$15.285 \pm  0.018$&$-0.875 \pm  0.102$\\  
Si II &1808&$15.285 \pm  0.018$\\  
Si IV &1393&$13.594 \pm  0.025$\\  
Cr II &2056&$13.291 \pm  0.019$&$13.243 \pm  0.016$&$-1.027 \pm  0.101$\\  
Cr II &2062&$13.219 \pm  0.030$\\  
Cr II &2066&$13.091 \pm  0.058$\\  
Fe II &1608&$>14.780$&$14.951 \pm  0.063$&$-1.149 \pm  0.118$\\  
Fe II &1611&$14.951 \pm  0.063$\\  
Co II &1574&$<13.513$&$<12.728$&$<-0.782$\\  
Co II &2012&$<12.728$\\  
Ni II &1370&$<13.964$&$13.632 \pm  0.020$&$-1.218 \pm  0.102$\\  
Ni II &1454&$<13.832$\\  
Ni II &1709&$13.657 \pm  0.031$\\  
Ni II &1741&$13.628 \pm  0.029$\\  
Ni II &1751&$13.590 \pm  0.049$\\  
Zn II &2026&$12.409 \pm  0.032$&$12.370 \pm  0.029$&$-0.900 \pm  0.104$\\  
Zn II &2062&$12.263 \pm  0.069$\\  
\tableline
\end{tabular}
\end{center}
\end{table}

\begin{figure*}
\begin{center}
\includegraphics[height=8.5in, width=6.0in]{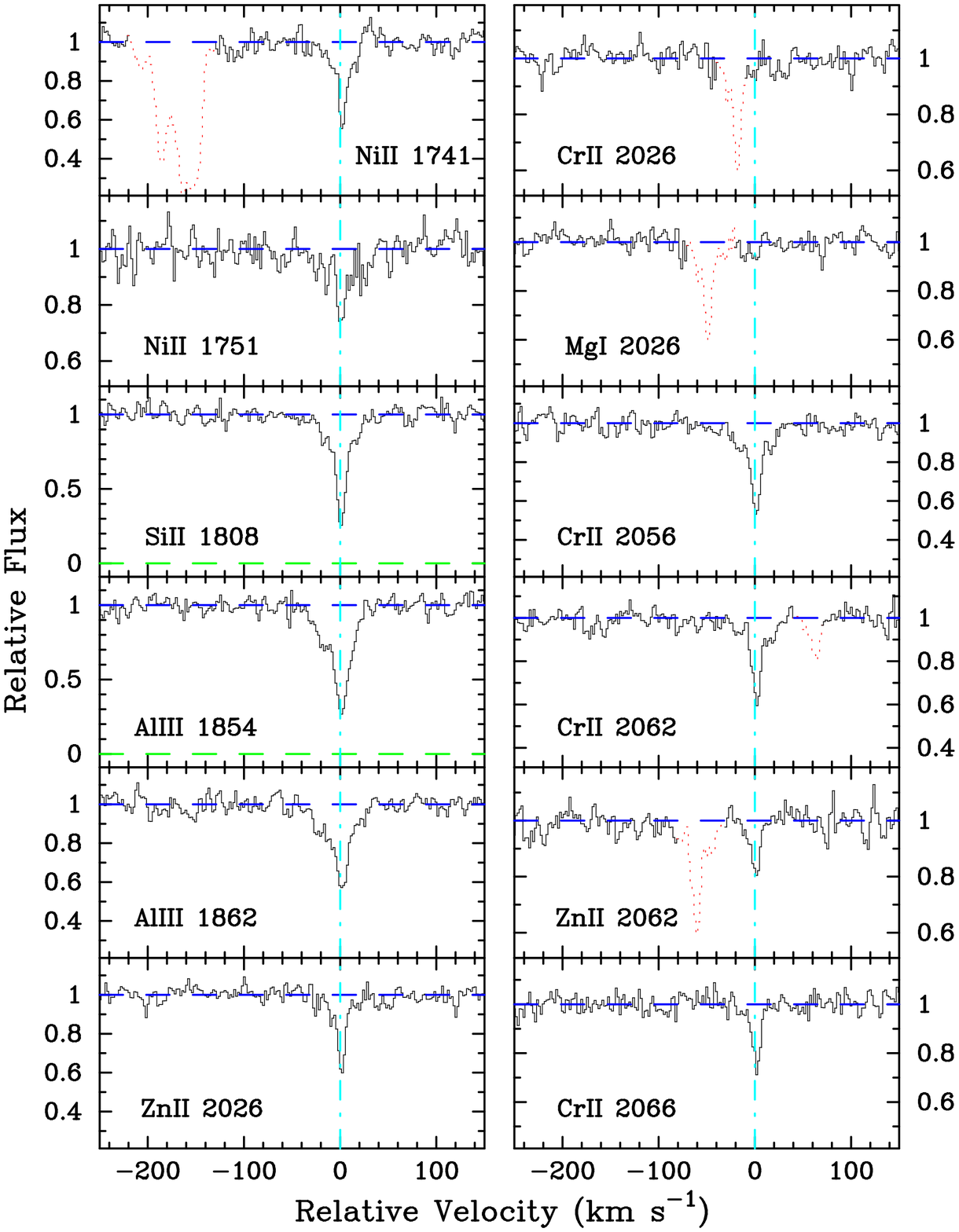}
\end{center}
\end{figure*}

\clearpage

\begin{figure}[ht]
\begin{center}
\includegraphics[height=4.3in, width=3.3in]{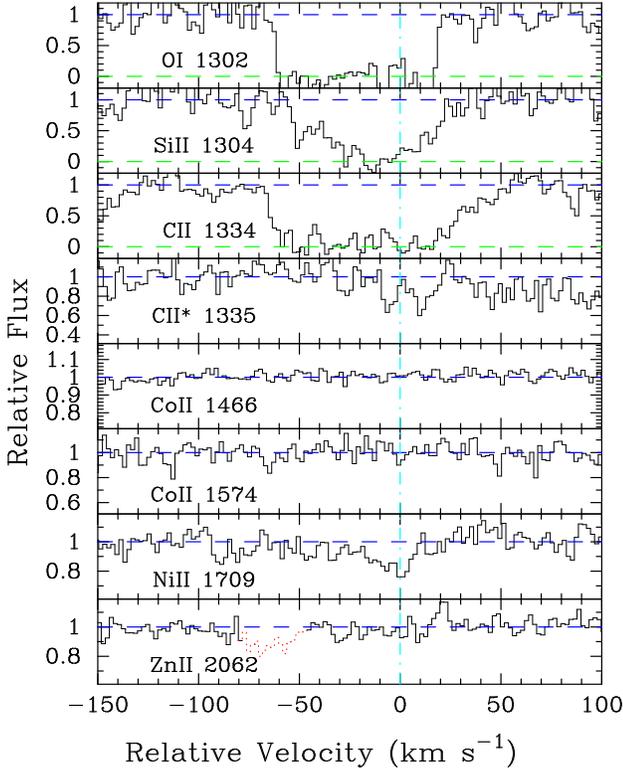}
\caption{Velocity plot of the new metal-line transitions for the 
damped \lya system at $z = 1.999$ toward Q1215+33. 
The vertical line at $v=0$ corresponds to $z = 1.9991$.}
\label{fig:1215}
\end{center}
\end{figure}

\subsection{Q1215$+$33, $z$ = 1.999}

Although we presented a full analysis of this damped system in PW99, 
a number of transitions were overlooked and we have revised the Fe
abundance.  Figure~\ref{fig:1215} plots the new transitions and
Table~\ref{tab:Q1215+33_1.999} lists the ionic column densities.  
We now report only a limit on $\N{Fe^+}$ because the Fe~II 1608 profile is 
saturated and the Fe~II 1611 transition 
is too weak to provide a reasonable measurement.
In the subsequent analysis, we assume a value based on 
an average of the two limits: $\N{Fe^+} = 10^{14.748 \pm 0.05} \cm{-2}$.
Finally, we also report an upper limit on $\N{Zn^+}$ based on the
Zn~II 2062 profile.  It is 0.2~dex lower than the value derived
from Zn~II 2026 which is difficult to understand aside from the
fact that the Zn~II 2026 profile is noisy and the continuum is 
poorly constrained in that region.  For now, we continue to adopt
the value from Zn~II 2026.

\begin{table}[ht]\footnotesize
\begin{center}
\caption{ {\sc
IONIC COLUMN DENSITIES: Q1215+33, $z = 1.999$ \label{tab:Q1215+33_1.999}}}
\begin{tabular}{lcccc}
\tableline
\tableline
Ion & $\lambda$ & AODM & $N_{\rm adopt}$ & [X/H] \\
\tableline
HI &1215 & $20.950  \pm 0.067  $ \\
C  II &1334&$>14.630$&$>14.630$&$>-2.870$\\  
C  II &1335&$<13.173$\\  
O  I  &1302&$>15.127$&$>15.127$&$>-2.693$\\  
O  I  &1355&$<18.065$\\  
Si II &1304&$>14.617$&$15.030 \pm  0.025$&$-1.480 \pm  0.072$\\  
Si II &1526&$>14.660$\\  
Si II &1808&$15.030 \pm  0.025$\\  
Fe II &1608&$>14.696$&$14.748 \pm  0.053$&$-1.702 \pm  0.085$\\  
Fe II &1611&$<14.800$\\  
Co II &1466&$<12.860$&$<12.860$&$<-1.000$\\  
Co II &1574&$<13.358$\\  
Ni II &1709&$13.579 \pm  0.061$&$13.594 \pm  0.027$&$-1.606 \pm  0.072$\\  
Ni II &1741&$13.602 \pm  0.039$\\  
Ni II &1751&$13.592 \pm  0.048$\\  
Zn II &2026&$12.330 \pm  0.049$&$12.330 \pm  0.049$&$-1.290 \pm  0.083$\\  
Zn II &2062&$<12.138$\\  
\tableline
\end{tabular}
\end{center}
\end{table}

\begin{figure*}
\begin{center}
\includegraphics[height=8.5in, width=6.0in]{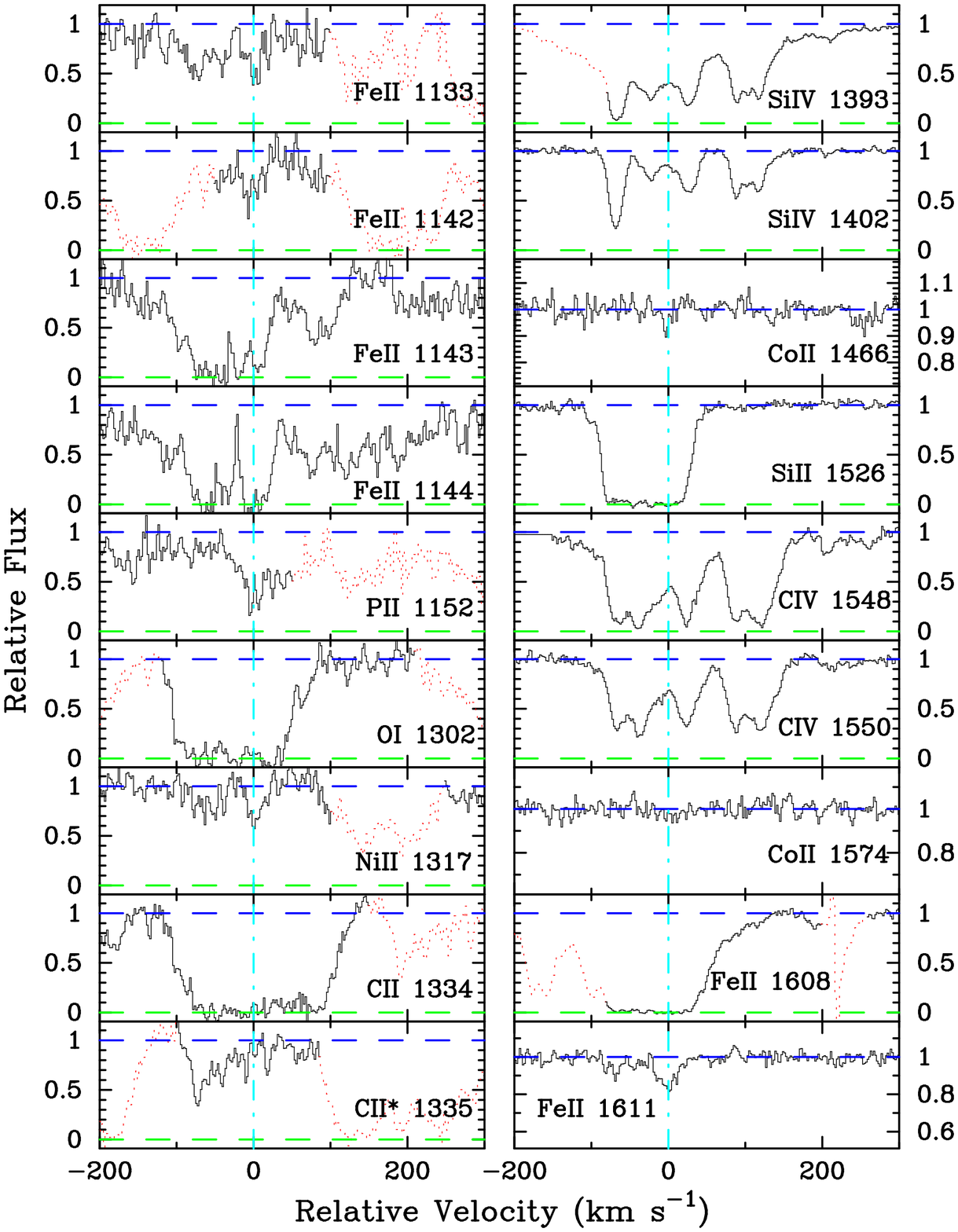}
\caption{Velocity plot of the metal-line transitions for the 
damped \lya system at $z = 2.466$ toward Q1223+17.
The vertical line at $v=0$ corresponds to $z = 2.466083$.}
\label{fig:1223}
\end{center}
\end{figure*}

\begin{figure*}
\begin{center}
\includegraphics[height=8.5in, width=6.0in]{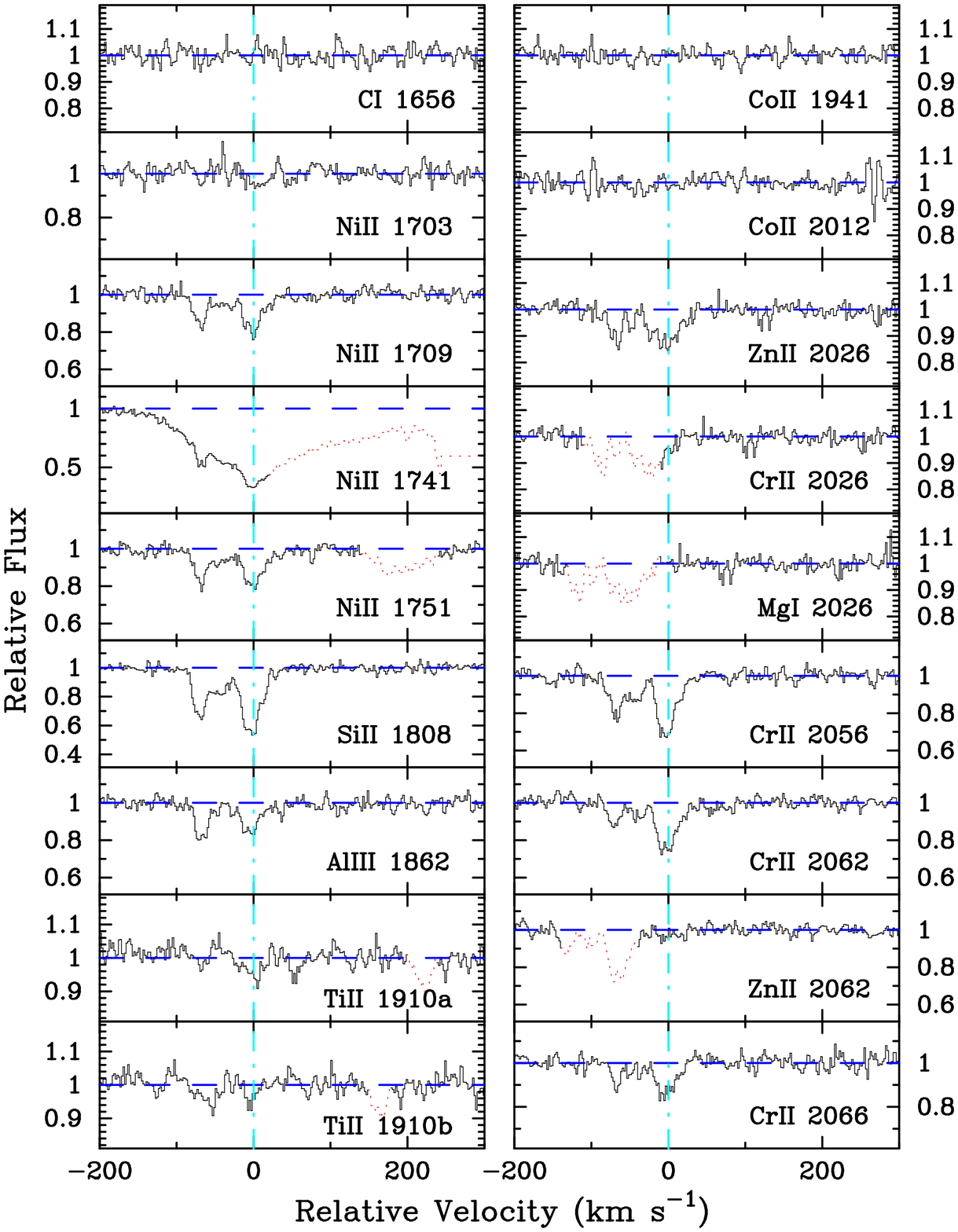}
\end{center}
\end{figure*}

\subsection{Q1223$+$17, $z$ = 2.466}

The combination of a very large $\N{HI}$ value and extensive wavelength
coverage allows for the analysis of a terrific number of transitions.
Figure~\ref{fig:1223} and Table~\ref{tab:Q1223+17_2.466} 
present over 20 transitions
including a large number of limits.  One of the most interesting ratios
is Ti/Fe whose upper limit is less than 1/2 the solar value
([Ti/Fe]~$< -0.4$).  In general, a subsolar Ti/Fe ratio implies 
significant dust depletion because Ti is more readily locked up into dust
grains, but the Zn/Fe ratio is not particularly large 
as one would expect in a significantly dust depleted region
([Zn/Fe] = +0.22).  
Our observations also place a tight constraint on Co/Fe which is described
in greater detail in \cite{ellison01}.

\begin{table}[ht]\footnotesize
\begin{center}
\caption{ {\sc
IONIC COLUMN DENSITIES: Q1223+17, $z = 2.466$ \label{tab:Q1223+17_2.466}}}
\begin{tabular}{lcccc}
\tableline
\tableline
Ion & $\lambda$ & AODM & $N_{\rm adopt}$ & [X/H] \\
\tableline
HI &1215 & $21.500  \pm 0.100  $ \\
C  I  &1656&$<12.426$\\  
C  II &1334&$>15.155$&$>15.155$&$>-2.895$\\  
C  II &1335&$<14.007$\\  
C  IV &1548&$>14.696$\\  
C  IV &1550&$14.658 \pm  0.004$\\  
O  I  &1302&$>15.477$&$>15.477$&$>-2.893$\\  
Al III&1862&$12.909 \pm  0.023$\\  
Si II &1526&$>15.037$&$15.468 \pm  0.008$&$-1.592 \pm  0.100$\\  
Si II &1808&$15.468 \pm  0.008$\\  
Si IV &1402&$13.891 \pm  0.004$\\  
P  II &1152&$<13.883$&$<13.883$&$<-1.147$\\  
Ti II &1910&$<12.252$&$<12.252$&$<-2.188$\\  
Cr II &2056&$13.521 \pm  0.013$&$13.493 \pm  0.010$&$-1.677 \pm  0.100$\\  
Cr II &2062&$13.480 \pm  0.018$\\  
Cr II &2066&$13.411 \pm  0.032$\\  
Fe II &1133&$15.098 \pm  0.059$&$15.157 \pm  0.022$&$-1.843 \pm  0.102$\\  
Fe II &1142&$15.258 \pm  0.051$\\  
Fe II &1611&$15.152 \pm  0.027$\\  
Co II &1466&$<13.174$&$<12.631$&$<-1.779$\\  
Co II &1574&$<13.122$\\  
Co II &1941&$<12.917$\\  
Co II &2012&$<12.631$\\  
Ni II &1317&$13.853 \pm  0.051$&$13.949 \pm  0.011$&$-1.801 \pm  0.101$\\  
Ni II &1703&$<13.817$\\  
Ni II &1709&$13.901 \pm  0.020$\\  
Ni II &1751&$14.000 \pm  0.014$\\  
Zn II &2026&$12.550 \pm  0.026$&$12.550 \pm  0.026$&$-1.620 \pm  0.103$\\  
Zn II &2062&$>11.785$\\  
\tableline
\end{tabular}
\end{center}
\end{table}

\begin{figure*}
\begin{center}
\includegraphics[height=8.5in, width=6.0in]{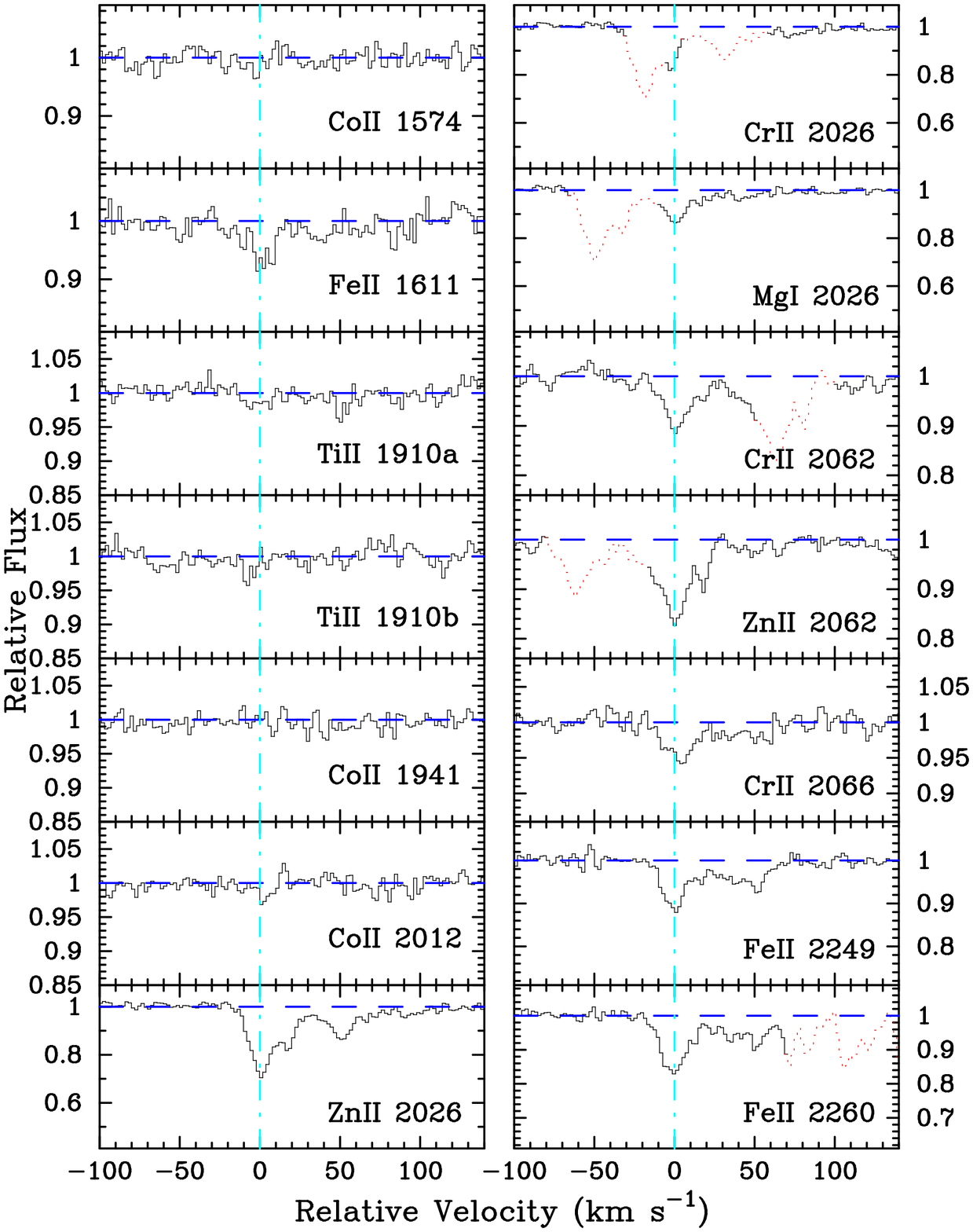}
\caption{Velocity plot of the new metal-line transitions for the 
damped \lya system at $z = 1.776$ toward Q1331+17. 
The vertical line at $v=0$ corresponds to $z = 1.77636.$}
\label{fig:1331}
\end{center}
\end{figure*}

\subsection{Q1331$+$17, $z$ = 1.776}

An analysis of the
damped system toward the very bright quasar Q1331+17 was given
by PW99 but a number of transitions were missed (notably Ti~II 1910
and Co~II 2012).  The new transitions are plotted in Figure~\ref{fig:1331}
and the ionic column densities are given in Table~\ref{tab:Q1331+17_1.776}.  

With respect to the $\N{Zn^+}$ value presented by PW99, 
this system exhibits one of the largest Zn/Fe ratios of any damped system.
This $\N{Zn^+}$ value is nearly 0.3~dex higher than
the value reported in \cite{ptt94}, however, because we did not correct for 
possible contamination from the Mg~I 2026 transition.  
Although a significant feature is apparent at $v=0 \mkms$ of the Mg~I 2026
profile, this feature is perfectly aligned with an absorption feature
at $v = +52 \mkms$ in the unsaturated Fe~II profiles. 
Furthermore, the $\N{Zn^+}$ value from Zn~II 2062 is identical to
the value derived from Zn~II 2026 using the AODM method.  We suspect,
however, that this is a coincidence resulting from blending between
the Zn~II 2062 and Cr~II 2062 profiles.  Performing a detailed line-profile
analysis of the Zn and Cr lines and including Mg~I 2026, we find 
$\log \N{Zn^+} = 12.542 \pm 0.029$ and 
$\log \N{Mg^0} = 12.419 \pm 0.048$. 
We discuss this issue further and its impact on studies of Zn
in Paper~II, $\S$~\secZn.

\begin{table}[ht]\footnotesize
\begin{center}
\caption{ {\sc
IONIC COLUMN DENSITIES: Q1331+17, $z = 1.776$ \label{tab:Q1331+17_1.776}}}
\begin{tabular}{lcccc}
\tableline
\tableline
Ion & $\lambda$ & AODM & $N_{\rm adopt}$ & [X/H] \\
\tableline
HI &1215 & $21.176  \pm 0.041  $ \\
C  I  &1560&$13.573 \pm  0.013$\\  
C  I  &1656&$13.312 \pm  0.012$\\  
Mg I  &2026&$$\\  
Ti II &1910&$11.836 \pm  0.118$&$11.836 \pm  0.118$&$-2.280 \pm  0.125$\\  
Cr II &2056&$12.957 \pm  0.017$&$12.874 \pm  0.012$&$-1.972 \pm  0.043$\\  
Cr II &2066&$12.834 \pm  0.034$\\  
Fe II &1608&$14.630 \pm  0.003$&$14.618 \pm  0.001$&$-2.058 \pm  0.041$\\  
Fe II &1611&$14.709 \pm  0.046$\\  
Fe II &2249&$14.595 \pm  0.015$\\  
Fe II &2260&$14.647 \pm  0.010$\\  
Fe II &2344&$>14.723$\\  
Fe II &2374&$14.616 \pm  0.002$\\  
Fe II &2382&$>14.461$\\  
Co II &1574&$<12.659$&$<12.306$&$<-1.780$\\  
Co II &1941&$<12.367$\\  
Co II &2012&$<12.306$\\  
Zn II &2026&$12.542 \pm  0.029$&$12.542 \pm  0.029$&$-1.304 \pm  0.050$\\  
\tableline
\end{tabular}
\end{center}
\end{table}

In addition to the large Zn/Fe ratio,
this system shows rarely observed C~I absorption
and a significant subsolar Ti/Fe ratio.  Altogether the 
chemical abundances of this system represent the most compelling evidence for 
dust depletion in any damped \lya system.  
It is particularly important to note,
therefore, that it is one of the brightest (apparent magnitude)
quasars observed in our sample.  In Paper~II, $\S$~\secdustobsc\ we consider
the obscuration of this quasar due to this damped system and the implications
for dust obscuration in general.

This damped \lya system is one of the few cases where one can derive
$\N{Fe^+}$ values from both Fe~II 1608 and 1611.  Furthermore, our
observations also cover several of the Fe~II transitions longward of 2000\AA,
including Fe~II 2249 and 2260 which are the principal diagnostics of Fe$^+$
in the Galactic ISM.  Examining Table~\ref{tab:Q1331+17_1.776}, one notes that
nearly all of the $\N{Fe^+}$ values are consistent at the $2 \sigma$ level
and all are in accordance at $3 \sigma$.  One also notes that 
$\N{Fe~II~1611}$ exceeds all of the other measurements suggesting it is
unlikely the \cite{raassen98} analysis overestimated the Fe~II 1611
oscillator strength, at
least relative to the other Fe~II transitions.

\begin{figure*}
\begin{center}
\includegraphics[height=8.5in, width=6.0in]{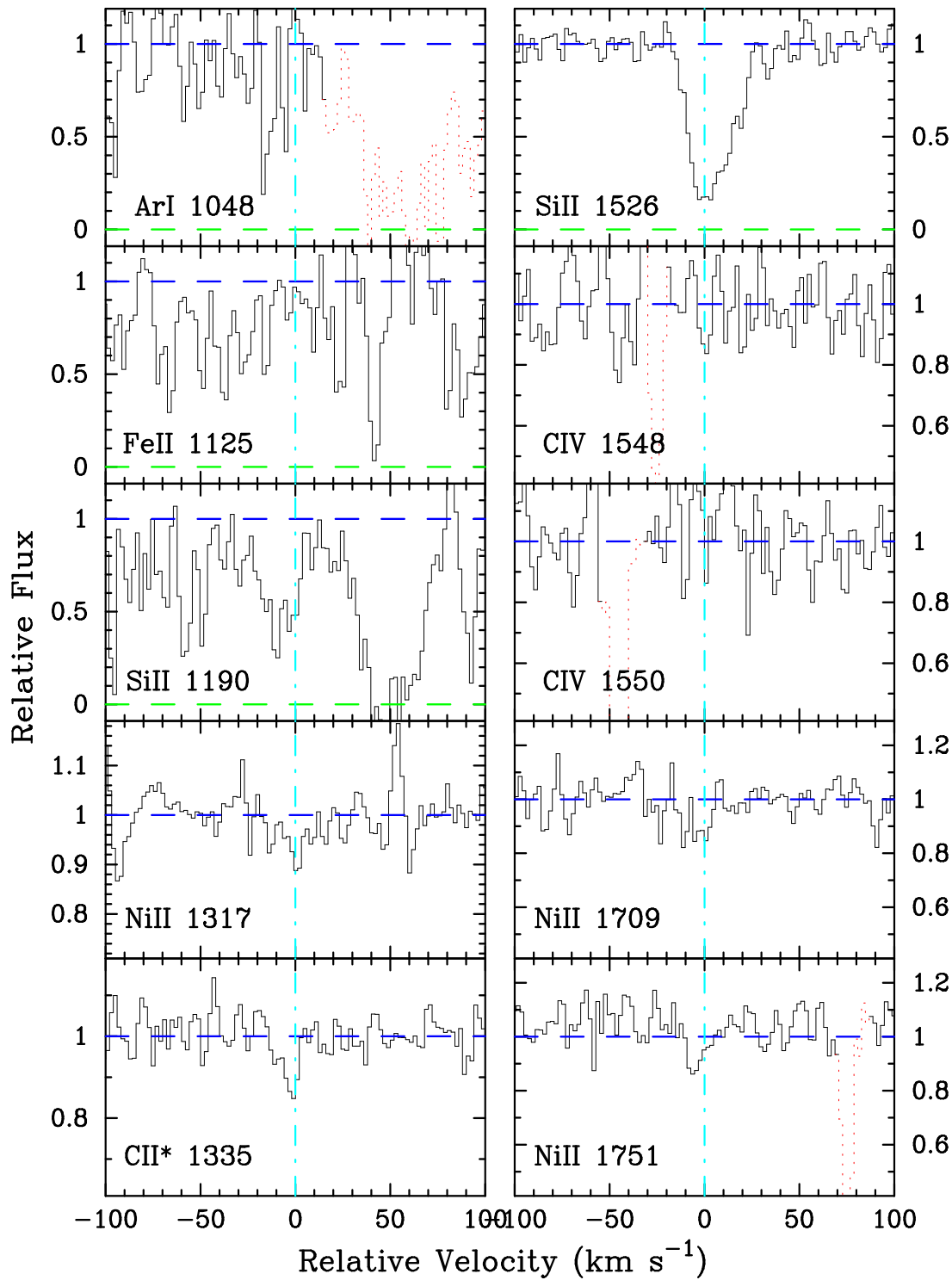}
\caption{Velocity plot of the new metal-line transitions for the 
damped \lya system at $z = 3.736$ toward BRI1346--03.
The vertical line at $v=0$ corresponds to $z = 3.73583.$}
\label{fig:1346}
\end{center}
\end{figure*}

\subsection{BRI1346$-$03, $z$ = 3.736}

Our additional observations blueward of the data presented in 
PW99 provide coverage of a few new transitions 
(Figure~\ref{fig:1346}, Table~\ref{tab:BRI1346-03_3.736}).  Unfortunately, we
still do not have coverage of a single unsaturated Fe~II profile or
any other Fe-peak metal transition.  Therefore, we have
adopted Al as a proxy for Fe (i.e.\ assume [Fe/H] = [Al/H]) and 
in this manner include 
the system in the metallicity and relative abundance analyses of Paper~II.
The implied Si/Fe ratios match typical values. 
If the feature at $v=0 \mkms$ in the Ni~II 1317 profile is not noise
or a coincident metal line, it implies a very large Ni/Al ratio
indicating we might be underestimating $\N{Fe^+}$.  For the moment,
we consider it as an upper limit. 

Our new observations also cover the C~IV doublet at 1550\AA.  
Although the spectra is particularly noisy over this region,
there is no obvious C~IV absorption.  This marks the first DLA
system with no detectable C~IV absorption and we note in passing
a possible trend of weaker C~IV absorption at $z>3$.

\begin{table}[ht]\footnotesize
\begin{center}
\caption{ {\sc
IONIC COLUMN DENSITIES: BRI1346-03, $z = 3.736$ \label{tab:BRI1346-03_3.736}}}
\begin{tabular}{lcccc}
\tableline
\tableline
Ion & $\lambda$ & AODM & $N_{\rm adopt}$ & [X/H] \\
\tableline
HI &1215 & $20.720  \pm 0.100  $ \\
C  II &1335&$12.550 \pm  0.113$&$>14.486$&$>-2.784$\\  
C  IV &1548&$<12.717$\\  
C  IV &1550&$<13.146$\\  
O  I  &1302&$>15.018$&$>15.019$&$>-2.571$\\  
Si II &1190&$13.430 \pm  0.122$&$13.948 \pm  0.009$&$-2.332 \pm  0.100$\\  
Si II &1304&$13.983 \pm  0.009$\\  
Si II &1526&$13.880 \pm  0.026$\\  
Ar I  &1048&$<13.114$&$<13.113$&$<-2.127$\\  
Fe II &1125&$<14.126$&$<14.126$&$<-2.094$\\  
Co II &1574&$<13.260$&$<13.260$&$<-0.370$\\  
Ni II &1317&$<12.759$&$<12.760$&$<-2.210$\\  
Ni II &1370&$<12.876$\\  
Ni II &1709&$<13.345$\\  
Ni II &1741&$<13.284$\\  
Ni II &1751&$<13.428$\\  
\tableline
\end{tabular}
\end{center}
\end{table}

\begin{figure*}
\begin{center}
\includegraphics[height=8.5in, width=6.0in]{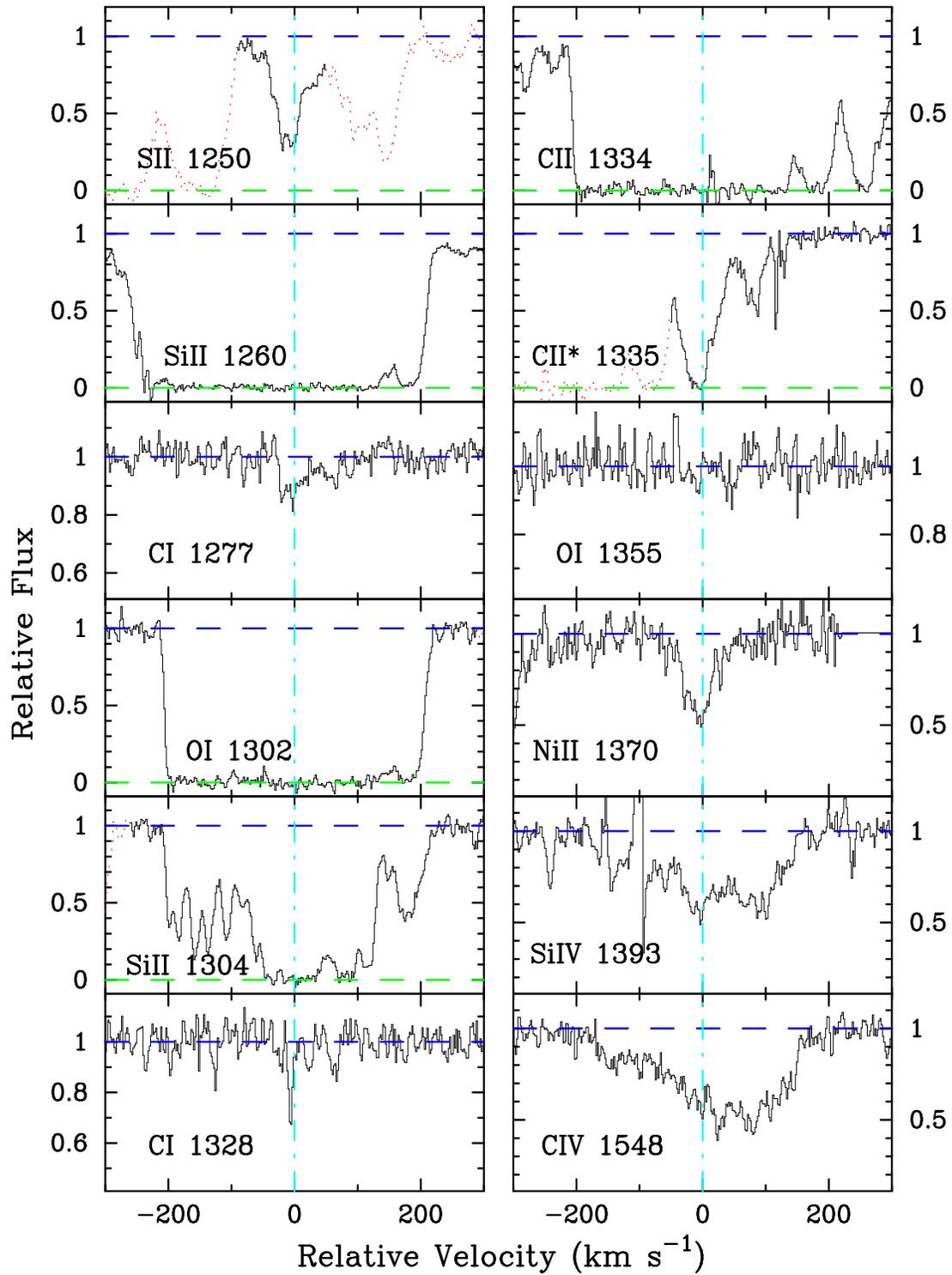}
\caption{Velocity plot of the metal-line transitions for the 
damped \lya system at $z = 4.224$ toward PSS1443+27.
The vertical line at $v=0$ corresponds to $z = 4.224099$.}
\label{fig:1443}
\end{center}
\end{figure*}

\begin{figure*}
\begin{center}
\includegraphics[height=8.5in, width=6.0in]{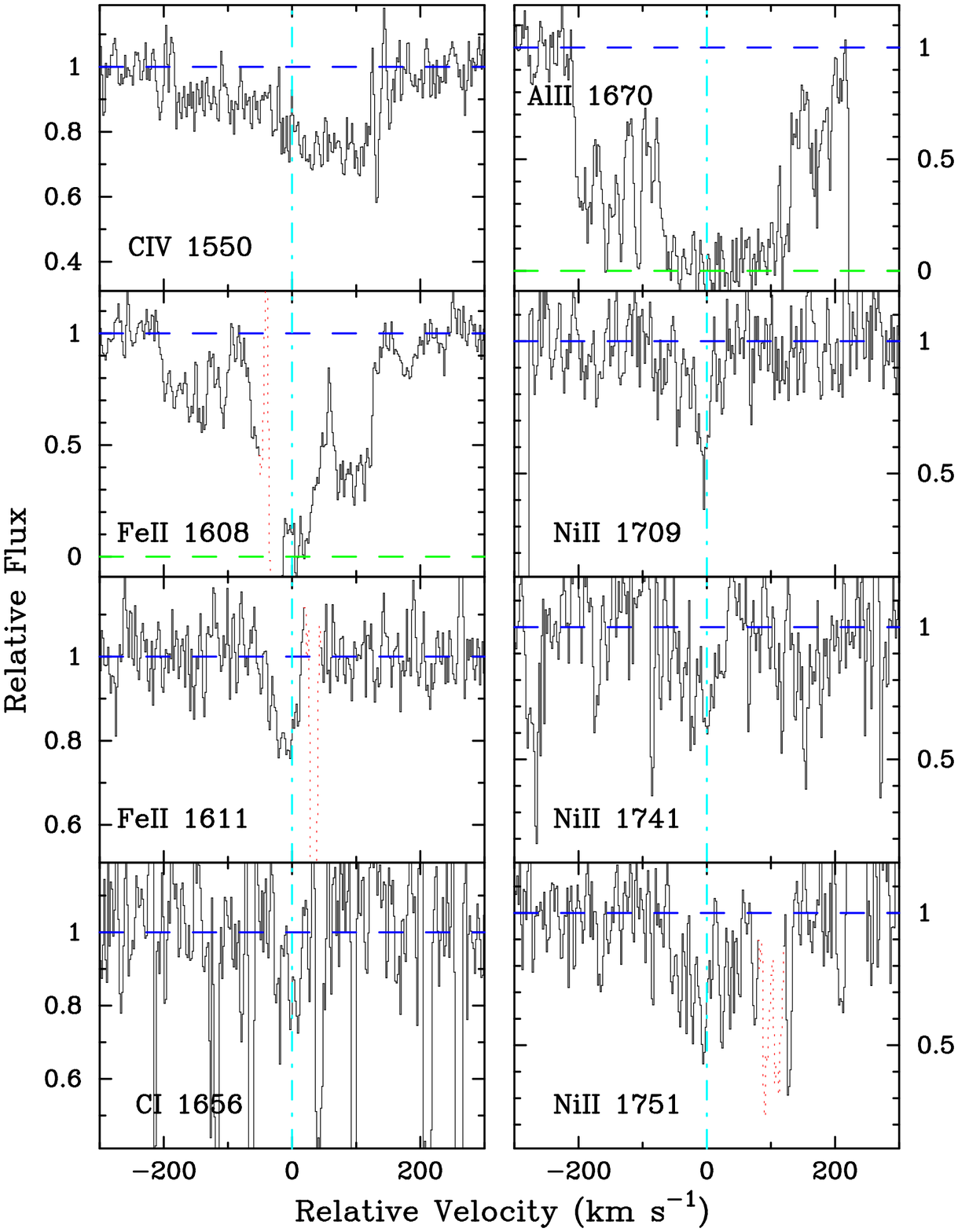}
\end{center}
\end{figure*}

\subsection{PSS1443$+$27, $z$ = 4.224}

This $z>4$ damped \lya systems was discovered by \cite{storr00} 
who determined the 
$\N{HI}$ value from a spectrum obtained using LRIS at Keck observatories.
Its very high metallicity was first reported in \cite{pro00}
We have since acquired further observations of this system which 
confirm the [Fe/H] metallicity. 
In particular, we observed the Ni~II 1370, 1709, and 1741 
transitions at reasonably
high S/N and found [Ni/H]~$\approx$~[Fe/H].  Figure~\ref{fig:1443} presents
the transitions observed for this system and
the ionic column densities are presented in Table~\ref{tab:PSS1443+27_4.224}.

In passing, we note the remarkable C~II$^*$ 1335 profile which is 
heavily saturated and suggests a large star formation rate for
this system \citep{wp01}.  Also, we identify possible absorption from
two C~I profiles which we expect is not due to coincident metal-line
systems.  Unfortunately our observations did not cover the stronger
C~II 1556 profile and strongest C~I 1656 profile is located within
a forest of sky lines.

\begin{table}[ht]\footnotesize
\begin{center}
\caption{ {\sc
IONIC COLUMN DENSITIES: PSS1443+27, $z = 4.224$ \label{tab:PSS1443+27_4.224}}}
\begin{tabular}{lcccc}
\tableline
\tableline
Ion & $\lambda$ & AODM & $N_{\rm adopt}$ & [X/H] \\
\tableline
HI &1215 & $20.800  \pm 0.100  $ \\
C  I  &1277&$13.446 \pm  0.037$\\  
C  I  &1328&$13.367 \pm  0.090$\\  
C  I  &1656&$13.041 \pm  0.133$\\  
C  II &1334&$>15.613$&$>15.612$&$>-1.738$\\  
C  II &1335&$>14.709$\\  
C  IV &1548&$14.245 \pm  0.009$\\  
C  IV &1550&$14.213 \pm  0.017$\\  
O  I  &1302&$>16.048$&$>16.048$&$>-1.622$\\  
O  I  &1355&$<17.734$\\  
Al II &1670&$>13.959$&$>13.958$&$>-1.332$\\  
Si II &1304&$>15.434$&$>15.434$&$>-0.926$\\  
Si IV &1393&$13.706 \pm  0.011$\\  
S  II &1253&$$\\  
Fe II &1608&$>15.101$&$15.204 \pm  0.056$&$-1.096 \pm  0.115$\\  
Fe II &1611&$15.204 \pm  0.056$\\  
Co II &1574&$<13.508$&$<13.509$&$<-0.201$\\  
Ni II &1370&$14.079 \pm  0.025$&$14.091 \pm  0.024$&$-0.959 \pm  0.103$\\  
Ni II &1709&$14.229 \pm  0.069$\\  
Ni II &1741&$13.877 \pm  0.074$\\  
\tableline
\end{tabular}
\end{center}
\end{table}

\subsection{Q1759$+$75, $z$ = 2.625}

\begin{figure*}
\begin{center}
\includegraphics[height=8.5in, width=6.0in]{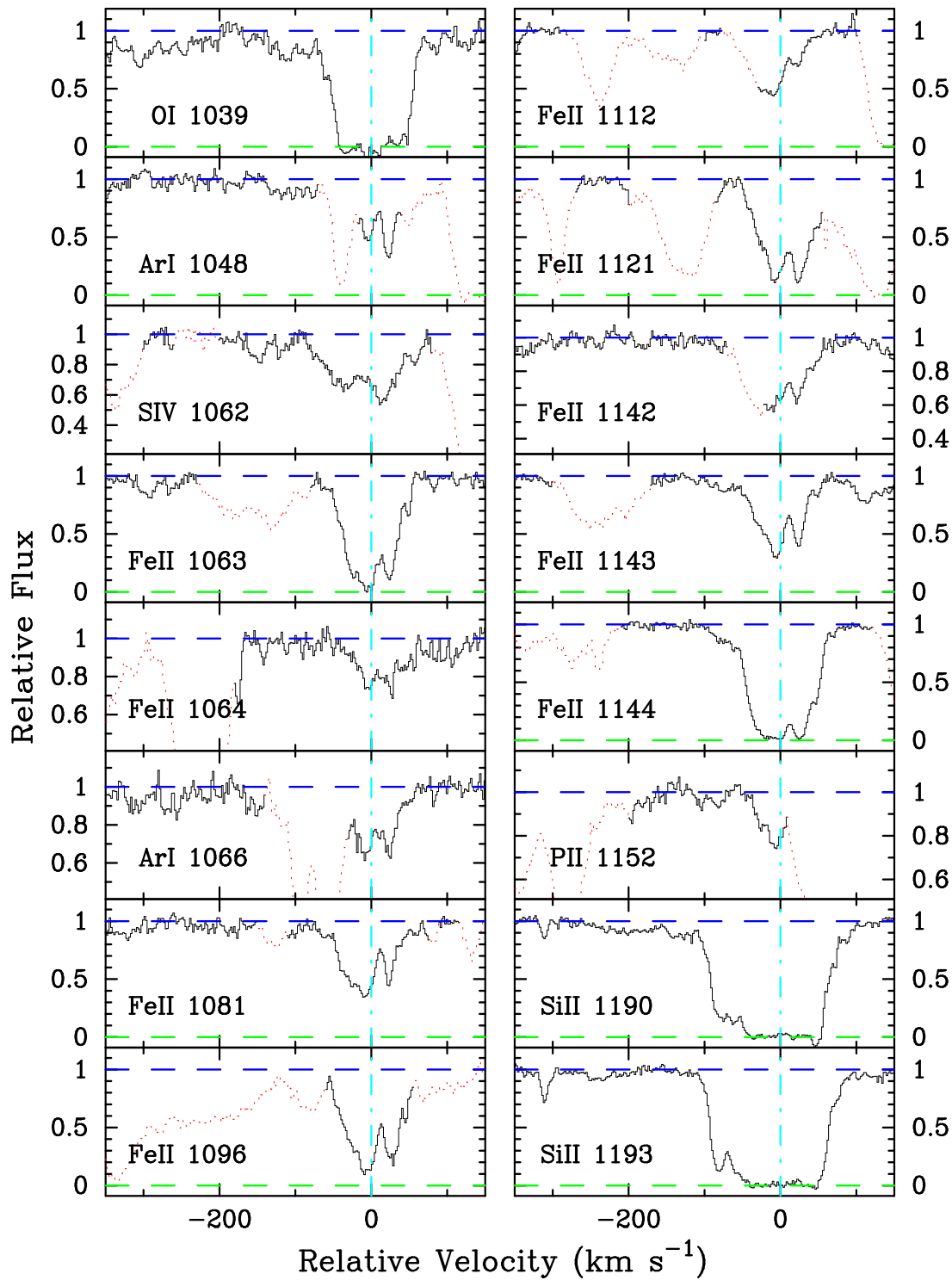}
\caption{Velocity plot of the new metal-line transitions for the 
damped \lya system at $z = 2.625$ toward Q1759+75.
The vertical line at $v=0$ corresponds to $z = 2.62530.$}
\label{fig:1759}
\end{center}
\end{figure*}

This system was presented in PW99 and has been subsequently analysed
by \cite{outram99}.  Here, we present an analysis of our spectrum blueward
of \lya emission.  Figure~\ref{fig:1759} presents the transitions and
Table~\ref{tab:Q1759+75_2.625} the column densities.

Our observations present measurements of a number of lines in the
\lya forest.  In particular, we have excellent coverage of the 
FUV Fe~II transitions, good measurements of the N~I triplets at 
1134 and 1200\rAA, moderate limits on Ar~I and O~I,
and an excellent measurement of $\N{S^+}$.  Regarding the Fe~II lines,
we find very good agreement between the many transitions which confirms
the $f$-values measured by \cite{howk00}.   The only exception is 
Fe~II 1062 (not analysed by Howk et al.) whose $f$-value appears to
be systematically high.  We recommend using a value $\approx 0.2$~dex
below the value reported by \cite{morton91}.  Finally, we point out
significant absorption at $v \approx -300$~km/s in the Si~II 1190 and
1193 transitions which coincide with a strong feature in C~IV and a 
weaker feature in Al~II 1670 (PW99).  We suspect this metal-line system
corresponds to a nearby Lyman limit system although there is no
significant evidence for asymmetry in the \lya profile.

\begin{table}[ht]\footnotesize
\begin{center}
\caption{ {\sc
IONIC COLUMN DENSITIES: Q1759+75, $z = 2.625$ \label{tab:Q1759+75_2.625}}}
\begin{tabular}{lcccc}
\tableline
\tableline
Ion & $\lambda$ & AODM & $N_{\rm adopt}$ & [X/H] \\
\tableline
HI &1215 & $20.800  \pm 0.100  $ \\
C  I  &1656&$<12.336$\\  
C  II &1334&$>15.300$&$>15.300$&$>-2.050$\\  
C  II &1335&$13.138 \pm  0.032$\\  
O  I  &1039&$>16.261$&$>16.261$&$>-1.409$\\  
O  I  &1302&$>15.759$\\  
Si II &1190&$>14.928$&$15.536 \pm  0.008$&$-0.824 \pm  0.100$\\  
Si II &1193&$>14.614$\\  
Si II &1260&$>14.396$\\  
Si II &1304&$>15.198$\\  
Si II &1808&$15.536 \pm  0.008$\\  
P  II &1152&$>13.046$&$>13.047$&$>-1.283$\\  
S  II &1250&$15.243 \pm  0.009$&$15.243 \pm  0.010$&$-0.757 \pm  0.100$\\  
S  II &1253&$<15.486$\\  
S  II &1259&$<15.335$\\  
Ar I  &1048&$<13.714$&$<13.714$&$<-1.606$\\  
Ar I  &1066&$<14.053$\\  
Fe II &1062&$14.860 \pm  0.037$&$15.091 \pm  0.004$&$-1.209 \pm  0.100$\\  
Fe II &1063&$>15.002$\\  
Fe II &1063&$15.287 \pm  0.020$\\  
Fe II &1081&$15.182 \pm  0.007$\\  
Fe II &1096&$15.059 \pm  0.007$\\  
Fe II &1112&$<15.389$\\  
Fe II &1121&$<15.260$\\  
Fe II &1142&$<15.565$\\  
Fe II &1143&$15.079 \pm  0.005$\\  
Fe II &1144&$>15.051$\\  
Fe II &1608&$>15.077$\\  
Fe II &1611&$14.923 \pm  0.034$\\  
Co II &1466&$<13.066$&$<13.019$&$<-0.691$\\  
Co II &1466&$<13.019$\\  
Co II &1574&$<13.108$\\  
Ni II &1317&$<14.248$&$13.802 \pm  0.007$&$-1.248 \pm  0.100$\\  
Ni II &1370&$13.766 \pm  0.010$\\  
Ni II &1454&$13.791 \pm  0.039$\\  
Ni II &1703&$<13.908$\\  
Ni II &1709&$13.929 \pm  0.020$\\  
Ni II &1741&$13.841 \pm  0.017$\\  
Ni II &1751&$13.868 \pm  0.021$\\  
\tableline
\end{tabular}
\end{center}
\end{table}

\begin{figure*}
\begin{center}
\includegraphics[height=8.5in, width=6.0in]{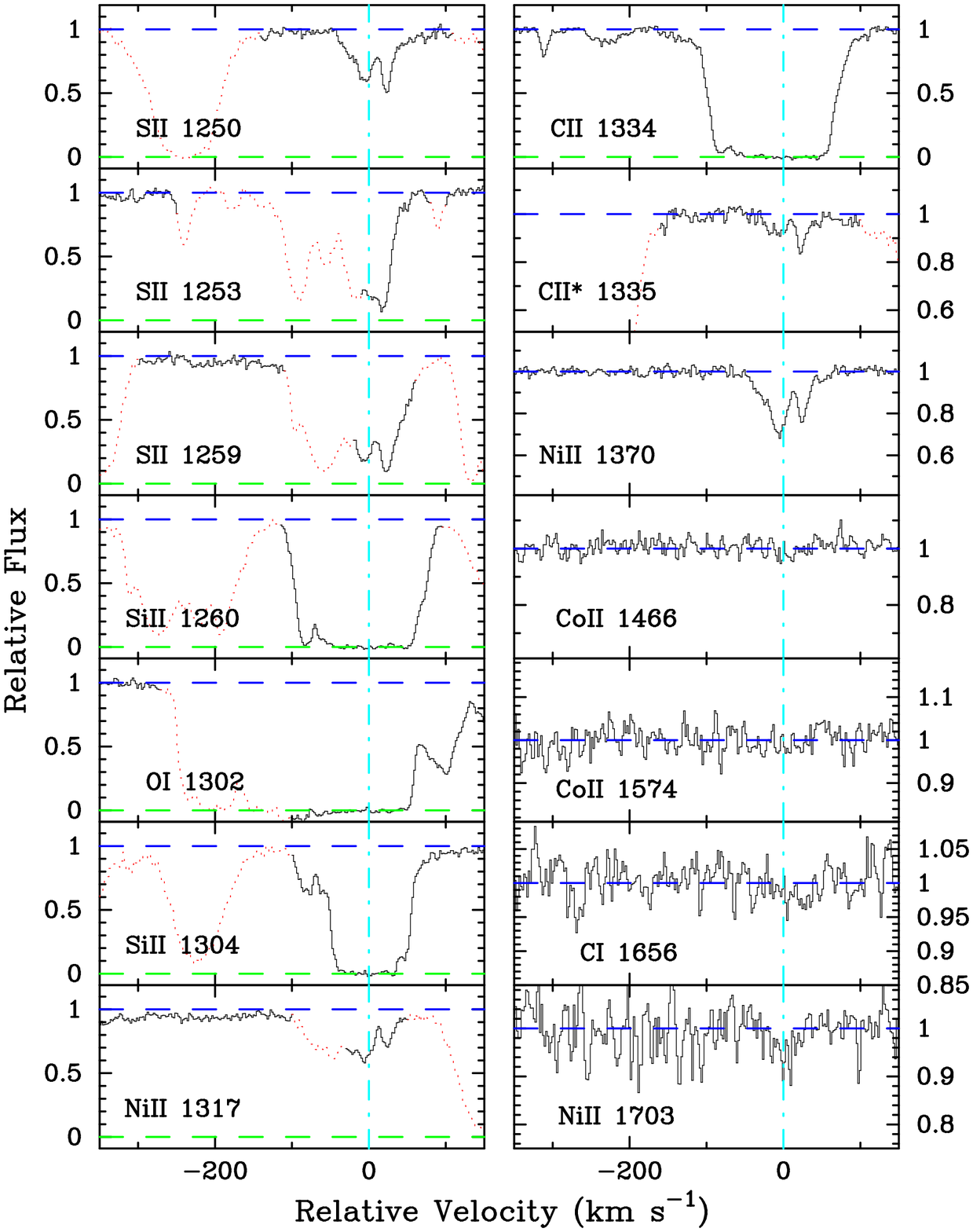}
\end{center}
\end{figure*}

\begin{figure*}
\begin{center}
\includegraphics[height=8.5in, width=6.0in]{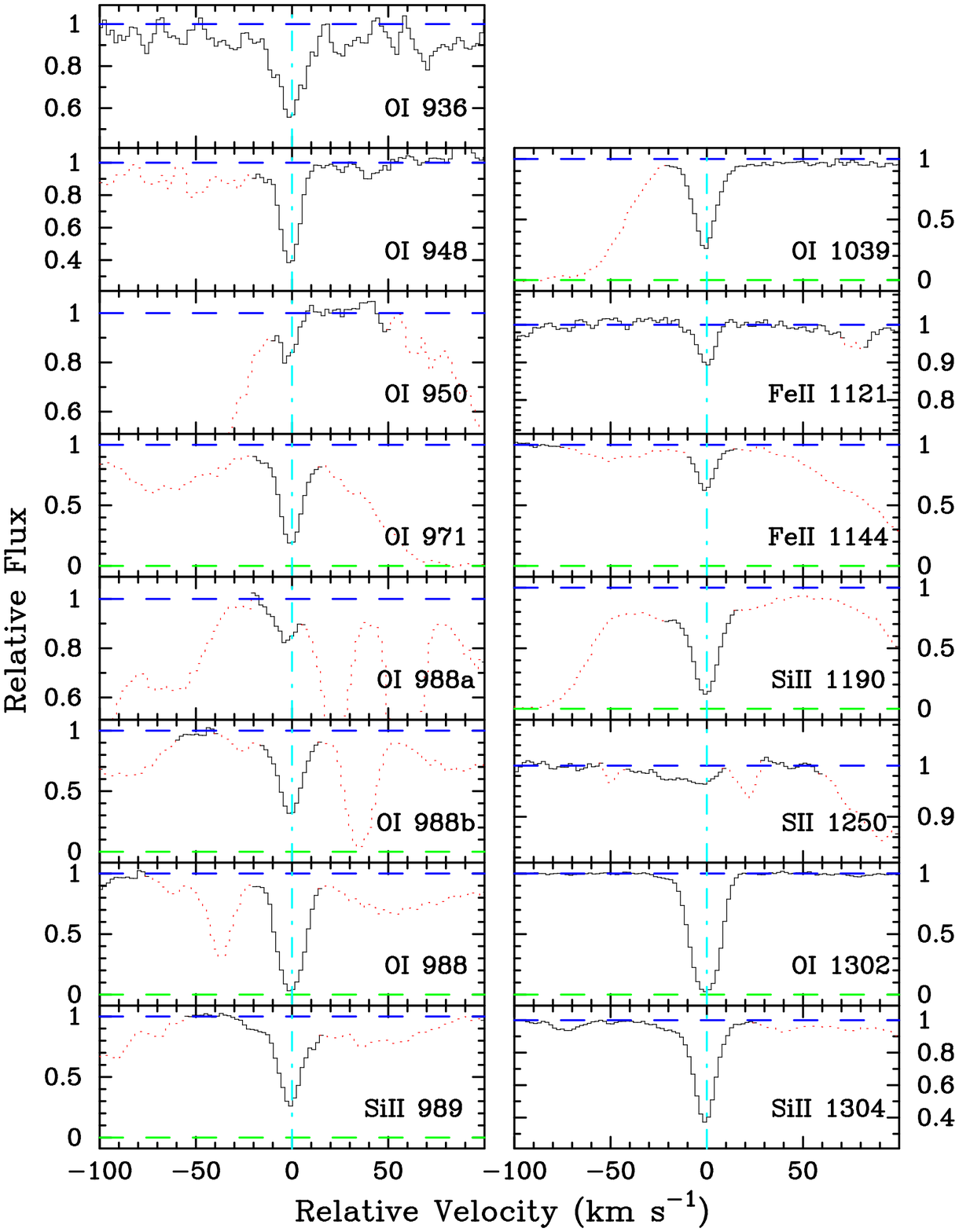}
\caption{Velocity plot of the metal-line transitions for the 
damped \lya system at $z = 2.844$ toward Q1946+76.
The vertical line at $v=0$ corresponds to $z = 2.8443.$}
\label{fig:1946a}
\end{center}
\end{figure*}

\subsection{Q1946+76, $z$ = 2.844}

\cite{kirk97} analysed this very high S/N spectrum of Q1946+76
to describe the \lya forest at $z \sim 2.8$.  Here, we analyse
the metal-line transitions for the system at $z=2.844$, ignoring the
probably damped \lya 
system at $z=1.73$ because we have no measure of its HI column density.
Figure~\ref{fig:1946a} presents the metal-line profiles for the
$z=2.844$ system and Table~\ref{tab:Q1946+7658_2.844} summarizes the
column density measurements.  For the HI column density we adopt the
value presented in L96.
This systems is notable for providing
one of the few cases where one can accurately determine $\N{O^0}$.
The observed O/Fe ratio is enhanced relative to solar, but at a
lower level than metal-poor halo stars with comparable metallicity.
Interestingly, the implied O/Si ratio is sub-solar which is almost
never observed in metal-poor stars.  Nevertheless, we believe the
$\N{O^0}$ value is accurate.
In a separate paper, we examine the N/O ratio of this system.

\begin{table}[ht]\footnotesize
\begin{center}
\caption{ {\sc
IONIC COLUMN DENSITIES: Q1946+7658, $z = 2.844$ \label{tab:Q1946+7658_2.844}}}
\begin{tabular}{lcccc}
\tableline
\tableline
Ion & $\lambda$ & AODM & $N_{\rm adopt}$ & [X/H] \\
\tableline
HI &1215 & $20.270  \pm 0.060  $ \\
O  I  & 936&$15.036 \pm  0.030$&$14.819 \pm  0.007$&$-2.321 \pm  0.060$\\  
O  I  & 948&$14.835 \pm  0.025$\\  
O  I  & 971&$<14.725$\\  
O  I  & 988&$<15.244$\\  
O  I  & 988&$<14.862$\\  
O  I  & 988&$>14.627$\\  
O  I  &1039&$14.811 \pm  0.008$\\  
O  I  &1302&$>14.587$\\  
Si II &1190&$<13.579$&$13.602 \pm  0.005$&$-2.228 \pm  0.060$\\  
Si II &1304&$13.602 \pm  0.005$\\  
S  II &1250&$<13.491$&$<13.491$&$<-1.979$\\  
Fe II &1121&$13.241 \pm  0.057$&$13.238 \pm  0.009$&$-2.532 \pm  0.061$\\  
Fe II &1144&$13.238 \pm  0.009$\\  
\tableline
\end{tabular}
\end{center}
\end{table}

\break

\begin{figure}[ht]
\begin{center}
\includegraphics[height=4.3in, width=3.3in]{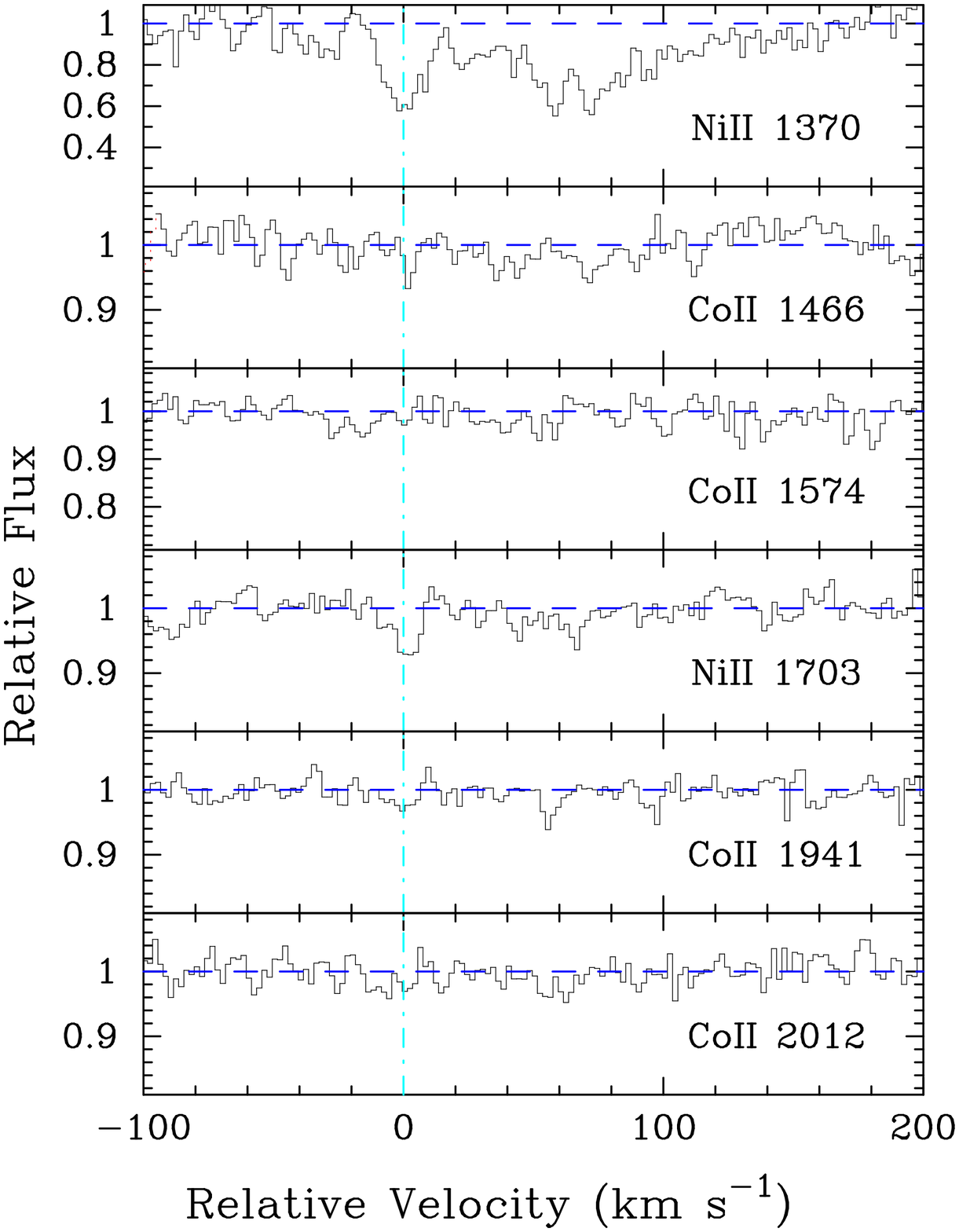}
\caption{Velocity plot of the new metal-line transitions for the 
damped \lya system at $z = 1.920$ toward Q2206--19.
The vertical line at $v=0$ corresponds to $z = 1.920.$}
\label{fig:2206a}
\end{center}
\end{figure}

\begin{figure}[ht]
\begin{center}
\includegraphics[height=4.3in, width=3.3in]{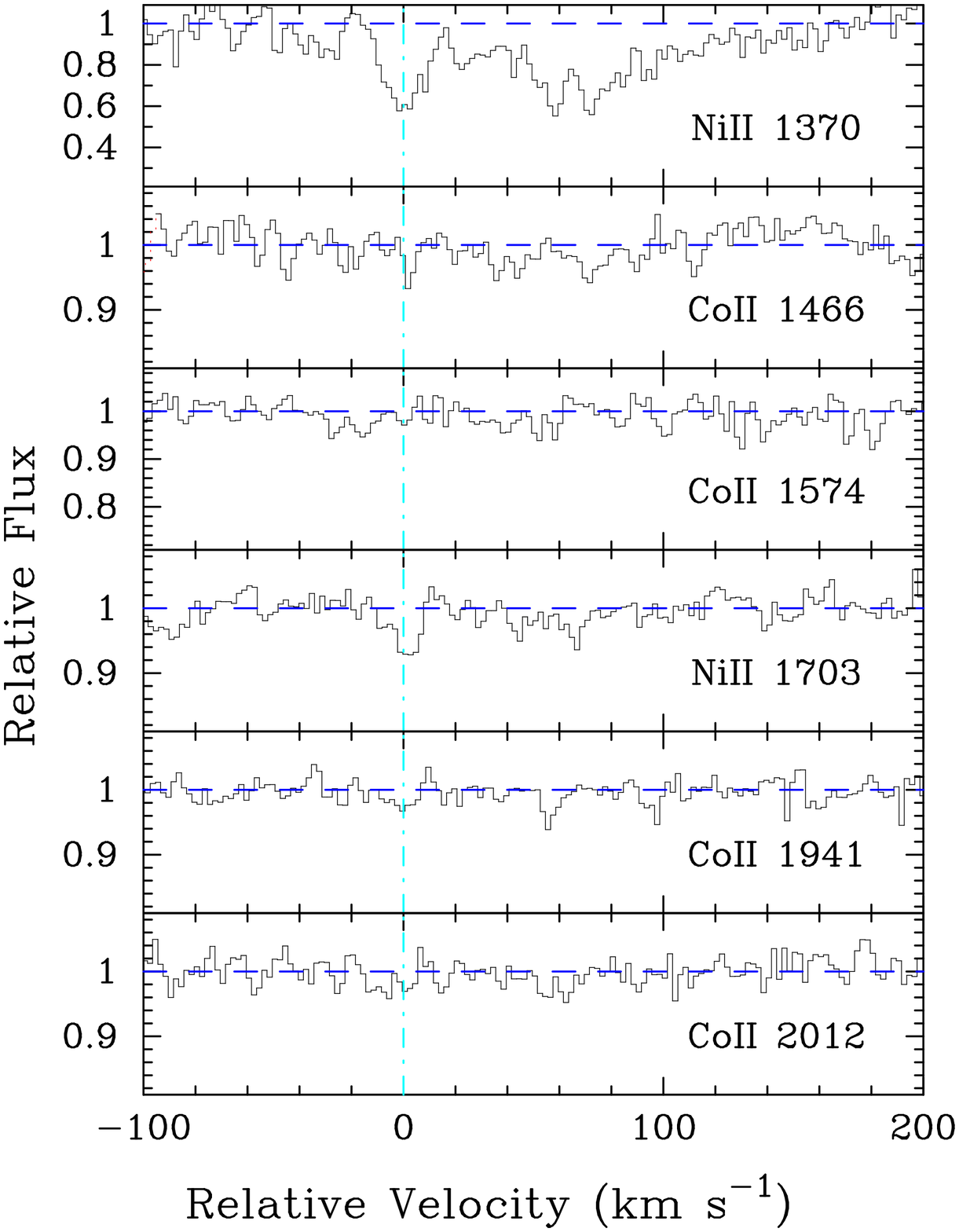}
\caption{Velocity plot of the new metal-line transitions for the 
damped \lya system at $z = 2.076$ toward Q2206--19.
The vertical line at $v=0$ corresponds to $z = 2.07623.$}
\label{fig:2206b}
\end{center}
\end{figure}

\begin{table}[ht]\footnotesize
\begin{center}
\caption{ {\sc
IONIC COLUMN DENSITIES: Q2206-19, $z = 1.920$ \label{tab:Q2206-19_1.920}}}
\begin{tabular}{lcccc}
\tableline
\tableline
Ion & $\lambda$ & AODM & $N_{\rm adopt}$ & [X/H] \\
\tableline
HI &1215 & $20.653  \pm 0.071  $ \\
Al II &1670&$>14.070$&$>14.070$&$>-1.073$\\  
Si II &1526&$>15.275$&$15.796 \pm  0.005$&$-0.417 \pm  0.071$\\  
Si II &1808&$15.796 \pm  0.005$\\  
Ti II &1910&$12.768 \pm  0.040$&$12.768 \pm  0.040$&$-0.825 \pm  0.081$\\  
Cr II &2056&$13.627 \pm  0.009$&$13.638 \pm  0.007$&$-0.685 \pm  0.071$\\  
Cr II &2066&$13.665 \pm  0.013$\\  
Fe II &1608&$>15.376$&$15.296 \pm  0.018$&$-0.857 \pm  0.073$\\  
Fe II &1611&$15.296 \pm  0.018$\\  
Co II &1574&$12.960 \pm  0.139$&$12.960 \pm  0.140$&$-0.603 \pm  0.157$\\  
Co II &1941&$<12.814$\\  
Co II &2012&$<12.832$\\  
Ni II &1370&$14.154 \pm  0.022$&$14.232 \pm  0.005$&$-0.671 \pm  0.071$\\  
Ni II &1703&$13.807 \pm  0.116$\\  
Ni II &1709&$14.221 \pm  0.009$\\  
Ni II &1741&$14.239 \pm  0.006$\\  
Ni II &1751&$14.266 \pm  0.010$\\  
Zn II &2026&$12.914 \pm  0.009$&$12.914 \pm  0.009$&$-0.409 \pm  0.072$\\  
\tableline
\end{tabular}
\end{center}
\end{table}

\begin{table}[ht]\footnotesize
\begin{center}
\caption{ {\sc
IONIC COLUMN DENSITIES: Q2206-19, $z = 2.076$ \label{tab:Q2206-19_2.076}}}
\begin{tabular}{lcccc}
\tableline
\tableline
Ion & $\lambda$ & AODM & $N_{\rm adopt}$ & [X/H] \\
\tableline
HI &1215 & $20.431  \pm 0.060  $ \\
C  II &1334&$>14.207$&$>14.207$&$>-2.774$\\  
C  II &1335&$<13.157$\\  
C  IV &1548&$13.707 \pm  0.005$\\  
C  IV &1550&$13.739 \pm  0.008$\\  
O  I  &1302&$>14.540$&$>14.540$&$>-2.761$\\  
Al II &1670&$12.158 \pm  0.012$&$12.158 \pm  0.012$&$-2.763 \pm  0.061$\\  
Al III&1854&$11.515 \pm  0.098$\\  
Al III&1862&$11.719 \pm  0.103$\\  
Si II &1304&$13.682 \pm  0.035$&$13.682 \pm  0.035$&$-2.309 \pm  0.069$\\  
Si IV &1402&$12.845 \pm  0.016$\\  
Cr II &2056&$<11.911$&$<11.911$&$<-2.190$\\  
Cr II &2062&$<12.158$\\  
Fe II &1608&$13.325 \pm  0.017$&$13.325 \pm  0.017$&$-2.606 \pm  0.062$\\  
Fe II &1611&$<13.948$\\  
Ni II &1709&$<12.585$&$<12.585$&$<-2.096$\\  
Ni II &1751&$<12.591$\\  
Zn II &2026&$<11.199$&$<11.199$&$<-1.902$\\  
\tableline
\end{tabular}
\end{center}
\end{table}

\subsection{Q2206$-$19, $z$ = 1.920 and $z$ = 2.076}

In Figures~\ref{fig:2206a} and \ref{fig:2206b}
we show a number of 
transitions left unanalysed by \cite{pro97a} and PW99 for the 
two damped \lya systems toward Q2206--19.  Furthermore, we now
consider only ionic column densities measured with the apparent optical
depth method in order to coincide with the rest of the database.
As we showed in \cite{pro97a}, there is very little difference between the
abundances derived from a Voigt profile analysis and the AODM.
All of the values are listed in 
Tables~\ref{tab:Q2206-19_1.920} and \ref{tab:Q2206-19_2.076}.

\break

\begin{figure}[ht]
\begin{center}
\includegraphics[height=4.0in, width=3.0in]{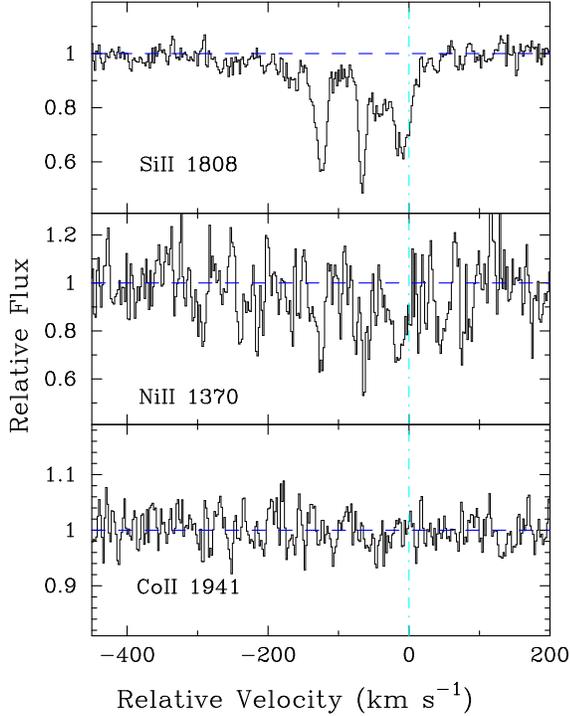}
\caption{Velocity plot of CII$^*$ 1335 transition for the
damped \lya system at $z = 1.864$ toward Q2230+02.
For comparison, we plot the Si~II 1304 and C~II 1334 profiles.
The vertical line at $v=0$ corresponds to $z = 1.864388.$}
\label{fig:2230}
\end{center}
\end{figure}

\subsection{Q2230$+$02, $z$ = 1.864}

This system was extensively analysed in PW99.  We simply add a limit
on $\N{Co^+}$ from the Co~II 1941 transition and a measurement
for Ni~II 1370 (Figure~\ref{fig:2230}).
We also include the new values for $\N{Fe^+}$ as the extensive wavelength
coverage provides a comparison between the Fe~II 1611, 2249, and 2260
transitions.  All three values are in good agreement which indicates the
relative $f$-values are reasonably accurate.

\begin{table}[ht]\footnotesize
\begin{center}
\caption{ {\sc
IONIC COLUMN DENSITIES: Q2230+02, $z = 1.864$ \label{tab:Q2230+02_1.864}}}
\begin{tabular}{lcccc}
\tableline
\tableline
Ion & $\lambda$ & AODM & $N_{\rm adopt}$ & [X/H] \\
\tableline
HI &1215 & $20.850  \pm 0.084  $ \\
Fe II &1608&$>15.160$&$15.184 \pm  0.016$&$-1.166 \pm  0.086$\\  
Fe II &1611&$15.148 \pm  0.084$\\  
Fe II &2249&$15.119 \pm  0.036$\\  
Fe II &2260&$15.210 \pm  0.019$\\  
Fe II &2344&$>15.039$\\  
Fe II &2374&$>15.213$\\  
Fe II &2382&$>14.744$\\  
Co II &1941&$<13.118$&$<13.118$&$<-0.642$\\  
Ni II &1370&$14.161 \pm  0.052$&$14.128 \pm  0.011$&$-0.972 \pm  0.085$\\  
Ni II &1709&$14.171 \pm  0.014$\\  
Ni II &1741&$14.097 \pm  0.023$\\  
Ni II &1751&$14.049 \pm  0.028$\\  
\tableline
\end{tabular}
\end{center}
\end{table}

\begin{figure}[ht]
\begin{center}
\includegraphics[height=4.0in, width=3.0in]{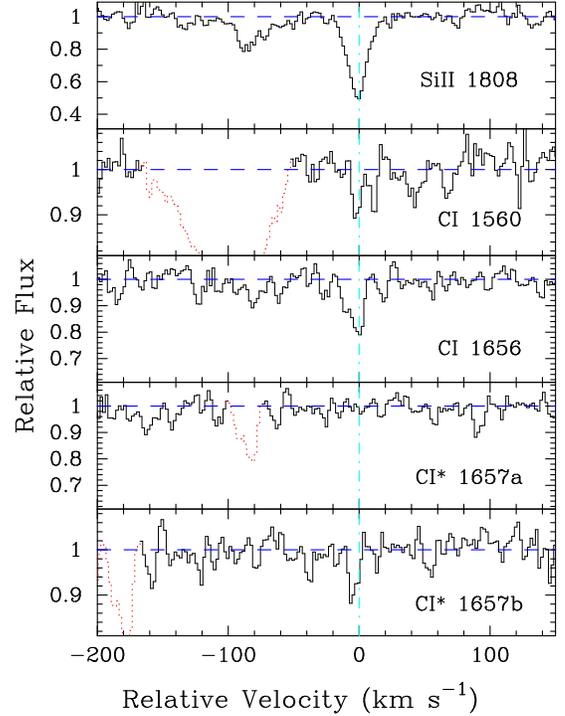}
\caption{Velocity plot of the new metal-line transitions for the 
damped \lya system at $z = 2.066$ toward Q2231--00.
For comparison, we also plot the Si~II 1808 profile.
The vertical line at $v=0$ corresponds to $z = 2.066150.$}
\label{fig:2231}
\end{center}
\end{figure}

\subsection{Q2231$-$00, $z$ = 2.066}

This damped system was analysed in PW99.  At the time we considered possible
absorption from the C~I 1656 and C~I$^*$ 1657 transitions but
were unconvinced that the profiles were associated with the damped
\lya system. Figure~\ref{fig:2231} presents the two transitions and 
the Si~II 1808 profile for comparison. 
We are now reasonably confident that these profiles arise in the damped
\lya system and their relative strengths place constraints
on the temperature of the CMB at this redshift \citep{pro01c}.

\begin{table}[ht]\footnotesize
\begin{center}
\caption{ {\sc
IONIC COLUMN DENSITIES: Q2231-002, $z = 2.066$ \label{tab:Q2231-002_2.066}}}
\begin{tabular}{lcccc}
\tableline
\tableline
Ion & $\lambda$ & AODM & $N_{\rm adopt}$ & [X/H] \\
\tableline
HI &1215 & $20.560  \pm 0.100  $ \\
C  I  &1656&$12.701 \pm  0.035$\\  
C  I  &1657&$12.662 \pm  0.108$\\  
Co II &1941&$<12.816$&$<12.816$&$<-0.654$\\  
\tableline
\end{tabular}
\end{center}
\end{table}

\begin{figure*}
\begin{center}
\includegraphics[height=8.5in, width=6.0in]{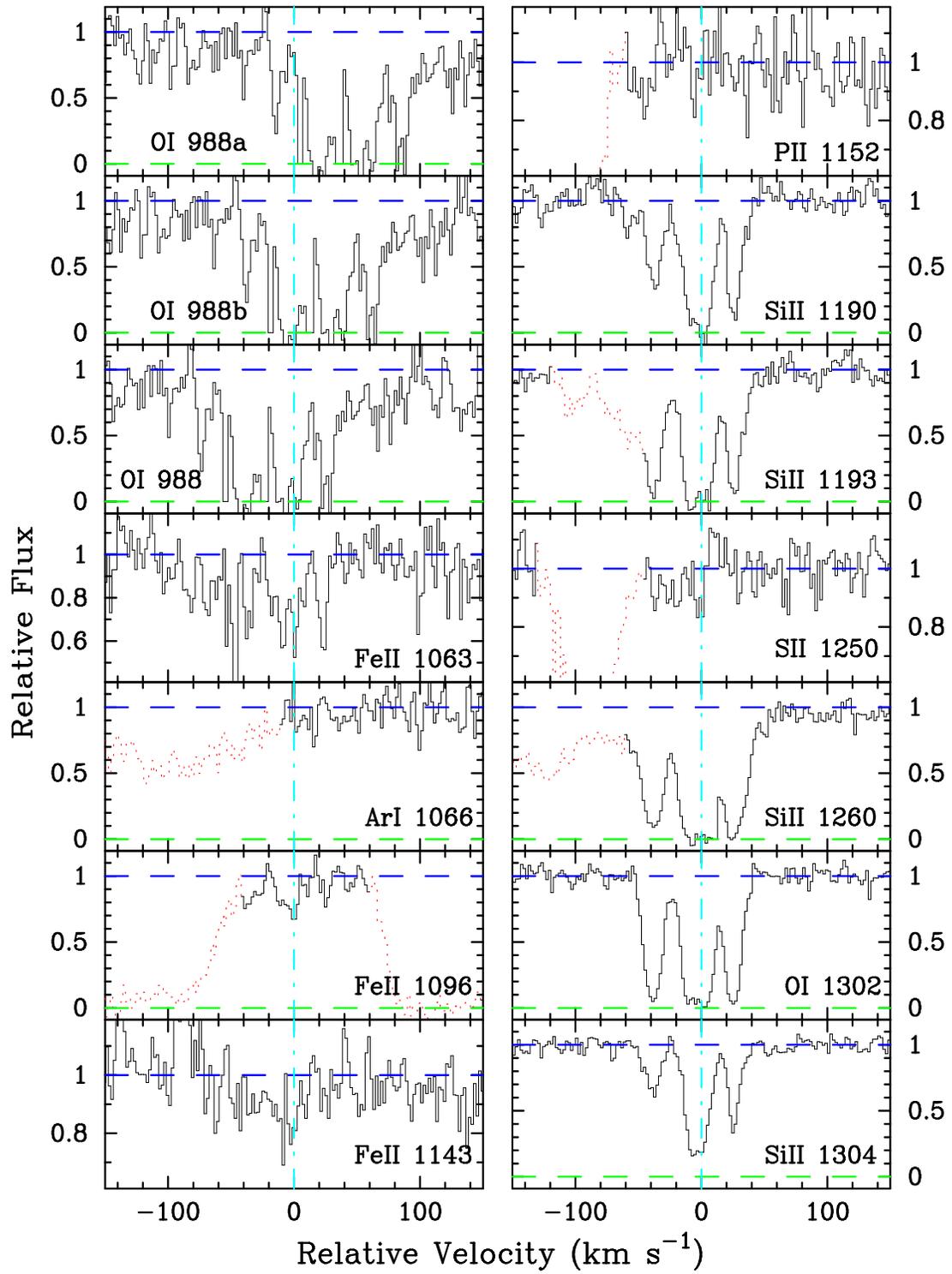}
\caption{Velocity plot of the metal-line transitions for the 
damped \lya system at $z = 2.538$ toward Q2344+12.
The vertical line at $v=0$ corresponds to $z = 2.53790$.}
\label{fig:2344}
\end{center}
\end{figure*}

\begin{figure}[ht]
\begin{center}
\includegraphics[height=4.5in, width=3.5in]{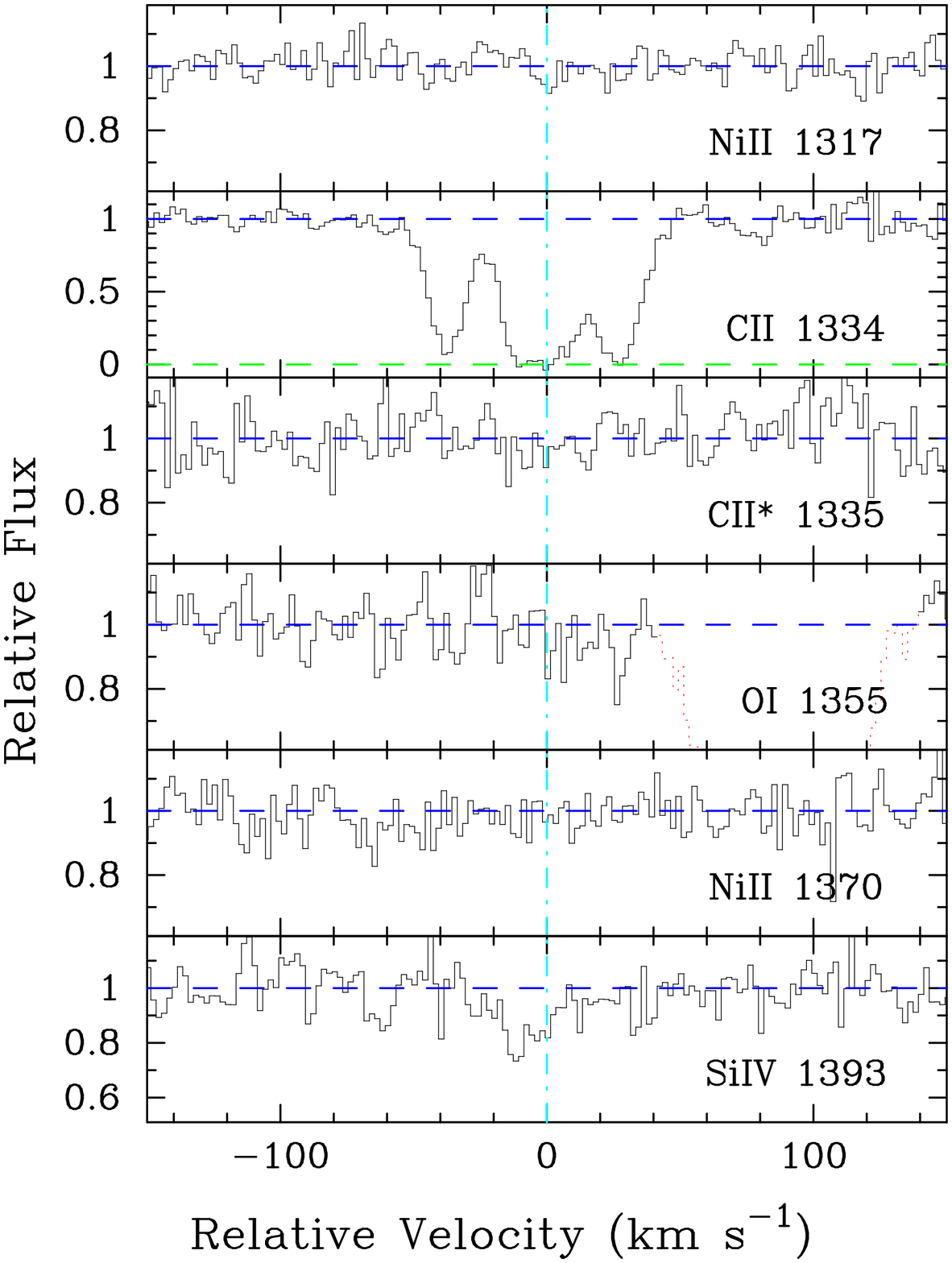}
Figure 34 -- cont
\end{center}
\end{figure}

\subsection{Q2344+12, $z$ = 2.538}

This system has been observed previously by \cite{lu99} and they
presented an [Fe/H] metallicity and $\N{HI}$ value.  We adopt their
measurement of the HI column density and have an independent measurement
of $\N{Fe^+}$ from several FUV Fe~II transitions.  In addition, our
blue spectra cover a number of transitions in the \lya forest 
(Figure~\ref{fig:2344}, Table~\ref{tab:Q2344+12_2.538}).

\begin{table}[ht]\footnotesize
\begin{center}
\caption{ {\sc
IONIC COLUMN DENSITIES: Q2344+12, $z = 2.538$ \label{tab:Q2344+12_2.538}}}
\begin{tabular}{lcccc}
\tableline
\tableline
Ion & $\lambda$ & AODM & $N_{\rm adopt}$ & [X/H] \\
\tableline
HI &1215 & $20.360  \pm 0.100  $ \\
C  II &1334&$>14.646$&$>14.645$&$>-2.265$\\  
C  II &1335&$<12.831$\\  
O  I  & 988&$>15.031$&$>15.031$&$>-2.199$\\  
O  I  &1302&$>15.020$\\  
O  I  &1355&$<17.814$\\  
Si II &1190&$>14.131$&$14.179 \pm  0.012$&$-1.741 \pm  0.101$\\  
Si II &1193&$>14.007$\\  
Si II &1260&$>13.838$\\  
Si II &1304&$14.179 \pm  0.012$\\  
Si IV &1393&$12.569 \pm  0.087$\\  
P  II &1152&$<12.744$&$<12.744$&$<-1.146$\\  
S  II &1250&$<14.201$&$<14.201$&$<-1.359$\\  
Ar I  &1066&$<13.262$&$<13.262$&$<-1.618$\\  
Fe II &1063&$14.021 \pm  0.046$&$14.030 \pm  0.032$&$-1.830 \pm  0.105$\\  
Fe II &1096&$14.007 \pm  0.053$\\  
Fe II &1143&$14.147 \pm  0.077$\\  
Ni II &1317&$<12.814$&$<12.814$&$<-1.796$\\  
Ni II &1370&$<12.999$\\  
\tableline
\end{tabular}
\end{center}
\end{table}

\begin{figure*}
\begin{center}
\includegraphics[height=8.5in, width=6.0in]{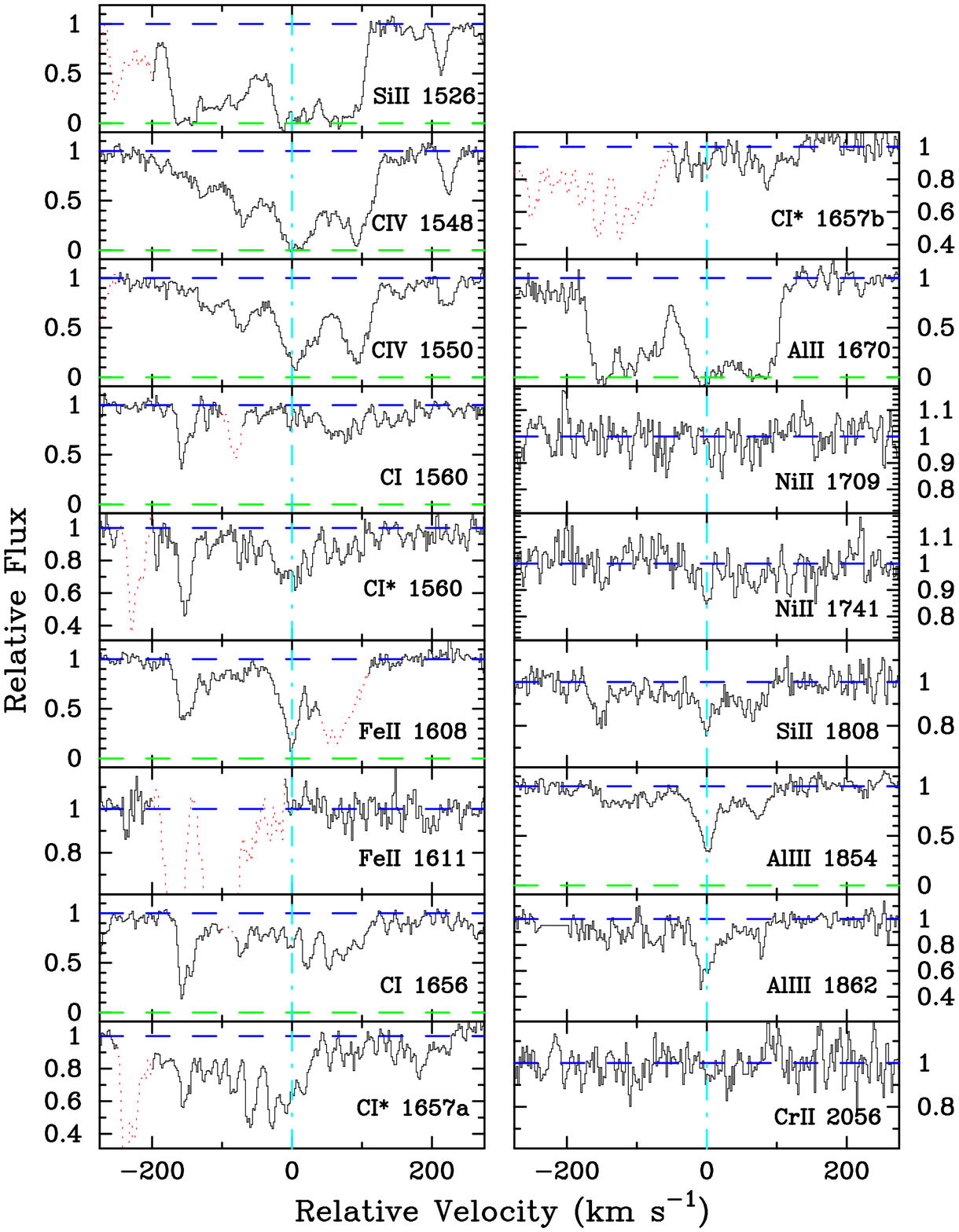}
\caption{Velocity plot of the metal-line transitions for the 
damped \lya system at $z = 2.426$ toward Q2348--01.
The vertical line at $v=0$ corresponds to $z = 2.426301$.}
\label{fig:2348-01a}
\end{center}
\end{figure*}

\begin{figure*}
\begin{center}
\includegraphics[height=8.5in, width=6.0in]{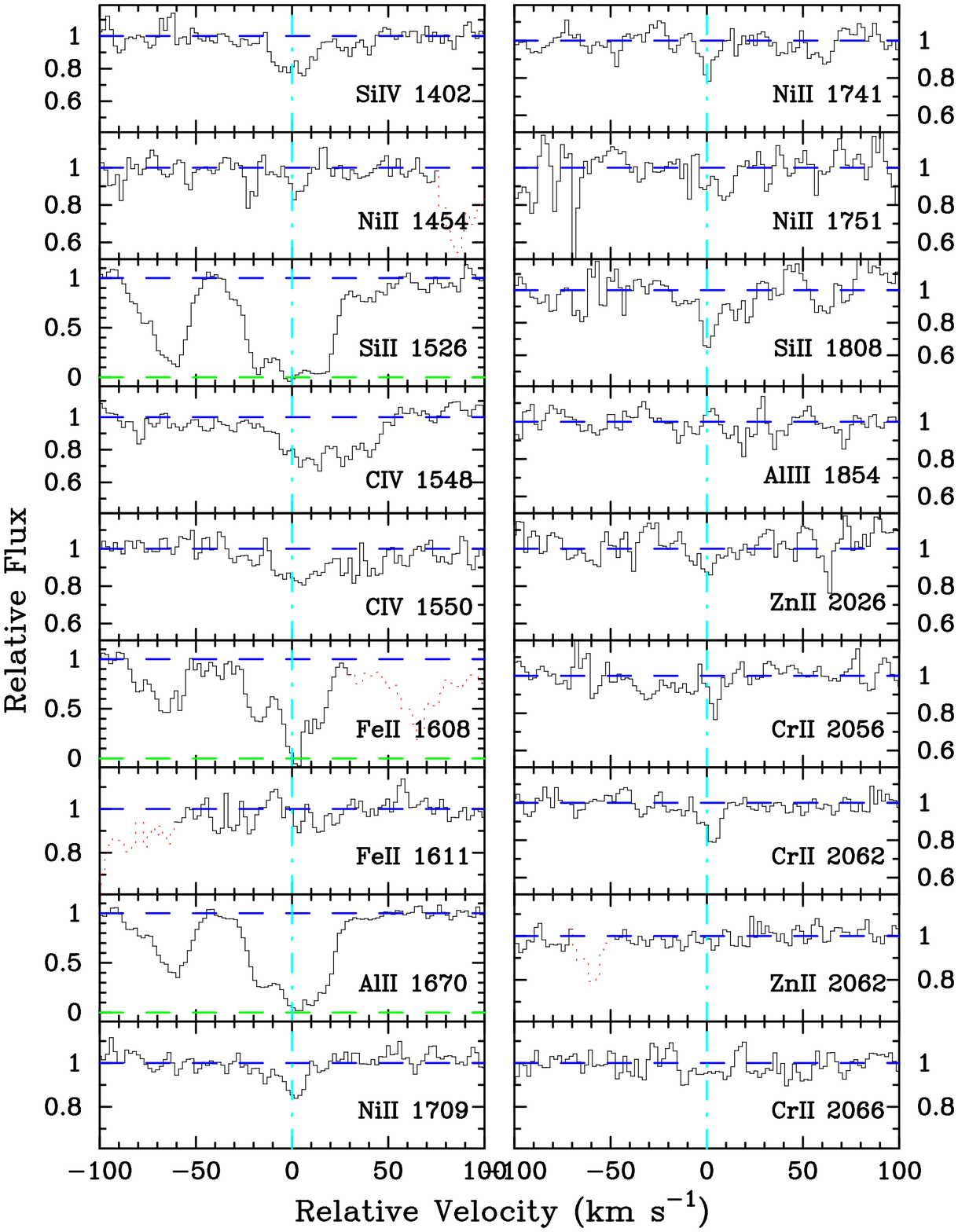}
\caption{Velocity plot of the metal-line transitions for the 
damped \lya system at $z = 2.615$ toward Q2348--01.
The vertical line at $v=0$ corresponds to $z = 2.614714$.}
\label{fig:2348-01b}
\end{center}
\end{figure*}

\subsection{Q2348$-$01, $z$ = 2.426 and $z$ = 2.615}

The two damped systems along this sightline were first identified by
\cite{tshk89} 
and are both part of the LBQS statistical sample.  This quasar is very
faint and the S/N of our 4.5hr spectrum is relatively poor.  
Figures~\ref{fig:2348-01a},\ref{fig:2348-01b} and 
Tables~\ref{tab:Q2348-01_2.426},\ref{tab:Q2348-01_2.615} 
present the transitions and 
column densities for the two systems.  With respect to the system at
$z=2.426$, the Fe~II 1608 profile is blended at $v>40$~km/s and we
estimate the $\N{Fe^+}$ value by integrating this profile at $v<40$~km/s.
Therefore the value is strictly a lower limit, although the Fe~II 1611
indicates that the column density at $v>40$~km/s is less than 
$10^{14.46} \cm{-2}$.
The system at $z=2.426$ is special for showing absorption from 
C~I and C~I$^*$.  In a companion paper, we analyse these transitions
to place a limit on the temperature of the 
cosmic microwave background radiation at $z=2.4$ \citep{pro01c}.
The system at $z=2.615$
is notable for exhibiting a very low metallicity 
([Fe/H]~$\sim -2.2$, [Ni/H]~$\sim -2.5$).  
In fact, this is the only system with
$\N{HI} > 10^{21} \cm{-2}$ which also has a metallicity less than 1/100 solar.
For the Fe$^+$ column density, we have averaged the lower and upper
limits established by Fe~II 1608 and 1611 respectively.

\begin{table}[ht]\footnotesize
\begin{center}
\caption{ {\sc
IONIC COLUMN DENSITIES: Q2348-01, $z = 2.426$ \label{tab:Q2348-01_2.426}}}
\begin{tabular}{lcccc}
\tableline
\tableline
Ion & $\lambda$ & AODM & $N_{\rm adopt}$ & [X/H] \\
\tableline
HI &1215 & $20.500  \pm 0.100  $ \\
C  IV &1548&$>14.705$\\  
C  IV &1550&$14.767 \pm  0.008$\\  
Al II &1670&$>13.939$&$>13.939$&$>-1.051$\\  
Al III&1854&$13.379 \pm  0.011$\\  
Al III&1862&$13.515 \pm  0.021$\\  
Si II &1526&$>15.160$&$15.365 \pm  0.022$&$-0.695 \pm  0.102$\\  
Si II &1808&$15.365 \pm  0.022$\\  
Cr II &2056&$<12.713$&$<12.713$&$<-1.457$\\  
Fe II &1608&$14.614 \pm  0.012$&$14.614 \pm  0.012$&$-1.386 \pm  0.101$\\  
Ni II &1709&$13.434 \pm  0.109$&$13.350 \pm  0.104$&$-1.400 \pm  0.144$\\  
Ni II &1741&$13.350 \pm  0.104$\\  
\tableline
\end{tabular}
\end{center}
\end{table}

\begin{table}[ht]\footnotesize
\begin{center}
\caption{ {\sc
IONIC COLUMN DENSITIES: Q2348-01, $z = 2.615$ \label{tab:Q2348-01_2.615}}}
\begin{tabular}{lcccc}
\tableline
\tableline
Ion & $\lambda$ & AODM & $N_{\rm adopt}$ & [X/H] \\
\tableline
HI &1215 & $21.300  \pm 0.100  $ \\
C  IV &1548&$13.291 \pm  0.024$\\  
C  IV &1550&$13.336 \pm  0.046$\\  
Al II &1670&$>13.139$&$>13.139$&$>-2.651$\\  
Al III&1854&$<12.203$\\  
Si II &1526&$>14.562$&$14.892 \pm  0.072$&$-1.968 \pm  0.123$\\  
Si II &1808&$14.892 \pm  0.072$\\  
Si IV &1402&$12.899 \pm  0.051$\\  
Cr II &2056&$12.619 \pm  0.100$&$12.674 \pm  0.060$&$-2.296 \pm  0.117$\\  
Cr II &2062&$12.718 \pm  0.075$\\  
Fe II &1608&$>14.483$&$14.573 \pm  0.088$&$-2.227 \pm  0.133$\\  
Fe II &1611&$<14.663$\\  
Ni II &1454&$13.121 \pm  0.110$&$13.193 \pm  0.074$&$-2.357 \pm  0.124$\\  
Ni II &1709&$13.296 \pm  0.097$\\  
Ni II &1741&$<13.100$\\  
Zn II &2026&$<11.871$&$<11.871$&$<-2.099$\\  
\tableline
\end{tabular}
\end{center}
\end{table}

\clearpage

\begin{figure}[ht]
\begin{center}
\includegraphics[height=4.0in, width=3.0in]{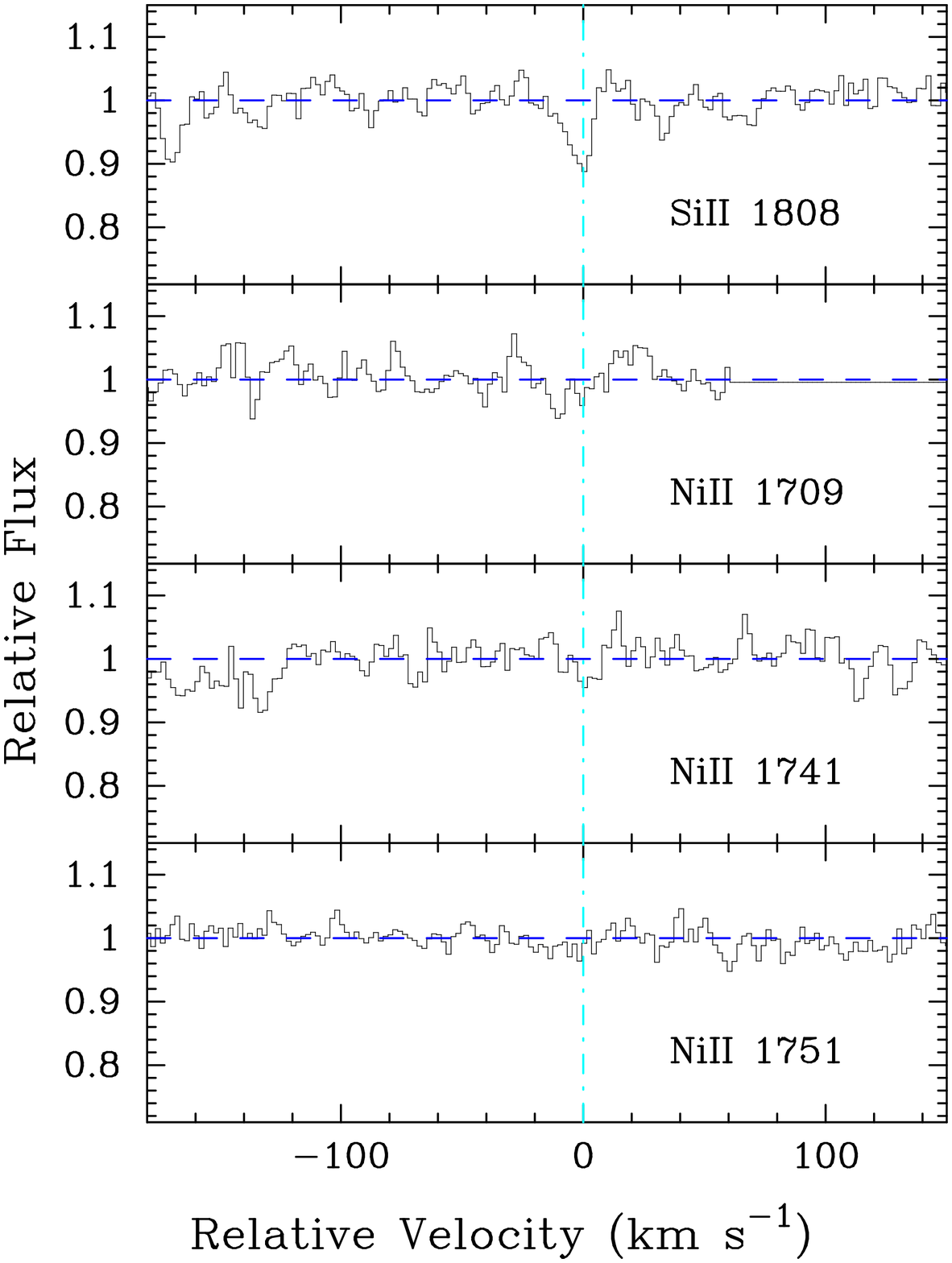}
\caption{Velocity plot of the new metal-line transitions for the 
damped \lya system at $z = 2.066$ toward Q2348--14.
For comparison, we also plot the Si~II 1808 profile.
The vertical line at $v=0$ corresponds to $z = 2.066150.$}
\label{fig:2348-14}
\end{center}
\end{figure}

\subsection{Q2348$-$14, $z$ = 2.279}

The abundances for this damped system were first measured by \cite{ptt95}
and subsequently by PW99.  We now include a limit on $\N{Ni^+}$ from
the non-detection of two Ni~II transitions (Figure~\ref{fig:2348-14},
Table~\ref{tab:Q2348-14_2.279}).

\begin{table}[ht]\footnotesize
\begin{center}
\caption{ {\sc
IONIC COLUMN DENSITIES: Q2348-14, $z = 2.279$ \label{tab:Q2348-14_2.279}}}
\begin{tabular}{lcccc}
\tableline
\tableline
Ion & $\lambda$ & AODM & $N_{\rm adopt}$ & [X/H] \\
\tableline
HI &1215 & $20.560  \pm 0.075  $ \\
Ni II &1709&$<12.752$&$<12.583$&$<-2.227$\\  
Ni II &1741&$<12.583$\\  
Ni II &1751&$<12.704$\\  
\tableline
\end{tabular}
\end{center}
\end{table}

\begin{figure}[ht]
\begin{center}
\includegraphics[height=4.3in, width=3.3in]{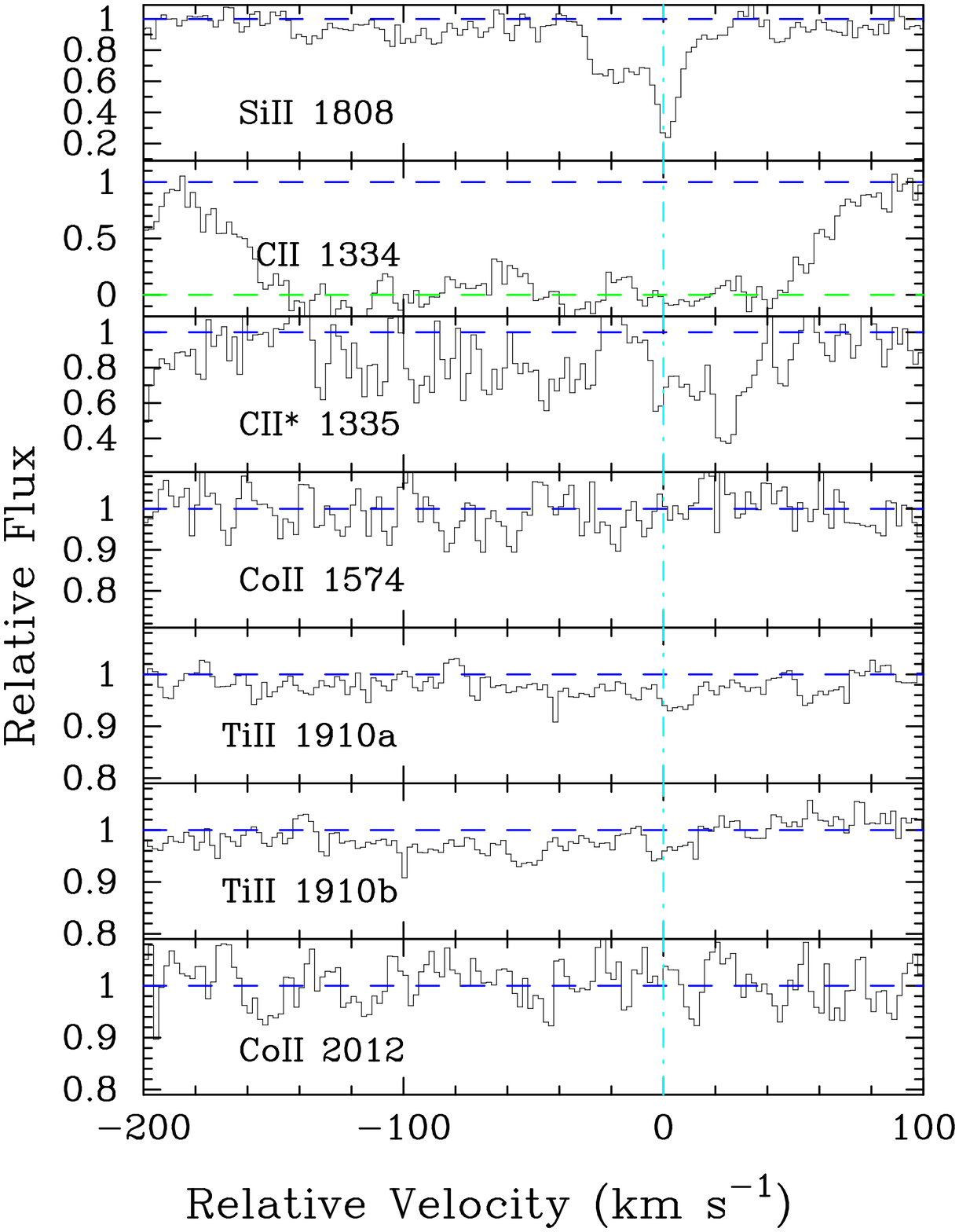}
\caption{Velocity plot of the new metal-line transitions for the 
damped \lya system at $z = 2.095$ toward Q2359--02.
For comparison, we also plot the Si~II 1808 profile.
The vertical line at $v=0$ corresponds to $z = 2.095067.$}
\label{fig:2359a}
\end{center}
\end{figure}

\begin{figure}[ht]
\begin{center}
\includegraphics[height=4.3in, width=3.3in]{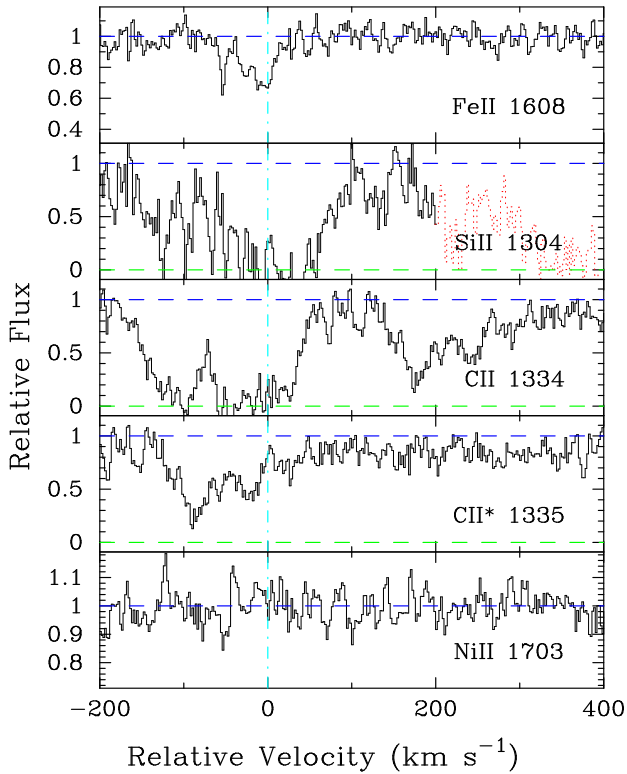}
\caption{Velocity plot of the new metal-line transitions for the 
damped \lya system at $z = 2.154$ toward Q2359--02.
For comparison, we also plot the Fe~II 1608 profile.
The vertical line at $v=0$ corresponds to $z = 2.153934.$}
\label{fig:2359b}
\end{center}
\end{figure}

\subsection{Q2359$-$02, $z$=2.095 \& $z$=2.154}

Although these two systems were analysed in PW99, we have now 
identified several new profiles
including measurements for C~II$^*$ 1335 for the two systems.
The C~II$^*$ column density for the $z=2.154$ is very high and
as we discuss in \cite{wp01} may indicate a high star formation rate
in this damped \lya system.  The
C~II$^*$ profile is within the \lya forest, however, and 
may be significantly blended with a \lya forest cloud.
We also report a tentative measurement
of Ti~II 1910 for the system at $z=2.095$.  Given the
very large Zn/Fe ratio for this system, the enhanced Ti/Fe ratio is
strongly suggestive of Type~II SN enrichment.
Figures~\ref{fig:2359a},\ref{fig:2359b} present all of the new transitions
and Tables~\ref{tab:Q2359-02_2.095},\ref{tab:Q2359-02_2.154} 
list the column densities.

\begin{table}[ht]\footnotesize
\begin{center}
\caption{ {\sc
IONIC COLUMN DENSITIES: Q2359-02, $z = 2.095$ \label{tab:Q2359-02_2.095}}}
\begin{tabular}{lcccc}
\tableline
\tableline
Ion & $\lambda$ & AODM & $N_{\rm adopt}$ & [X/H] \\
\tableline
HI &1215 & $20.700  \pm 0.100  $ \\
C  II &1334&$>15.147$&$>15.147$&$>-2.103$\\  
C  II &1335&$13.704 \pm  0.061$\\  
Ti II &1910&$12.330 \pm  0.055$&$12.330 \pm  0.055$&$-1.310 \pm  0.114$\\  
Co II &1574&$<13.398$&$<12.828$&$<-0.782$\\  
Co II &2012&$<12.828$\\  
\tableline
\end{tabular}
\end{center}
\end{table}

\begin{table}[ht]\footnotesize
\begin{center}
\caption{ {\sc
IONIC COLUMN DENSITIES: Q2359-02, $z = 2.154$ \label{tab:Q2359-02_2.154}}}
\begin{tabular}{lcccc}
\tableline
\tableline
Ion & $\lambda$ & AODM & $N_{\rm adopt}$ & [X/H] \\
\tableline
HI &1215 & $20.300  \pm 0.100  $ \\
C  II &1334&$>14.991$&$>14.991$&$>-1.859$\\  
C  II &1335&$14.475 \pm  0.020$\\  
Si II &1304&$>15.098$&$14.277 \pm  0.015$&$-1.583 \pm  0.101$\\  
Ni II &1703&$<14.171$&$<13.208$&$<-1.342$\\  
\tableline
\end{tabular}
\end{center}
\end{table}

\break

\section{SUMMARY}

Tables~\ref{tab:XHsum} and \ref{tab:XFesum} present a summary of the
absolute and relative abundances of the \ndla\ damped \lya systems
in our complete database.  Regarding Table~\ref{tab:XFesum} where
we present abundances relative to Fe, in a few cases we have 
considered Ni, Cr, or Al as a proxy for Fe, as noted.

We have presented ionic column density measurements for our complete
sample of damped \lya systems.  With the exception of a few important
transitions which exhibit blends with other profiles, we have measured
each column density with the apparent optical depth method.  In 
general, therefore,
all of the data has been reduced and analysed with an identical
approach.  We
have used the most up to date atomic data and will continue to update
the database as new information becomes available.  Visit 
http://kingpin.ucsd.edu/$\sim$hiresdla for tables, figures and updated
measurements.  A series of companion papers, in particular Paper~II
present new scientific results based on this database.

\acknowledgments

The authors wish to extend special thanks to those of Hawaiian ancestry 
on whose sacred mountain we are privileged to be guests.  Without 
their generous hospitality, none of the observations presented 
herein would have been possible.
We thank T. Barlow for providing the HIRES reduction package.
We also thank Jim Lawler and Chris Howk for helpful discussions.
We acknowledge the Keck support staff for their efforts
in performing these observations.  We thank R. Carswell and J. Webb
for providing the VPFIT software package.
J.X.P. acknowledges support from a
Carnegie postdoctoral fellowship. AMW was partially supported by 
NASA grant NAGW-2119 and NSF grant AST 86-9420443.

\begin{sidewaystable*} \footnotesize
\begin{center}
\caption{
{\sc ABUNDANCE SUMMARY \label{tab:XHsum}}}
\begin{tabular}{lccrrrrrrrrrrrrrr}
\tableline
\tableline \tskip
Name & $z_{abs}$ & $\N{HI}$ & [C/H] & [O/H] & [Al/H] & [Si/H]& [P/H] & [S/H] 
& [Ar/H] & [Ti/H] & [Cr/H] & [Fe/H] & [Co/H] & [Ni/H] & [Zn/H] \\
\tableline
Q0000-2619&3.390&21.41&&&&$-1.884$&&&&&&$-2.160$&&$-2.295$&\\  
BR0019-15&3.439&20.92&&&&$-1.057$&&&&&&$>-1.631$&&$-1.487$&\\  
PH957&2.309&21.40&&$<-0.596$&&$>-2.241$&&&&$<-2.133$&$-1.686$&$-1.929$&$<-1.146$&$-1.778$&$-1.623$\\  
Q0149+33&2.141&20.50&$>-2.406$&$< 0.428$&$>-2.017$&$-1.488$&&&&$<-1.271$&$-1.450$&$-1.770$&&$-1.581$&$-1.674$\\  
Q0201+36&2.463&20.38&&&$>-0.737$&$-0.406$&&&&$<-1.124$&$-0.802$&$-0.870$&$<-0.333$&$-0.608$&$-0.286$\\  
J0255+00&3.253&20.70&&&$>-1.311$&$-0.937$&&&&$<-0.835$&&$-1.436$&$<-0.352$&$-1.342$&\\  
J0255+00&3.915&21.30&$>-3.118$&$>-3.003$&&$>-2.667$&&$-1.779$&&&&$-2.050$&$<-0.998$&$-2.279$&\\  
Q0336-01&3.062&21.20&$>-2.792$&$>-1.130$&&$>-1.619$&$-1.597$&$-1.406$&$-1.374$&&&$-1.795$&&$-1.981$&\\  
Q0347-38&3.025&20.80&$>-2.285$&$>-1.717$&$>-1.882$&$-1.343$&&$<-1.240$&$-1.038$&&&$-1.797$&$<-0.514$&$-1.667$&\\  
Q0458-02&2.040&21.65&&$>-3.110$&$>-2.377$&$>-1.167$&&&&$<-2.095$&$-1.501$&$-1.767$&$<-1.467$&$-1.719$&$-1.186$\\  
HS0741+4741&3.017&20.48&$>-2.166$&$>-1.639$&$-2.146$&$-1.686$&$<-1.930$&$-1.680$&$-1.834$&&&$-1.928$&$<-0.432$&$-1.972$&\\  
Q0836+11&2.465&20.58&$>-2.104$&$>-1.965$&&$-1.153$&&$<-1.120$&&$<-0.982$&$<-1.352$&$-1.403$&$<-0.107$&$-1.442$&$<-1.131$\\  
Q0841+12&2.375&20.95&&&$>-2.105$&$-1.271$&&&&&$-1.548$&&$<-0.870$&$-1.677$&$-1.507$\\  
Q0841+12&2.476&20.78&&&$>-2.054$&$>-1.845$&&&&$<-1.562$&$-1.608$&$-1.750$&$<-0.964$&$-1.675$&$-1.401$\\  
BRI0951-04&3.857&20.60&&&$-1.814$&$>-1.535$&&&&&&$-1.997$&$< 0.087$&$<-1.873$&\\  
BRI0951-04&4.203&20.40&&$>-2.674$&&$-2.618$&&&&&&$<-2.591$&&$-1.730$&\\  
BRI0952-01&4.024&20.55&$>-1.788$&&&&&&&&&$-1.863$&$< 0.290$&$<-1.361$&\\  
PSS0957+33&3.279&20.32&&&$>-1.488$&$-1.000$&&&&&&$-1.453$&$< 0.055$&$-1.252$&$<-0.863$\\  
PSS0957+33&4.178&20.50&&$>-2.026$&$>-1.734$&$-1.504$&&$-1.308$&&&&$-1.871$&$< 0.238$&$<-1.840$&\\  
BRI1108-07&3.608&20.50&$>-2.375$&$>-2.497$&$-2.168$&$-1.798$&&&&&&$-2.116$&&$<-1.614$&\\  
Q1210+17&1.892&20.60&&&$>-1.650$&$-0.875$&&&&&$-1.027$&$-1.149$&$<-0.782$&$-1.218$&$-0.900$\\  
Q1215+33&1.999&20.95&$>-2.870$&$>-2.693$&$>-2.039$&$-1.480$&&&&&$-1.516$&$-1.702$&$<-1.000$&$-1.606$&$-1.290$\\  
Q1223+17&2.466&21.50&$>-2.895$&$>-2.893$&&$-1.592$&$<-1.147$&&&$<-2.188$&$-1.677$&$-1.843$&$<-1.779$&$-1.801$&$-1.620$\\  
Q1331+17&1.776&21.18&&&$>-1.927$&$-1.451$&&&&$-2.280$&$-1.972$&$-2.058$&$<-1.780$&$-1.890$&$-1.304$\\  
BRI1346-03&3.736&20.72&$>-2.784$&$>-2.571$&$-2.634$&$-2.332$&&&$<-2.127$&&&$<-2.094$&$<-0.370$&$<-2.210$&\\  
PSS1443+27&4.224&20.80&$>-1.738$&$>-1.622$&$>-1.332$&$>-0.926$&&&&&&$-1.096$&$<-0.201$&$-0.959$&\\  
Q1759+75&2.625&20.80&$>-2.050$&$>-1.409$&&$-0.824$&$>-1.283$&$-0.757$&$<-1.606$&&$-1.259$&$-1.209$&$<-0.691$&$-1.248$&$>-1.782$\\  
Q1946+7658&2.844&20.27&&$-2.321$&&$-2.228$&&$<-1.979$&&&&$-2.532$&&&\\  
Q2206-19&1.920&20.65&&&$>-1.073$&$-0.417$&&&&$-0.825$&$-0.685$&$-0.857$&$-0.603$&$-0.671$&$-0.409$\\  
Q2206-19&2.076&20.43&$>-2.774$&$>-2.761$&$-2.763$&$-2.309$&&&&&$<-2.190$&$-2.606$&&$<-2.096$&$<-1.902$\\  
Q2230+02&1.864&20.85&&&&$-0.754$&&&&$-1.175$&$-1.117$&$-1.166$&$<-0.642$&$-0.972$&$-0.720$\\  
Q2231-002&2.066&20.56&&$>-1.877$&&$-0.873$&&&&$-1.022$&$-1.065$&$-1.402$&$<-0.654$&$-1.206$&$-0.882$\\  
Q2344+12&2.538&20.36&$>-2.265$&$>-2.199$&&$-1.741$&$<-1.146$&$<-1.359$&$<-1.618$&&&$-1.830$&&$<-1.796$&\\  
Q2348-01&2.426&20.50&&&$>-1.051$&$-0.695$&&&&&$<-1.457$&$-1.386$&&$-1.400$&\\  
Q2348-01&2.615&21.30&&&$>-2.651$&$-1.968$&&&&&$-2.296$&$-2.227$&&$-2.357$&$<-2.099$\\  
Q2348-14&2.279&20.56&$>-2.459$&$>-2.581$&$-2.393$&$-1.917$&&$-2.035$&&&&$-2.238$&&$<-2.227$&\\  
Q2359-02&2.095&20.70&$>-2.103$&&$>-1.476$&$-0.777$&&&&$-1.310$&$-1.550$&$-1.655$&$<-0.782$&$-1.526$&$-0.775$\\  
Q2359-02&2.154&20.30&$>-1.859$&&$-1.625$&$-1.583$&&&&&$-1.368$&$-1.877$&&$<-1.342$&$<-1.069$\\  
\tskip \tableline
\end{tabular}
\end{center}
\end{sidewaystable*}

\begin{sidewaystable*} \footnotesize
\begin{center}
\caption{
{\sc RELATIVE ABUNDANCE SUMMARY \label{tab:XFesum}}}
\begin{tabular}{lccrrrrrrrrrrrrrr}
\tableline
\tableline \tskip
Name & $z_{abs}$ & $\N{HI}$ & [C/Fe] & [O/Fe] & [Al/Fe] & [Si/Fe] & [P/Fe] 
& [S/Fe] & [Ar/Fe] & [Ti/Fe] & [Cr/Fe] & [Co/Fe] & [Ni/Fe] & [Zn/Fe]\\
\tableline
Q0000-2619$^a$&3.390&21.41&&&&$+ 0.411$&&&&&&&&\\  
BR0019-15$^a$&3.439&20.92&&&&$+ 0.430$&&&&&&&&\\  
PH957&2.309&21.40&&$<+ 1.333$&&$>-0.312$&&&&$<-0.204$&$+ 0.243$&$<+ 0.783$&$+ 0.151$&$+ 0.306$\\  
Q0149+33&2.141&20.50&$>-0.636$&$<+ 2.198$&$>-0.247$&$+ 0.282$&&&&$<+ 0.499$&$+ 0.320$&&$+ 0.189$&$+ 0.096$\\  
Q0201+36&2.463&20.38&&&$>+ 0.133$&$+ 0.464$&&&&$<-0.254$&$+ 0.068$&$<+ 0.537$&$+ 0.262$&$+ 0.584$\\  
J0255+00&3.253&20.70&&&$>+ 0.125$&$+ 0.499$&&&&$<+ 0.601$&&$<+ 1.084$&$+ 0.094$&\\  
J0255+00&3.915&21.30&$>-1.068$&$>-0.953$&&$>-0.617$&&$+ 0.271$&&&&$<+ 1.052$&$-0.229$&\\  
Q0336-01&3.062&21.20&$>-0.997$&$>+ 0.665$&&$>+ 0.176$&$+ 0.198$&$+ 0.389$&$+ 0.421$&&&&$-0.186$&\\  
Q0347-38&3.025&20.80&$>-0.488$&$>+ 0.080$&$>-0.085$&$+ 0.454$&&$<+ 0.557$&$+ 0.759$&&&$<+ 1.283$&$+ 0.130$&\\  
Q0458-02&2.040&21.65&&$>-1.343$&$>-0.610$&$>+ 0.600$&&&&$<-0.328$&$+ 0.266$&$<+ 0.300$&$+ 0.048$&$+ 0.581$\\  
HS0741+4741&3.017&20.48&$>-0.238$&$>+ 0.289$&$-0.218$&$+ 0.242$&$<-0.002$&$+ 0.248$&$+ 0.094$&&&$<+ 1.496$&$-0.044$&\\  
Q0836+11&2.465&20.58&$>-0.701$&$>-0.562$&&$+ 0.250$&&$<+ 0.283$&&$<+ 0.421$&$<+ 0.051$&$<+ 1.296$&$-0.039$&$<+ 0.272$\\  
Q0841+12$^a$&2.375&20.95&&&$>-0.428$&$+ 0.406$&&&&&$+ 0.129$&$<+ 0.807$&&$+ 0.170$\\  
Q0841+12&2.476&20.78&&&$>-0.304$&$>-0.095$&&&&$<+ 0.188$&$+ 0.142$&$<+ 0.786$&$+ 0.075$&$+ 0.349$\\  
BRI0951-04&3.857&20.60&&&$+ 0.183$&$>+ 0.462$&&&&&&$<+ 2.084$&$<+ 0.124$&\\  
BRI0951-04&4.203&20.40\\  
BRI0952-01&4.024&20.55&$>+ 0.075$&&&&&&&&&$<+ 2.153$&$<+ 0.502$&\\  
PSS0957+33&3.279&20.32&&&$>-0.035$&$+ 0.453$&&&&&&$<+ 1.508$&$+ 0.201$&$<+ 0.590$\\  
PSS0957+33&4.178&20.50&&$>-0.155$&$>+ 0.137$&$+ 0.367$&&$+ 0.563$&&&&$<+ 2.109$&$<+ 0.031$&\\  
BRI1108-07&3.608&20.50&$>-0.259$&$>-0.381$&$-0.052$&$+ 0.318$&&&&&&&$<+ 0.502$&\\  
Q1210+17&1.892&20.60&&&$>-0.501$&$+ 0.274$&&&&&$+ 0.122$&$<+ 0.367$&$-0.069$&$+ 0.249$\\  
Q1215+33&1.999&20.95&$>-1.168$&$>-0.991$&$>-0.337$&$+ 0.222$&&&&&$+ 0.186$&$<+ 0.702$&$+ 0.096$&$+ 0.412$\\  
Q1223+17&2.466&21.50&$>-1.052$&$>-1.050$&&$+ 0.251$&$<+ 0.696$&&&$<-0.345$&$+ 0.166$&$<+ 0.064$&$+ 0.042$&$+ 0.223$\\  
Q1331+17&1.776&21.18&&&$>+ 0.131$&$+ 0.607$&&&&$-0.222$&$+ 0.086$&$<+ 0.278$&$+ 0.168$&$+ 0.754$\\  
BRI1346-03$^c$&3.736&20.72&$>-0.150$&$>+ 0.063$&&$+ 0.302$&&&$<+ 0.507$&&&$<+ 2.264$&$<+ 0.424$&\\  
PSS1443+27&4.224&20.80&$>-0.642$&$>-0.526$&$>-0.236$&$>+ 0.170$&&&&&&$<+ 0.895$&$+ 0.137$&\\  
Q1759+75&2.625&20.80&$>-0.841$&$>-0.200$&&$+ 0.385$&$>-0.074$&$+ 0.452$&$<-0.397$&&$-0.050$&$<+ 0.518$&$-0.039$&$>-0.573$\\  
Q1946+7658&2.844&20.27&&$+ 0.211$&&$+ 0.304$&&$<+ 0.553$&&&&&&\\  
Q2206-19&1.920&20.65&&&$>-0.216$&$+ 0.440$&&&&$+ 0.032$&$+ 0.172$&$+ 0.254$&$+ 0.186$&$+ 0.448$\\  
Q2206-19&2.076&20.43&$>-0.168$&$>-0.155$&$-0.157$&$+ 0.297$&&&&&$<+ 0.416$&&$<+ 0.510$&$<+ 0.704$\\  
Q2230+02&1.864&20.85&&&&$+ 0.412$&&&&$-0.009$&$+ 0.049$&$<+ 0.524$&$+ 0.194$&$+ 0.446$\\  
Q2231-002&2.066&20.56&&$>-0.475$&&$+ 0.529$&&&&$+ 0.380$&$+ 0.337$&$<+ 0.748$&$+ 0.196$&$+ 0.520$\\  
Q2344+12&2.538&20.36&$>-0.435$&$>-0.369$&&$+ 0.089$&$<+ 0.684$&$<+ 0.471$&$<+ 0.212$&&&&$<+ 0.034$&\\  
Q2348-01&2.426&20.50&&&$>+ 0.335$&$+ 0.691$&&&&&$<-0.071$&&$-0.014$&\\  
Q2348-01$^a$&2.615&21.30&&&$>-0.294$&$+ 0.389$&&&&&$+ 0.061$&&&$<+ 0.258$\\  
Q2348-14&2.279&20.56&$>-0.221$&$>-0.343$&$-0.155$&$+ 0.321$&&$+ 0.203$&&&&&$<+ 0.011$&\\  
Q2359-02&2.095&20.70&$>-0.448$&&$>+ 0.179$&$+ 0.878$&&&&$+ 0.345$&$+ 0.105$&$<+ 0.873$&$+ 0.129$&$+ 0.880$\\  
Q2359-02&2.154&20.30&$>+ 0.018$&&$+ 0.252$&$+ 0.294$&&&&&$+ 0.509$&&$<+ 0.535$&$<+ 0.808$\\  
\tskip \tableline
\end{tabular}
\end{center}
$^a$Ni is serving as a proxy for Fe \\
$^b$Cr is serving as a proxy for Fe \\
$^c$Al is serving as a proxy for Fe
\end{sidewaystable*}


\clearpage 

\begin{thebibliography}{}

\bibitem[Bergeson \& Lawler (1993)]{bergs93}      
Bergeson, S.D. \& Lawler, J.E. 1993, \apj, 408, 382 

\bibitem[Bergeson \& Lawler (1993b)]{bergs93b}      
Bergeson, S.D. \& Lawler, J.E. 1993, \apj, 414, L137

\bibitem[Bergeson, Mullman, \& Lawler (1994)]{bergs94}      
Bergeson, S.D., Mullman, K.L., \& Lawler, J.E. 1993, \apj, 435, L157

\bibitem[Bergeson et al.\ (1996)]{bergs96}      
Bergeson, S.D., Mullman, K.L.,  Wickliffe, M.W., Lawler, J.E.,  
Litzen, U., and Johansson, S. 1996, \apj 464, 1044







\bibitem[Djorgovski et al.\ (1998)]{djg98}		
Djorgovski, S.G., Gal, R.R., Odewahn, S.C., de Carvalho, R.R.,
Brunner, R., Longo, G., \& Scaramella, R. 1998, in ``Wide Field
Surveys in Cosmology'', eds.\ S. Colombi \& Y. Mellier, (astro-ph/9809187)



\bibitem[Ellison et al.\ (2001)]{ellison01}	
Ellison, S.L., Ryan, S., \& Prochaska, J.X. 2001, \mnras, in press



\bibitem[Fan et al.\ (1999)]{fan99}
Fan, X., SDSS collaboration, 1999, AJ, 118, 1 (SDSS Collaboration)

\bibitem[Fedchak, \& Lawler (1999)]{fedchak99}  
Fedchak, J. A. \& Lawler, J. E. 1999, \apj, 523, 734

\bibitem[Fedchak, Wiese, \& Lawler (2000)]{fedchak00}  
Fedchak, J. A., Wiese, L. M., \& Lawler, J. E. 2000, \apj, 538, 773


\bibitem[Grevesse et al.\ (1996)]{grvss96}	
Grevesse, N., Noels, A., \& Sauval, A.J. 1996, In: Cosmic Abundances,
S. Holt and G. Sonneborn (eds.), ASPCS, V. 99, (BookCrafters: San Fransisco),
p. 117


\bibitem[Hagen et al.\ (1999)]{hagen99}		
Hagen, H.J., Engels, D., \& Reimers, D. 1999, A\&AS, 134, 483


\bibitem[Howk, Savage, \& Fabian (1999)]{howk99}     
Howk, J.C., Savage, B.D., \& Fabian, D. 1999, \apj, 525, 253

\bibitem[Howk et al.\ (2000)]{howk00}		
Howk, J.C., Sembach, K.R., Roth, K.C., \& Kruk, J.W. 2000, \apj, 544, 867


\bibitem[Kirkman \& Tytler (1997)]{kirk97}   	
Kirkman, D. \& Tytler, D. 1997, \apj, 484, 672

\bibitem[Lanzetta et al.\ (1995)]{lzwt95}
Lanzetta, K. M., Wolfe, A. M.,\&  Turnshek
1995, \apj, 440, 435


\bibitem[Lu et al.\ (1996)]{lu96}
Lu, L., Sargent, W.L.W., Barlow, T.A.,
Churchill, C.W., \& Vogt, S. 1996, \apjsupp, 107, 475 (L96)

\bibitem[Lu, Sargent, \& Barlow (1999)]{lu99}   
Lu, L., Sargent, W.L.W., \& Barlow, T.A. 1999, 
{\it ASP Conference Series: 
  Highly Redshifted Radio Lines }, ed.\ C. Carilli, S. Radford,
  K. Menten, \& G. Langston (San Fransisco: ASP), p. XXX,
(astro-ph/9711298)  






\bibitem[Molaro et al.\ (2000)]{molaro00}	
Molaro, P., Bonifacio, P., Centuri${\rm \acute o}$n, M.,
D'Odorico, S., Vladilo, G., Santin, P., \& Di Marcantonio, P. 2000,
\apj, 541, 54

\bibitem[Molaro et al.\ (2001)]{molaro01}	
Molaro, P. et al.\ 2001, \apj, in press

\bibitem[Morton (1991)]{morton91}               
Morton, D.C. 1991, \apjsupp, 77, 119

\bibitem[Morton (2001)]{morton01}		
Morton, D. 2001, in prep

\bibitem[Mullman et al.\ (1998)]{mullman98}          
Mullman, K. L., Lawler, J. E., Zsargo, J., \& Federman, S. R.
1998, \apj, 500, 1064


\bibitem[Outram, Chaffee, \& Carswell (1999)]{outram99}	
Outram, P.J., Chaffee, F.H., \& Carswell, R.F.
1999, \mnras, 310, 289



\bibitem[Pettini et al.\ (1994)]{ptt94}
Pettini, M., Smith, L. J., Hunstead, R. W., and King,
D. L. 1994, \apj, 426, 79

\bibitem[Pettini et al.\ (1995)]{ptt95}		
Pettini, M., Lipman, K., \& Hunstead, R.W. 1995, \apj, 451, 100

\bibitem[Pettini et al.\ (1997)]{ptt97}
Pettini, M., Smith, L.J., King, D.L., \& Hunstead, R.W. 1997,
\apj, 486, 665


\bibitem[Pettini et al.\ (2000)]{ptt00}	
Pettini, M., Ellison, S.L., Steidel, C.C., Shapley, A.E., \& Bowen, D.V. 
2000, \apj, 532, 65


\bibitem[Prochaska \& Wolfe (1996)]{pro96}
Prochaska, J. X. \& Wolfe, A. M. 1996, \apj, 470, 403

\bibitem[Prochaska \& Wolfe (1997)]{pro97a}		
Prochaska, J. X. \& Wolfe, A. M. 1997, \apj, 474, 140

\bibitem[Prochaska \& Wolfe (1999)]{pro99}
Prochaska, J. X. \& Wolfe, A. M. 1999, \apjs, 121, 369 (PW99)
 
\bibitem[Prochaska \& Wolfe (2000)]{pro00}	
Prochaska, J.X. \& Wolfe, A.M., 2000, \apj, 533, L5

\bibitem[Prochaska et al.\ (2000)]{pro00b}      
Prochaska, J. X., Naumov, S.O., Carney, B.W., McWilliam, A., 
\& Wolfe, A.M. 2000, \aj, 120, 2513

\bibitem[Prochaska, Gawiser, \& Wolfe (2001)]{pro01}  
Prochaska, J.X., Gawiser, E., \& Wolfe, A.M. 2001, \apj, in press (PGW01)

\bibitem[Prochaska \& Wolfe (2001a)]{pro01a}  
Prochaska, J.X.  \& Wolfe, A.M. 2001, \apj, in press, (Paper II)

\bibitem[Prochaska \& Wolfe (2001b)]{pro01b}  
Prochaska, J.X.  \& Wolfe, A.M. 2001, \apj, in prep

\bibitem[Prochaska, O'Meara, \& Wolfe (2001c)]{pro01c}  
Prochaska, J.X., O'Meara, J.M.,  \& Wolfe, A.M. 2001, in prep

\bibitem[Raassen \& Uylings (1998)]{raassen98}  
Raassen, A.J.J. \& Uylings, P.H.M. 1998, \aap, 340, 300

\bibitem[Savage and Sembach (1991)]{sav91}
Savage, B. D. and Sembach, K. R. 1991, \apj, 379, 245



\bibitem[Schectman et al.\ (1998)]{schect98}  		
Schectman, R.M., Povolny, H.S., \& Curtis, L.J. 1998, \apj, 504, 921




\bibitem[Storrie-Lombardi et al.\ (1996)]{storr96}
Storrie-Lombardi, L.J., Irwin, M.J. 1996, \&
McMahon, R.G. \mnras, 282, 1330

\bibitem[Storrie-Lombardi and Wolfe (2000)]{storr00}  
Storrie-Lombardi, L.J. \& Wolfe, A.M. 2000, \apj, 543, 552



\bibitem[Tripp et al.\ (1996)]{trp96}
Tripp, T. M., Lu L., \& Savage B.D. 1996, \apjsupp, 102, 239

\bibitem[Turnshek et al.\ (1989)]{tshk89}       
Turnshek, D.A., Wolfe, A.M., Lanzetta, K.M., Briggs, F.H.,
Cohen, R.D., Foltz, C.B., Smith, H.E., \& Wilkes, B.J. 1989, \apj, 344, 567

\bibitem[Verner et al.\ (1994)]{verner94}       
Verner, D. A., Barthel, P. D., Tytler, D. 1994, A\&AS, 108, 287

\bibitem[Verner (1996)]{verner96}       
Verner, D. A. 1996, Atomic Data, Nuc. Data Tables, 64, 1



\bibitem[Vogt et al.\ (1994)]{vogt94}		
Vogt, S.S., Allen, S.L., Bigelow, B.C., Bresee, L., Brown, B., et al.\ 1994,
SPIE, 2198, 362



\bibitem[Wiese, Fedchak, \& Lawler (2001)]{wiese01}  
Wiese, L.M., Fedchak, J. A., \& Lawler, J. E. 2001, \apj, 547, 1178

\bibitem[Wolfe et al.\ (1986)]{wol86}
Wolfe, A.M., Turnshek, D.A., Smith, H.E., \& Cohen, R.D.
1986, \apjs, 61, 249

\bibitem[Wolfe et al.\ (1994)]{wol94}       	
Wolfe, A. M., Fan, X-M., Tytler, D., Vogt, S. S., Keane, M. J.,
\&  and Lanzetta, K. M. 1994, \apj, 435, L101

\bibitem[Wolfe et al.\ (1995)]{wol95}
Wolfe, A. M., Lanzetta, K. M., Foltz, C. B., and
Chaffee, F. H. 1995, \apj, 454, 698

\bibitem[Wolfe et al.\ (2001)]{wolfe01}
Wolfe, A.M. et al.\ 2001, in prep


\bibitem[Wolfe \& Prochaska (2000)]{wol00a}	
Wolfe, A. M. \& Prochaska, J.X. 2000a, \apj, 545, 591

\bibitem[Wolfe, Prochaska, \& Gawiser (2001)]{wp01}	
Wolfe, A. M., Prochaska, J.X., \& Gawiser, E. 2001, in preparation



\end{thebibliography}
\end{document}